\documentclass[preprint, 3p, onecolumn]{elsarticle} 
\usepackage{pgfplots,tikz}
\usetikzlibrary{intersections}
\usepackage{amsmath,amssymb}
\usepackage{mathtools}
\usepackage{subfig}
\usepackage{rotating}
\usepackage{xspace}
\usetikzlibrary{fillbetween}
\usepackage{caption}
\usepackage{rotating}
\usepackage{booktabs}
\usepackage{multirow}
\usepackage{stackengine}
\usepackage{stmaryrd}
\usepackage{textcomp}
\usepackage{threeparttable}
\usepackage{csvsimple}
\usepackage{afterpage}
\usepackage{longtable}
\usepackage{pgfplotstable}
\usepackage{paralist}
\pgfplotsset{compat=1.13}
\usepackage{siunitx}
\usepackage{fmtcount}
\usepackage{datatool}
\usepackage{xstring}
\DTLsetseparator{;}
\DTLsetdelimiter{;}
\DTLloaddb[keys={Property,Type,SubType}]{cs}{CS.csv}
\newcounter{DA}
\newcounter{SDA}
\newcounter{SSPK}

\newcounter{SFT}
\newcounter{DAT}
\newcounter{FRSH}
\newcounter{RESP}
\newcounter{OSC}
\newcounter{SPK}
\newcounter{Total}
\newcounter{TotalDA}
\newcounter{TotalREL}
\newcounter{TotalDAS}
\newcounter{rownumbers}
\newcommand\rownumber{\stepcounter{rownumbers}\arabic{rownumbers}}

\newcounter{counterDA}
\newcounter{counterSpk}
\newcounter{counterOsc}
\newcounter{counterRFT}
\newcounter{counterOUSH}
\newcounter{counterRSH}
\newcommand{\rt}{rise time\xspace}
\newcommand{\ft}{fall time\xspace}
\newcommand{\osh}{overshoot\xspace}
\newcommand{\ush}{undershoot\xspace}
\newcommand{\signals}{$s_1$ plotted with a thick line (\veryThickLine) and $s_2$ 
plotted with a thin line (\thinLine)\xspace}
\newcommand{\signalss}{$s_1$ plotted with a continuous line (\frshF) and $s_2$ plotted with a dash-dotted 
line (\frshS)\xspace}

\newcommand{\so}{$s_1$ (\veryThickLine) \xspace}
\newcommand{\st}{$s_2$ (\thinLine) \xspace}

\newcommand{\veryThickLine}{\raisebox{2pt}{\tikz{\draw[very thick,line width = 0.9pt](0,0) -- (3mm,0);}}}

\newcommand{\thinLine}{\raisebox{2pt}{\tikz{\draw[line width = 0.2pt](0,0) -- (3mm,0);}}}

\newcommand{\frshF}{\raisebox{2pt}{\tikz{\draw[thick, line width = 0.5pt](0,0) -- (3mm,0);}}}

\newcommand{\frshS}{\raisebox{2pt}{\tikz{\draw[thick, line width = 0.5pt, dash dot](0,0) -- (3mm,0);}}}

\DeclareMathOperator{\abs}{abs}

\DeclareSIUnit \tu {tu}

\newcommand{\stl}{\emph{STL}\xspace}

\newcommand{\sfo}{\emph{SFO}\xspace}

\newcommand{\stlstar}{\emph{STL*}\xspace}

\usepackage[many]{tcolorbox}
\newtcbox{\logiclbl}[1]{enhanced,nobeforeafter,tcbox raise base,boxrule=0.4pt,top=0mm,bottom=0mm,
	right=0mm,left=4mm,arc=1pt,boxsep=2pt,before upper={\vphantom{dlg}},
	colframe=green!50!black,coltext=green!25!black,colback=green!10!white,
	overlay={\begin{tcbclipinterior}\fill[green!75!blue!50!white] (frame.south west)
			rectangle node[text=white,font=\sffamily\bfseries\tiny,rotate=90] {#1} ([xshift=4mm]frame.north 
			west);\end{tcbclipinterior}}}

\newdefinition{definition}{Definition}                  

\newcommand{\rev}[1]{\textcolor{black}{#1}}
\usepackage[sharp]{easylist}

\begin{document}
\begin{filecontents*}{CS.csv}
prop;type;subtype
The modulus of signal \emph{sat\_init\_angular\_velocity\_degree} shall be less than or equal to \SI{3}{\degree/s}                         ;FRSH;DA
After \SI{2000}{s}, the modulus of signal \emph{sat\_real\_angular\_velocity} shall be less than or equal to \SI{1.5}{\degree/s} ;FRSH;DA
The modulus of signal \emph{sat\_target\_attitude} shall be equal to 1 ;FRSH;DA
After \SI{2000}{s}, the modulus of signal \emph{sat\_target\_angular\_velocity} shall be less than or equal to \SI{1.5}{\degree/s} ;FRSH;DA
The modulus of signal \emph{sat\_estimated\_attitude} shall be equal to 1 ;FRSH;DA
After \SI{2000}{s}, the modulus of signal \emph{sat\_estimated\_angular\_velocity} shall be less than or equal to	\SI{1.5}{\degree/s} ;FRSH;DA
The modulus of signal \emph{sat\_angular\_velocity\_measured} shall be less than or equal to \SI{1.5}{\degree/s} ;FRSH;DA
The modulus of signal \emph{earth\_mag\_field\_in\_body\_measured} shall be less than or equal to \SI{60000} {\nano\tesla};FRSH;DA
The modulus of signal \emph{sun\_direction\_ECI} shall be equal to 1;FRSH;DA
The value of signal \emph{currentADCSMode} shall be equal to \emph{NMC}, \emph{NMF} or \emph{SM};DA;x
The value of signal \emph{pointing\_error\_above\_20} shall be equal to 0 or 1;DA;x
The value of signal \emph{pointing\_error\_under\_15} shall be equal to 0 or 1;DA;x
After \SI{2000}{s}, the modulus of signal \emph{sat\_target\_angular\_velocity\_safe\_spin\_mode} shall be	less than or equal to \SI{1.5}	{\degree/s} ;FRSH;DA
The modulus of signal \emph{RWs\_torque} shall be less than or equal to \SI{0.015}{\newton\meter};FRSH;DA
The value of signal \emph{RWs\_angular\_velocity} shall be equal to \SI{816.814}{\radian/s} ;DA;x
The elements sum of vector \emph{sun\_sensor\_availability} shall be at most 3 ;FRSH;DA
Starting from \SI{2000}{s}, the value of signal \emph{pointing\_error} shall be less than \ang{2} ;DAT;x
Between \SI{1500}{s} and \SI{2000}{s}, the value of signal \emph{RWs\_angular\_momentum} shall be less than \SI{0.35}{\newton\meter\s};DAT;x
At \SI{2000}{s} the value of signal \emph{pointing\_error} shall be between \ang{0} and $\delta$\si{\degree};DAT;x
At \SI{2000}{s}, the angular difference between signals \emph{q\_real} and \emph{q\_estimate\_attitude} shall be between \ang{0} and $\delta$\si{\degree};FRSH;DA
At \SI{2000}{s}, the angular difference between signals \emph{q\_target\_attitude} and	\emph{q\_estimate} shall be between \ang{0} and $\delta\si{\degree}$;FRSH;DA
The difference between signal \emph{sat\_estimated\_angular\_velocity} and signal \emph{sat\_real\_angular\_velocity} shall be between \SI{0}{\degree/s} and $\delta\si{\degree/s}$ ;FRSH;DA
If the value of signal \emph{not\_Eclipse} is equal to 0, then the value of signal	\emph{sun\_currents} shall be equal to 0;RESP;DA-DA
The difference between signal \emph{sat\_angular\_velocity\_measured} and signal \emph{sat\_real\_angular\_velocity} shall be between \SI{0}{\degree/s} and $\delta\si{\degree/s}$ ;FRSH;DA
If the value of signal \emph{pointing\_error\_under\_15} is equal to 1, then the value of signal \emph{pointing\_error\_above\_20} shall be different from 1;RESP;DA-DA
If the value of signal \emph{pointing\_error\_above\_20} is equal to 1, then the value of signal \emph{pointing\_error\_under\_15} shall be different from 1;RESP;DA-DA
The difference between signal \emph{RWs\_torque} and the derivative of signal \emph{RWs\_angular\_momentum} shall be equal to \SI{0}{\newton\meter} ;FRSH;DA
If the value of signal \emph{RWs\_command} is equal to 0, then the value of signal	\emph{RWs\_angular\_velocity} shall monotonically decrease to \SI{0}{\radian/s} within \SI{60}{s} ;RESP;DA-FT
If the value of signal \emph{RWs\_angular\_momentum} is greater than \SI{0.35}{\newton\meter\s}, then the value of signal \emph{RWs\_torque} shall be equal to \SI{0}{\newton\meter};RESP;DA-DA
If the value of signal \emph{currentADCSMode} is equal to \emph{NMC}, then the value of signal \emph{control\_error} shall be greater than or equal to \ang{10} ;RESP;DA-DA
If the value of signal \emph{control\_error} is less than \ang{10}, then the value of signal \emph{currentADCSMode} shall be equal to \emph{NMF};RESP;DA-DA
If the value of signal \emph{currentADCSMode} is equal to \emph{NMF}, then the value of signal \emph{control\_error} shall be less than or equal to \ang{15};RESP;DA-DA
If the value of signal \emph{currentADCSMode} is equal to \emph{NMF}, then if the value of signal \emph{RWs\_command} becomes greater than 0, then the value of signal \emph{pointing\_error} shall be less than \ang{2} within \SI{180}{s}   ;RESP;DA-DA-DA
If the value of signal \emph{currentADCSMode} is equal to \emph{NMF}, then if the value of signal \emph{RWs\_command} becomes greater than 0, then the value of signal \emph{control\_error} shall be less than \ang{0.5} within \SI{180}{s} ;RESP;DA-DA-DA
If the value of signal \emph{currentADCSMode} is equal to \emph{NMF}, then if the value of signal \emph{Not\_eclipse} becomes 1, then the value of signal \emph{knowledge\_error} shall be less than 1 within at most \SI{900}{s};RESP;DA-DA-DA
If the value of signal \emph{currentADCSMode} is equal to \emph{SM}, then if the value of signal \emph{RWs\_command} becomes greater than 0, then the value of signal \emph{RWs\_angular\_momentum} shall be less than \SI{0.25}{\newton\meter\s} within at most \SI{900}{s};RESP;DA-DA-DA
If the value of signal \emph{currentADCSMode} is equal to \emph{SM}, then the difference	between signal \emph{real\_Omega} and signal \emph{target\_Omega} shall be equal to 0  within at most \SI{10799}{s} ;RESP;DA-DA
If the value of signal \emph{not\_Eclipse} is equal to 1, then the value of signal \emph{sun\_angle} shall be less than \ang{45};RESP;DA-DA
If, starting from \SI{16200}{s}, the value of signal \emph{pointing\_error} goes below the pointing	accuracy threshold of \ang{2}, then in signal \emph{pointing\_error} there shall exist a spike with a maximum width of \SI{600}{s} in an interval of \SI{5400}{s} 	;RESP;DA-SPK
Between \SI{2000}{s} and \SI{7400}{s}, in signal \emph{pointing\_error} there shall exist a spike with a maximum width of \SI{20}{s};SPK;x
Between \SI{2000}{s} and \SI{7400}{s}, signal \emph{pointing\_error} shall exhibit oscillations with a period greater than or equal to \SI{0.01}{s};OSC;x
\end{filecontents*}

\begin{frontmatter}
\title{Signal-Based Properties of Cyber-Physical Systems: Taxonomy and Logic-based
  Characterization}

\author[1]{Chaima Boufaied\corref{cor1}}
\ead{chaima.boufaied@uni.lu}

\author{Maris Jukss\fnref{fn1}}
\ead{maris.jukss@gmail.com}

\author[1]{Domenico Bianculli}
\ead{domenico.bianculli@uni.lu}

\author[1,2]{Lionel Claude Briand}
\ead{lionel.briand@uni.lu}

\author[3]{Yago Isasi Parache}
\ead{Isasi@luxspace.lu}

\cortext[cor1]{Corresponding author}
\fntext[fn1]{This work was done while the author was
  affiliated with the Interdisciplinary Centre for Security, Reliability and
  Trust (SnT), University of Luxembourg, Luxembourg.}
\address[1]{Interdisciplinary Centre for Security, Reliability and
  Trust (SnT), University of Luxembourg, Luxembourg}
\address[2]{School of EECS, University of Ottawa, Canada}
\address[3]{LuxSpace Sàrl}

\begin{abstract}
The behavior of a cyber-physical system (CPS) is usually defined in terms
of the input and output signals processed by sensors
and actuators. Requirements specifications of CPSs
are typically expressed using signal-based temporal
properties. Expressing such requirements is challenging, because of
\begin{inparaenum}[(1)]
\item the many features that can be used to characterize a signal
  behavior;
\item the broad variation in expressiveness of the specification languages
  (i.e., temporal logics) used for defining
signal-based temporal properties.
\end{inparaenum}
Thus, system and software engineers need effective guidance on
selecting appropriate signal behavior types and an adequate specification language, based on the type of requirements
they have to define.

In this paper, we present a taxonomy of the various types of
signal-based properties and provide, for each type, a comprehensive
and detailed description as well as a formalization in a temporal
logic. Furthermore, we review the expressiveness of state-of-the-art
signal-based temporal logics in terms of the property types identified
in the taxonomy.  Moreover, we report on the application of our
taxonomy to classify the requirements specifications of an industrial
case study in the aerospace domain, in order to assess the feasibility
of using the property types included in our taxonomy and the
completeness of the latter.

 \end{abstract}


\end{frontmatter}

\DTLforeach*{cs}{\pt=Property,\ty=Type,\sty=SubType}{\IfStrEqCase{\ty}{{DA}{\stepcounter{DA}}{DAT}{\stepcounter{DAT}}{SPK}{\stepcounter{SPK}}{OSC}{\stepcounter{OSC}}{FRSH}{\stepcounter{FRSH}}{RESP}{\stepcounter{RESP}}}\IfStrEqCase{\sty}{{DA}{\stepcounter{SDA}}{DA-DA}{\stepcounter{SDA}\stepcounter{SDA}}{DA-DA-DA}{\stepcounter{SDA}\stepcounter{SDA}\stepcounter{SDA}}{DA-SPK}{\stepcounter{SSPK}\stepcounter{SDA}}{DA-FT}{\stepcounter{SFT}\stepcounter{SDA}}}}
\setcounter{Total}{\numexpr\value{DA}+\value{DAT}+\value{SPK}+\value{OSC}+\value{FRSH}+\value{RESP}\relax}
\setcounter{TotalDA}{\numexpr\value{DA}+\value{DAT}\relax}
\setcounter{TotalDAS}{\numexpr\value{DA}+\value{DAT}+\value{SDA}\relax}
\setcounter{TotalREL}{\numexpr\value{FRSH}+\value{RESP}\relax}

\section{Introduction}
\label{sec:introduction}

Cyber-physical systems (CPSs) are systems characterized by a complex
interweaving of hardware and
software~\cite{Lee:2016:IES:3086978}. They are widely used in many
safety-critical domains (e.g., aerospace, automotive, medical) where
validation and verification (V\&V)
activities~\cite{bartocci2018specification} of the system's intended
functionality play a crucial role to guarantee the reliability and safety of the system.

A typical CPS consists of a mix of analog and digital components, such
as sensors, actuators, and control units, which process input and
output signals. System engineers specify the desired system behavior
by defining requirements in terms of the signals obtained from these
components.  Such requirements can be specified using
\emph{signal-based temporal properties}, which characterize the expected
behavior of signals.  For example, a property may require that a
signal must not exhibit an abrupt increase of amplitude (i.e., a spike
or bump) within a certain time interval, or that the signal shall
manifest an oscillatory behavior with a particular period.

Expressing requirements in terms of signal-based temporal properties
poses a number of challenges for system and software engineers. First,
a signal behavior (e.g., a spike) can be characterized using a number
of features (e.g., amplitude, slope, width); for example, a total of
16 different features (and eight parameters) have been identified in
the literature~\cite{adam2014feature} to detect (and thus
characterize) a spike in a signal.  Engineers may decide to choose
various subsets of features; without proper guidelines for selecting
the features most appropriate in a certain context and without their
precise characterization, the resulting specification of a signal
behavior may become ambiguous or inconsistent.
The second challenge is related to the
expressiveness of the specification languages used for defining
signal-based temporal properties. Starting from the seminal work on
\stl~\cite{maler2004monitoring} (Signal Temporal Logic), there have
been several proposals of languages that extend more traditional
temporal logics like \textit{LTL} (Linear Temporal Logic) to support the specification of signal-based
behaviors. Such languages have different levels of expressiveness when
it comes to describing certain signal behaviors. For example, \stl
cannot be used to express properties (like those related to
oscillatory behaviors) that require to reference the concrete value of
a signal at an instant in which a certain property was
satisfied~\cite{brim2014stl}. This means that engineers need 
guidance to carefully select the language to use for defining
signal-based properties, based on the type of requirements they are
going to define, the expressiveness of the candidate specification
languages, and the availability of suitable tools (e.g., trace
checker) for each language.

We remark that these challenges for the specification of signal-based
temporal properties have implications also in terms of V\&V. The lack
of precise descriptions of signal behaviors (and their features) and
the use of specification languages with limited expressiveness, may
lead engineers to resort to manual checking (e.g., visual inspection
of signal waveforms) of properties on signals. Although an anomalous
spike in amplitude can be easily spotted by visual inspection of the
waveform of a signal that is mostly stable, manually detecting complex
signal behaviors on waveforms with intricate shapes is a cumbersome
and error-prone process.

In this paper, we tackle these two challenges by proposing a taxonomy
of the most common types of signal-based temporal properties and a
logic-based characterization of such properties. Based on industrial
experience and a thorough review of the literature, our goal is to
provide system and software engineers, as well as researchers working
on CPSs, with a reference guide to systematically identify and
characterize signal behaviors, to support both requirements
specification and V\&V activities.  More specifically, we address the
first challenge by providing, through the taxonomy, a comprehensive
and detailed description of the different types of signal-based
behaviors, with each property type precisely characterized in terms of
a temporal logic. As a result, an engineer can be guided by the
precise characterization of the property types included in our
taxonomy, to derive---from an informal requirements specification---a
formal specification of a property, which can then be used in the
context of V\&V activities (e.g., as test oracle).  We take on the
second challenge by reviewing the expressiveness of the main temporal
logics that have been proposed in the literature for specifying
signal-based temporal properties (i.e., \stl,
\stlstar~\cite{brim2014stl}, \sfo~\cite{bakhirkin2018first} \rev{-
  Signal First-Order Logic}), in terms of
the property types identified in the taxonomy.  In this way, we can
guide engineers to choose a specification formalism based on their
needs in terms of property types to express.

We developed our taxonomy of signal-based properties based on
practical experience in analyzing temporal requirements in CPS domains
like the aerospace industry, and by reviewing the literature in the
area of verification of cyber-physical systems, starting from the
recent survey of specification formalisms in
reference~\cite{bartocci2018specification}. We identified and included
in our taxonomy the following property types:
\begin{itemize}
\item \textit{Data assertion}, which specifies constraints on 
the signal value;
\item \textit{Signal behavior}, representing a signal
  behavior in terms of a particular waveform, such as spikes and oscillations;
\item \textit{Relationship between signals}, a type that includes
  \textit{functional relationship} properties, based on the
  application of a transformation (e.g., differentiation) on signals,
  and \textit{order relationship} properties, stating constraints on
  the order of events/states related to signal behaviors. The
  \textit{order relationship} type also includes properties describing
  the \emph{transient behavior} of a signal when changing from the
  current value to a new target value (i.e., rising/falling,
  overshooting/undershooting behaviors).
\end{itemize}

For each of these types, we provide a logic-based characterization
using \sfo and also discuss alternative formalizations---when
applicable---using also \stl and \stlstar. In this way, we are able to
report on the expressiveness of state-of-the-art temporal logics with respect to the property types included in
our taxonomy: \sfo is the only language 
among the three we considered in which we can express \emph{all the property types of our
taxonomy}.

We also report on the application of our taxonomy to classify the
requirements specifications of an industrial case study in the
aerospace domain. Through this case study we show:
 \begin{itemize}
 \item The feasibility of expressing requirements specifications of a
   real-world CPS using the property types included in our
   taxonomy. Indeed, in the vast majority of the cases, the mapping
   from a specification written in English to its corresponding
   property type defined in the taxonomy was straightforward.
\item The completeness of our taxonomy:   all requirements
  specifications of the case study could be defined using the property
  types included in our taxonomy.
\end{itemize}

To summarize, the main contributions of this paper are:
\begin{itemize}
\item a taxonomy of signal-based properties;
\item a logic-based characterization of the various property types
  included in the taxonomy;
\item a discussion on the expressiveness of state-of-the-art temporal
  logics with respect to the property types included in our taxonomy;
\item the application of our taxonomy to classify the requirements
  specifications of an industrial case study in the aerospace domain.

\end{itemize}

The rest of the paper is structured as
follows. Section~\ref{sec:background} provides background concepts on
signals and temporal logics for signal-based
properties. Section~\ref{sec:taxonomy} illustrates our taxonomy of
signal-based properties and provides a logic-based characterization of
each property type.  In section~\ref{sec:expr} we discuss the expressiveness of
state-of-the-art temporal logics with respect to the property types
included in our taxonomy. Section~\ref{sec:casestudy} presents the application of our taxonomy
to an industrial case study.
\rev{Section~\ref{sec:discussion} discusses how the paper contributions can support the research community and practitioners.}
Section~\ref{sec:related-work} discusses
related work. Section~\ref{sec:concl-future-work} concludes the paper,
providing directions for future work.

 \section{Background}
\label{sec:background}

\subsection{Signals}

A finite length signal $s$ over a domain $\mathbb{D}$ is a function
$s\colon \mathbb{T} \to \mathbb{D}$, where $\mathbb{T}$ is the time
domain and $\mathbb{D}$ is an application-dependent value domain.
In the context of CPSs, we need to differentiate between
\emph{analog}, \emph{discrete}, and \emph{digital}
signals~\cite{QuantSTL}.

An analog signal is a signal that is
continuous both in the time and in the value domains. The time domain
$\mathbb{T}$ of an analog signal is thus the set of non-negative real
numbers $\mathbb{R}_{\geq 0}$ and the value domain $\mathbb{D}$ is the
set of real numbers $\mathbb{R}$. More formally, we define an analog
signal $s_a$ as $s_a\colon \mathbb{T} \to \mathbb{R}$.
The domain of definition of $s_a$ is the interval $I_{s_a}=[0,r)$,
with $r\in \mathbb{Q}_{\geq 0}$; the length of $s_a$ is defined as
$\lvert s_a \rvert =r$;  undefined signal values are denoted by
$s_a(t)=\bot, \forall t\geq |s_a|$.

In a discrete signal, the value domain is continuous whereas the time
domain is the set of natural numbers $\mathbb{N}$. More specifically,
a discrete signal can be obtained from an analog signal through
\emph{sampling}, which is the process of converting the
continuous-time domain of a signal to a discrete-time domain.
Throughout this process, the analog signal is read at a regular time
interval $\Delta$ called the \emph{sampling interval}. The resulting
discretized signal $s_{\mathit{dsc}}$ can be represented by the values
of an analog signal $s_a$ read at the following time points:
$0,\Delta, 2 \times \Delta, \dots, k\times \Delta $.
A digital signal has the set of natural numbers $\mathbb{N}$ as time
domain and a finite discrete set as value domain. Such a signal can be
obtained from a discrete signal by \emph{quantization}, which is the
process of transforming continuous values into their finite discrete
approximations.  

In the rest of the paper we will consider analog signals, simply
denoted by $s$, unless a specific signal type is explicitly
mentioned. This choice is motivated by the context in which this work
has been developed, which is the domain of
CPS~\cite{ShivaSimulink}. In such a domain, model-driven engineering
is used throughout the development process and \emph{simulation} is
used for design-time testing of system models; simulation models
(e.g., those defined in
\emph{Simulink}\textsuperscript{\textregistered}) capture both
continuous and discrete system behaviors and, when executed, produce
traces containing analog signals~\cite{gonzalez2018enabling}.

\subsection{Temporal Logics for Signal-based Properties}
In this section, we provide a brief introduction to the main temporal
logics that have been proposed in the literature for specifying
signal-based temporal properties.  They will be used in the next
section to present the formalization of signal-based properties.

\subsubsection{Signal Temporal Logic (\stl)}\label{bkgd:stl} 

\stl~\cite{maler2004monitoring} has been one of the first proposals of
a temporal logic for the specification of temporal properties over
dense-time (i.e., $\mathbb{T}=\mathbb{R}_{\geq 0}$), real-valued
signals.

Let $\Pi$ be a finite set of atomic propositions, $X$ be a finite set
of real variables, and
$\mathcal{I}$ be an interval\footnote{The restriction on the non-punctual
  interval $\mathcal{I}$ for \stl has been lifted in
  reference~\cite{maler2013monitoring}.} $[a,b]$
over $\mathbb{R}$ with $a, b \in \mathbb{Q}_{\geq 0}$ such that
$0 \leq a < b$. The syntax of \stl with both \emph{future} and
\emph{past} operators~\cite{maler2013monitoring} is defined by the
following grammar: \[\varphi ::= p \mid x \sim c \mid \neg \varphi \mid \varphi_1 \lor \varphi_2  \mid
\varphi_1 \mathbin{\mathsf{U}_{\mathcal{I}}} \varphi_2 \mid
\varphi_1 \mathbin{\mathsf{S}_{\mathcal{I}}} \varphi_2\] where $p \in \Pi$, $x \in X$, $\sim \in\{<,\le,=,\ge,>\}$, $c \in
\mathbb{R}$, $\mathsf{U}_{\mathcal{I}}$ is the metric
``\emph{Until}'' operator, and $\mathsf{S}_{\mathcal{I}}$ is the metric
``\emph{Since}'' operator.
Additional temporal operators can be derived using the usual
conventions; for example, 
``\emph{Eventually}'' 
$ \mathsf{F}_{\mathcal{I}} \varphi \equiv \top \mathbin{\mathsf{U}_{\mathcal{I}}}  \varphi$;
``\emph{Globally}''
$ \mathsf{G}_{\mathcal{I}} \varphi \equiv \neg \mathsf{F}_{\mathcal{I}} \neg 
\varphi$;
``\emph{Once (Eventually in the Past)}'' 
$ \mathsf{P}_{\mathcal{I}} \varphi \equiv \top \mathbin{\mathsf{S}_{\mathcal{I}}}  \varphi$;
``\emph{Historically}''
$ \mathsf{H}_{\mathcal{I}} \varphi \equiv \neg \mathsf{P}_{\mathcal{I}} \neg
\varphi$.

The semantics of \stl is defined through a satisfaction relation
$(s,t) \models_{\mathit{STL}} \varphi$, which indicates that signal $s$ satisfies
formula $\varphi$ starting from position $t$ in the signal. The
satisfaction relation is defined inductively as follows:
\begin{equation*}
\begin{aligned}
  (s,t) \models_{\mathit{STL}} p & \text{ iff } p \text{ holds on } s
  \text{ in } t, \text{ for } p \in \Pi\\
     (s,t) \models_{\mathit{STL}} x \sim c & \text{ iff } x \sim c \text{ holds on } s
  \text{ in } t, \text{ for } x \in X ~\text{and}~ c \in \mathbb{R} \\
  (s,t) \models_{\mathit{STL}} \neg \varphi & \text{ iff } (s,t) \not
  \models_{\mathit{STL}} \varphi \\
  (s,t) \models_{\mathit{STL}}  \varphi_1 \lor \varphi_2 & \text{ iff
  } (s,t) \models_{\mathit{STL}}  \varphi_1  \text{ or } (s,t)
  \models_{\mathit{STL}}  \varphi_2\\
 (s,t) \models_{\mathit{STL}}  \varphi_1 \mathbin{\mathsf{U}_{[a,b]}} \varphi_2 & \text{ iff }  \exists
t^\prime.(t^\prime \in [t+a, t+b] \text{ and } (s,t^\prime) \models_{\mathit{STL}}  \varphi_2 \\ &\phantom{\text{ iff}}\text{ and } \forall
t^{\prime\prime}.(t^{\prime\prime}  \in [t,t^\prime]  \text{ and }
(s,t^{\prime\prime}) \models_{\mathit{STL}} \varphi_1 ))\\
 (s,t) \models_{\mathit{STL}}  \varphi_1 \mathbin{\mathsf{S}_{[a,b]}} \varphi_2 & \text{ iff }  \exists
t^\prime.(t^\prime \in [t-a, t-b] \text{ and } (s,t^\prime) \models_{\mathit{STL}}  \varphi_2 \\ &\phantom{\text{ iff}}\text{ and } \forall
t^{\prime\prime}.(t^{\prime\prime}  \in [t,t^\prime]  \text{ and }
(s,t^{\prime\prime}) \models_{\mathit{STL}} \varphi_1 ))\\
\end{aligned}
\end{equation*}
We say that a signal $s$ satisfies an \stl formula $\varphi$ iff $ (s,0) \models_{\mathit{STL}}  \varphi$.

Several extensions of \stl have been proposed in the literature. For
example, STL/PSL~\cite{nickovic2007:amt:-a-property} adds an analog
layer to STL that enables the application of (low-level) signal
operations; xSTL~\cite{nickovic2018amt} adds support for Timed
Regular Expressions~\cite{2002:TRE:506147.506151}. The \stl
expressions that we will present in the rest of the paper can be
written in the same form also in STL/PSL or xSTL since they only rely
on the core operators of \stl.

\subsubsection{\stlstar}\label{bkgd:stlstar}  

\stlstar~\cite{brim2014stl} is an extension of \stl that adds a
signal-value \emph{freezing} operator that binds the value of a signal
to a precise instant of time.

Let $\mathcal{J}$ be a finite index set (e.g., the set
$\{1, \dots, n\}, n \in \mathbb{N}$) and let the function
$t^{*}\colon \mathcal{J} \to [0,|s|]$ be the \emph{frozen time
  vector}; the i-th frozen time can then be referred to with
$t^{*}_i=t^{*}(i)$. As in the case of \stl, let $\Pi$ be a finite set
of atomic propositions, $X$ be a finite set
of real variables, and $\mathcal{I}$ be an interval $[a,b]$ over $\mathbb{R}$ with
$a, b \in \mathbb{Q}_{\geq 0}$ such that $0 \leq a < b$.  The syntax
of \stlstar is defined by the following grammar:\[\varphi ::= p \mid x \sim c \mid \neg \varphi \mid \varphi_1 \lor \varphi_2 \mid
  \varphi_1 \mathbin{\mathsf{U}_{\mathcal{I}}} \varphi_2 \mid *_i [\varphi] \] where $p \in \Pi$,
$x \in X$, $\sim \in\{<,\le,=,\ge,>\}$, $c \in
\mathbb{R}$, $\mathsf{U}_{\mathcal{I}}$ is the metric
``\emph{Until}'' operator, and $*_i$ is the unary signal-value freezing
operator for all $ i \in \mathcal{J}$. Additional operators like
\emph{Eventually} and \emph{Globally} can be defined as done above for \stl.

The semantics of \stlstar is defined through a satisfaction relation
$(s,t,t^{*}) \models_{\mathit{STL*}} \varphi$, which indicates that
signal $s$ satisfies
formula $\varphi$ starting from position $t$ in the signal, taking
into account the  frozen time vector $t^{*} \in [0,|s|]^{\mathcal{J}}$.
The
satisfaction relation is defined inductively as follows:
\begin{equation*}
\begin{aligned}
(s,t,t^{*}) \models_{\mathit{STL*}} p &\text{ iff } p \text{ holds on } s
\text{ in } t , \text{ for } p \in \Pi, \text{ with the frozen time vector } t^{*}\\
(s,t,t^{*}) \models_{\mathit{STL*}} x \sim c &\text{ iff }  x \sim c
\text{ holds on } s \text{ in } t, \text{for} ~x \in X ~\text{and}~ c
\in \mathbb{R},  \text{with the frozen time vector } t^{*}\\
(s,t,t^{*}) \models_{\mathit{STL*}} \neg \varphi &\text{ iff }  (s,t,t^{*}) \not 
\models_{\mathit{STL*}} 
\varphi \\ 
(s,t,t^{*}) \models_{\mathit{STL*}} \varphi_1 \lor \varphi_2 &\text{ iff }  
(s,t,t^{*})\models_{\mathit{STL*}}  \varphi_1 \text{ 
or } 
(s,t,t^{*}) \models_{\mathit{STL*}} 
\varphi_2 \\
(s,t,t^{*}) \models_{\mathit{STL*}} \varphi_1 \mathbin{\mathsf{U}_{\mathcal{I}}} \varphi_2 &\text{ iff } \exists 
t^\prime . ( t^\prime
\in [t+a, t+b] \text{ and } (s,t,t^{*}) \models_{\mathit{STL*}}
\varphi_2 \\
&\phantom{\text{ iff}} \text{ and } \forall 
t^{\prime\prime}.( t^{\prime\prime} \in 
[t,t^\prime] \text { and } (s,t^{\prime\prime}, t^{*}) \models_{\mathit{STL*}} \varphi_1)) \\ 
(s,t,t^{*}) \models_{\mathit{STL*}} *_{i}[\varphi] &\text{ iff } (s,t,t^{*}[i \leftarrow t]) 
\models_{\mathit{STL*}} \varphi 
\end{aligned}
\end{equation*}
where $[i \leftarrow t]$ is the operator substituting $t$ with the i-th position in the frozen time 
vector, defined as 
$
  t^*  [i \leftarrow t] =
  \begin{cases}
    t, & i=j\\
    t^*(j), & i \neq j\\
  \end{cases}
$.  

We say  that a signal $s$ satisfies the \stlstar formula $\varphi$ iff $(s,0,\mathbf{0})\models_{\mathit{STL*}} \varphi$.

\subsubsection{Signal First-Order Logic (\sfo)}
\label{bkgd:sfo}

\sfo~\cite{bakhirkin2018first} is a formalism that combines first order
logic with linear real arithmetic and uninterpreted unary function
symbols; the latter represent real-valued signals evolving over time.

Let $F$ be a set of function symbols and let $X=T \cup R$ be a set of variables,
where $T$ is the set of \emph{time} variables and $R$ is the set of
\emph{value} variables. Let $\Sigma=\langle f_1, f_2,\dots,\mathbb{Z},
-,+,< \rangle$ be a (first-order) signature where $f_1,f_2,\dots \in
F$ are uninterpreted unary function symbols, $\mathbb{Z}$ are integer 
constants, and $-,+,< $ are the standard arithmetic functions and order relation.
The syntax of \sfo over $\Sigma$ is defined by the following grammar:\[
\begin{aligned}  
\varphi &::= \theta_1 < \theta_2 \mid \neg \varphi \mid \varphi_1 \lor
\varphi_2 \mid \exists r \colon \varphi\mid \exists t \in \mathcal{I} \colon \varphi\\
\theta &::= \tau \mid \rho\\
\tau &::= t \mid n\mid \tau_1- \tau_2 \mid \tau_1+ \tau_2 \\ 
\rho &::=r \mid f(\tau) \mid n \mid \rho_1- \rho_2 \mid \rho_1+ \rho_2\\
\end{aligned}
\]  
where $r \in R$, $t \in T$, $n \in \mathbb{Z}$,  $f \in F$,  $\mathcal{I}$ is a
time interval with bounds in $\mathbb{Z} \cup \{\pm\infty\}$. Notice
that a term $\theta$ can be either a  time term $\tau$ or a value term
$\rho$. Additional logical connectors can be derived using the usual
conventions; for example, $\forall r \colon \varphi \equiv \neg
\exists r \colon \neg \varphi$.

Let a trace $\omega$ be an interpretation of a function symbol
$f \in F$ as a signal, denoted by $\llbracket f \rrbracket_{\omega}$;
let a valuation $v$ be an interpretation of a variable $x \in X$ as a
real number, denoted by $\llbracket x \rrbracket_{v}$. The valuation
function for a term $\theta$ over the trace $\omega$ and the valuation
$v$, denoted as $ \llbracket \theta \rrbracket_{\omega,v}$ is defined
inductively as follows: $ \llbracket x \rrbracket_{\omega,v} = \llbracket x \rrbracket_v$, $ \llbracket n \rrbracket_{\omega,v} = n \text{ for all } n \in
\mathbb{Z}$, $ \llbracket f(\tau) \rrbracket_{\omega,v} = \left\llbracket f\left(
    \left \llbracket \tau \right \rrbracket_{\omega,v}\right)
\right\rrbracket_\omega$, $ \llbracket \theta_1 - \theta_2 \rrbracket_{\omega,v} = \llbracket
\theta_1 \rrbracket_{\omega,v} - \llbracket \theta_2
\rrbracket_{\omega,v}$, $ \llbracket \theta_1 + \theta_2 \rrbracket_{\omega,v} = \llbracket
\theta_1 \rrbracket_{\omega,v} + \llbracket \theta_2
\rrbracket_{\omega,v}$. The semantics of \sfo is defined through a satisfaction relation
$ (\omega, v) \models_{\mathit{\sfo}} \varphi$, which indicates the
satisfaction of formula $\varphi$ over the trace $\omega$ and the
valuation $v$. The satisfaction relation is defined inductively as
follows:
\begin{equation*}
  \begin{aligned}
    (\omega,v) \models_{\mathit{SFO}} \theta_{1} < \theta_{2} & \text{ iff } \llbracket  \theta_{1} \rrbracket_{\omega,v} <    \llbracket   \theta_{2} \rrbracket_{\omega,v} \\(\omega,v) \models_{\mathit{SFO}} \neg \varphi &\text{ iff }
    (\omega,v) \not \models_{\mathit{SFO}} \neg \varphi\\
    (\omega,v) \models_{\mathit{SFO}}  \varphi_1 \lor \varphi_2
    &\text{ iff }
    (\omega,v)  \models_{\mathit{SFO}} \varphi_1 \lor (\omega,v)
    \models_{\mathit{SFO}} \varphi_2 \\
     (\omega,v)  \models_{\mathit{SFO}} \exists r \colon \varphi &
     \text{ iff } (\omega,v[r \leftarrow a]) \models_{\mathit{SFO}} 
     \varphi \text{ for some } a \in \mathbb{R}\\
       (\omega,v)  \models_{\mathit{SFO}} \exists t \in \mathcal{I} \colon \varphi &
     \text{ iff } (\omega,v[t \leftarrow a]) \models_{\mathit{SFO}} 
 \varphi \text{ for some } a \in \mathbb{R}\\
  \end{aligned}
\end{equation*}
Variants of \sfo can be defined by opportunely changing the underlying
signature $\Sigma$.

 \section{Taxonomy of signal-based properties}
\label{sec:taxonomy}

One of the main challenges in using signal-based temporal properties
for expressing requirements of CPSs is the lack of precise
descriptions of signal behaviors.  First, a signal behavior (e.g., a
spike or an oscillation) can be ``described'' in different ways, i.e.,
it can be characterized using various features; for example, a total
of 16 different features (and eight parameters) have been identified
in the literature~\cite{adam2014feature} to detect a spike in a
signal. Given the large variety of options, (software and system)
engineers may choose various subsets of features for characterizing
the same type of signal behavior, leading to ambiguity and
inconsistency in the specifications.  In addition, slightly different
features may have similar names (e.g., ``peak amplitude'' and
``peak-to-peak amplitude''), potentially leading to mistakes when
writing specifications.  It is then important to define proper
guidelines for selecting the features most appropriate in a certain
context, and provide engineers with a precise characterization of such
features.

In this section, we tackle this challenge by proposing a taxonomy
of the most common types of signal-based temporal properties and a
logic-based characterization of such properties.  Our goal is to
provide system and software engineers, as well as researchers working
on CPSs, with a reference guide to systematically identify and
characterize signal behaviors, so that they can be defined precisely
and used correctly during the development process of CPSs, in
particular during the activities related to requirements specification
and V\&V.

Our taxonomy provides a comprehensive and detailed description of the
different types of signal-based behaviors, with each property type
precisely characterized in terms of a temporal logic. As a result, an
engineer can be guided by the precise characterization of the property
types included in our taxonomy, to derive---from an informal
requirements specification---a formal specification of a property,
which can be used in other development activities (e.g., V\&V).

We developed this taxonomy based on our general understanding of
temporal requirements in CPS domains like the aerospace industry, and
by reviewing the literature in the area of verification of
cyber-physical systems, starting from the recent survey in
reference~\cite{bartocci2018specification}. The taxonomy focuses on
properties specified in the time domain; we purportedly leave out
properties specified in the frequency
domain~\cite{nguyen2017abnormal,donze2012temporal} because in our
context (V\&V of CPS) the properties of
interest are mainly specified in the time domain. 

\begin{figure}
	\centering
	\begin{tikzpicture}[
 DA/.style={sibling distance=3cm, level distance=2cm, shape=rectangle, align=center},
 SB/.style={sibling distance=2.1cm, level distance=2cm, shape=rectangle, align=center},
 RSH/.style={sibling distance=5cm, level distance=2cm, shape=rectangle, align=center},
 catRSH/.style={sibling distance=3cm, level distance=2cm, shape=rectangle, align=center},
 orderRSH/.style={sibling distance=2cm, level distance=2cm, shape=rectangle, align=center}
  ]
  \node [draw]{Signal-based property}
    child[DA] { node[draw] {Data Assertion \\ (DA)} }
    child[SB] { node[draw] {Signal Behavior \\ (SB)}
      child[SB] { node[draw]  {Spike \\ (SPK)}}
      child[SB] { node[draw]  {Oscillatory \\ behavior\\ (OSC)}}
     }
    child[RSH] {node[draw] {Relationship \\ between signals\\ (RSH)} 
        child[catRSH] { node[draw] {Order\\ (RSH-O)}
        	 child[SB] { node[draw]  {Transient\\ behavior}
        		child[SB] { node[draw]  {Rise time\\(RT)}}
        		child[SB] { node[draw]  {Fall time\\(FT)}}
        		child[SB] { node[draw]  {Overshoot\\(OSH)}}
        		child[SB] { node[draw]  {Undershoot\\(USH)}}
        	}
              }
        child[catRSH] { node[draw] {Functional\\ (RSH-F)}}      
     }
     ;
   \end{tikzpicture}

 	\caption{Taxonomy of signal-based properties}
	\label{fig:taxonomy}
      \end{figure}
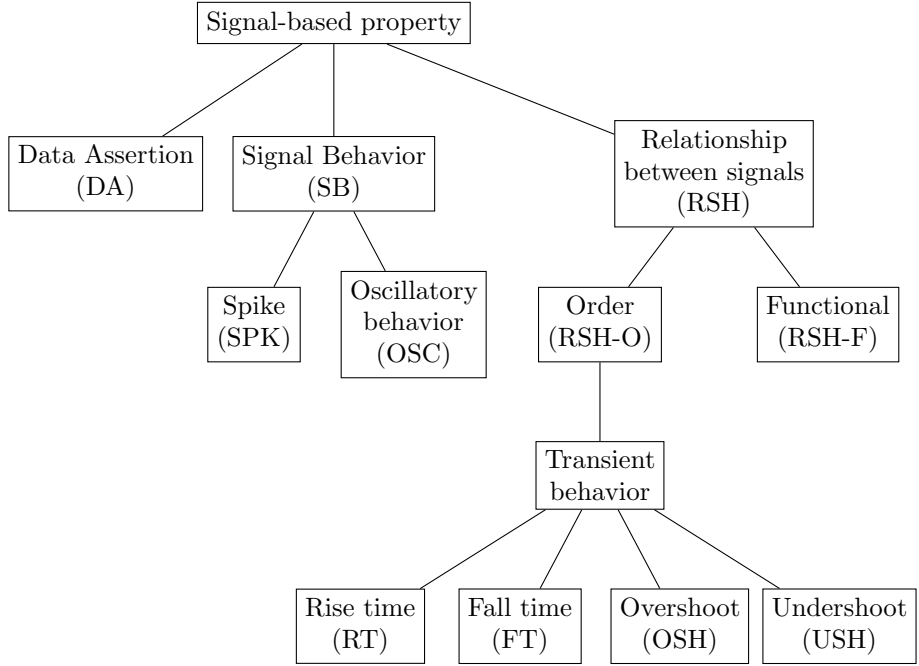

The taxonomy \rev(with the acronyms) of signal-based property types is shown
in figure~\ref{fig:taxonomy}. At the top level, it includes three  main signal-based property types:
\begin{description}
\item[Data assertion (DA):] properties expressing
  constraints on the value of a signal.
\item[Signal behavior (SB):] properties on the behavior represented by
    a signal shape. We further distinguish among two property subtypes:
    \begin{itemize}
    \item properties on signals exhibiting spikes \rev{(SPK)};
    \item properties on signals manifesting oscillatory behaviors \rev{(OSC)}.
    \end{itemize}      
  \item[Relationship between signals (RSH):] properties characterizing
relationships between signals.  This type includes two further property subtypes: 
    \begin{itemize}
    \item \emph{functional}, based on the application of a signal
      transforming function \rev{(RSH-F)};
    \item  \emph{order}, describing sequences of events/states related
      to signal behaviors \rev{(RSH-F)}. In this category we also include properties of transient behaviors of a signal when changing from  the current value to a new 
      target value, such as:
      \begin{itemize}
      	\item properties on signals exhibiting a \emph{rising}
          \rev{(Rise Time - RT)} or a
          \emph{falling} \rev{(Fall Time - FT)} behavior;
      	\item properties on signals exhibiting an \emph{\osh}
          \rev{(OSH)} or an \emph{\ush} \rev{(USH)} behavior.
      \end{itemize}
  \end{itemize}
\end{description}

In the following subsections we provide the detailed
description of each property type, including a mathematical
formalization and examples. We use (a variant of) \sfo to formalize the various
property types; anticipating the results of section~\ref{sec:expr}, the reason
for the adoption of \sfo is its expressiveness, which allows us to
express \emph{all the property types considered in this paper}. We
also provide examples of properties in \stl and \stlstar (when
applicable).The variant of \sfo we use for the formalization has the  following
signature  $\Sigma=\langle 
F,A,\mathit{Rel}, \mathbb{Z}, \mathbb{R}\rangle$, where:
\begin{itemize}
\item $F=\mathit{Sig} \cup \mathit{Aux}$ is the set of function symbols, composed of signal
  functions $\mathit{Sig}=\{s,s_1,s_2,s_{\mathit{tr}}\}$ and auxiliary
  functions and predicates 
  $\mathit{Aux}=\{\sigma^{\mathbb{B}e}_{s,P},\sigma^{\mathbb{B}s}_{s,P},\xi,\mathit{checkOsc},
  \mathit{local\_min},\mathit{local\_max}\}$;
\item $A$ is the set of (non-linear) arithmetic functions
  $A=\{+,-,\times,\div,\abs\}$, where $\abs$ represents the absolute
  value operator;
\item $\mathit{Rel}$ is the set of relational operators
  $\mathit{Rel}=\{<,>,\ge,\le, =, \neq\}$;
\item $\mathbb{Z}$ and $\mathbb{R}$ are integer and real constants, respectively.
\end{itemize}

\subsection{Data assertion}\label{sec:da}

A data assertion specifies a constraint on the value of a signal. This
constraint is expressed through a \emph{signal predicate} of the form
$s \bowtie \mathit{expr}$, where $\mathit{expr}$ is an \sfo value term
defined over the value domain of the signal $s$ and
$\bowtie{} \in \mathit{Rel}$. A data assertion property holds on the
signal if the assertion  predicate evaluates
to $\mathit{true}$. Data assertions can be combined to form more
complex expressions through the standard logical connectives.
We distinguish
between \emph{untimed} data assertions, which are evaluated through
the entire domain of definition $I_s$ of a signal $s$, and
\emph{time-constrained} data assertions, which are evaluated over one or
more distinct sub-intervals of the signal domain of definition.

More formally, let $H$ be a set of time intervals
$H=\{\mathcal{I}_1, \ldots, \mathcal{I}_K\}$, such that
$\mathcal{I}_k \subseteq I_s, 1 \leq k \leq K$, and for all
$i, j \in \{1, \ldots K\}, i \neq j $ implies
$\mathcal{I}_i \cap \mathcal{I}_j = \emptyset$. A data assertion
defined over the time intervals in $H$ holds on a signal $s$ if and
only if (iff) the \sfo formula
$\bigwedge_{h \in H} \forall i \in h \colon s(i) \bowtie
\mathit{expr}$ evaluates to \emph{true}. Notice that an \emph{untimed} data assertion over a
signal $s$ is defined by having $H=\{I_s\}$.

\begin{figure}[htb]
  \centering
  \begin{tikzpicture}
\begin{axis}[xmin=0, xmax=25, ymin = 0, ymax=5, xlabel={\emph{time (tu)}},
		ylabel={\emph{value}},
		every axis x label/.style={at={(current axis.right of origin)},anchor=west},
		every axis y label/.style={at={(current axis.north west)},above=2mm}, extra x ticks={2,6},
		extra x tick labels={$2$,$6$}]]
\draw [name path = vertical, dashed] (axis cs:2,0) -- (axis cs:2,5) ;
\draw [name path = vertical, dashed] (axis cs:6,0) -- (axis cs:6,5) ;
\draw [name path = vertical, dashed] (axis cs:10,0) -- (axis cs:10,5) ;
\draw [name path = vertical, dashed] (axis cs:15,0) -- (axis cs:15,5) ; 
\addplot[name path global=Threshold1,dashed,domain=0:25] {3} ; 

\node[inner sep=0pt] (n1) at (axis cs:0,3.5) {};
		\node[inner sep=0pt] (n2) at (axis cs:1,2.9) {};
		\node[inner sep=0pt] (n3) at (axis cs:3,1.5) {};
		\node[inner sep=0pt] (n4) at (axis cs:6,2.5) {};
		\node[inner sep=0pt] (n5) at (axis cs:10,1.4) {};
		\node[inner sep=0pt] (n6) at (axis cs:15,2.5) {};
		\node[inner sep=0pt] (n7) at (axis cs:17,2.5) {};
		\node[inner sep=0pt] (n8) at (axis cs:19,3.5) {};
		\node[inner sep=0pt] (n9) at (axis cs:21,2.5) {};
		\node[inner sep=0pt] (n10) at (axis cs:24,2.5) {};
\path[name path = GraphCurve, draw, color=black, very thick] plot [smooth, very thick] coordinates { (n1) (n2) (n3) (n4) (n5) (n6) (n7) (n8) (n9) (n10)}; 

\node[inner sep=0pt] (n1) at (axis cs:0,2.8) {};
		\node[inner sep=0pt] (n2) at (axis cs:2,4.5) {};
		\node[inner sep=0pt] (n3) at (axis cs:4,2.7) {};
		\node[inner sep=0pt] (n4) at (axis cs:6,3.5) {};
		\node[inner sep=0pt] (n5) at (axis cs:8,2.8) {};
		\node[inner sep=0pt] (n6) at (axis cs:10,2.8) {};
		\node[inner sep=0pt] (n7) at (axis cs:15,3.2) {};
		\node[inner sep=0pt] (n8) at (axis cs:18,2.3) {};
		\node[inner sep=0pt] (n9) at (axis cs:20,2.6) {};
		\node[inner sep=0pt] (n10) at (axis cs:23,2.6) {};
\path[name path = GraphCurve, draw, color=black] plot [smooth] coordinates { (n1) (n2) (n3) (n4) (n5) (n6) (n7) (n8) (n9) (n10)};

\end{axis}
\end{tikzpicture}    \caption{Two signals used to evaluate property \textit{pDA}: signal $s_1$ (\protect\veryThickLine) satisfies
    the property whereas signal $s_2$ (\protect\thinLine) violates it. }
  \label{fig:DA}
\end{figure}
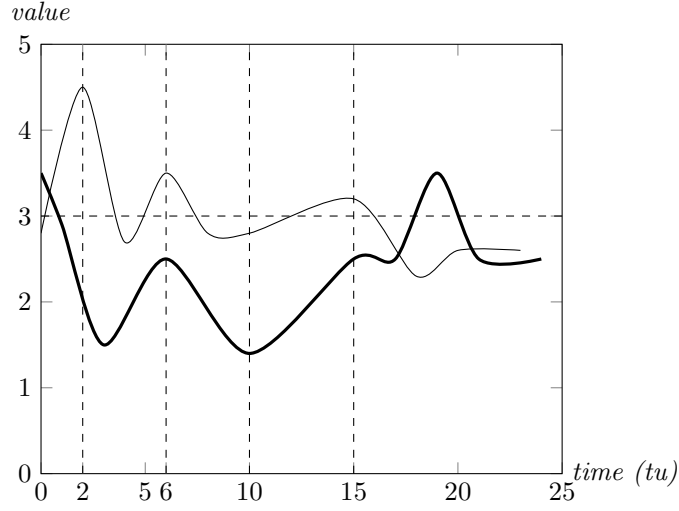

For example, let us consider the property \textit{pDA}: ``The signal
value shall be less than 3 between \SI{2}{\tu} and \SI{6}{\tu} and
between \SI{10}{\tu} and \SI{15}{\tu}'', where ``\si{\tu}'' is a
generic time unit (which has to be set according to the application
domain, e.g., seconds).  This property is a \emph{time-constrained}
data assertion over the two intervals $[2,6]$ and $[10,15]$; it can be
expressed in \sfo as:
\[
  \logiclbl{SFO}{\textit{pDA}}\quad  \forall t \in [2,6]: s(t) < 3 \land \forall t \in [10,15]: s(t) < 3
\]  
Figure~\ref{fig:DA} shows
two signals, $s_1$ plotted with a thick line (\veryThickLine), and $s_2$ plotted with a thin line (\thinLine); the threshold on the signal value specified by the
property is represented with a dashed horizontal line.
Property \textit{pDA} does not hold for $s_2$ as its 
value is 
above the threshold of 3 in the intervals $[2,6]$ and $[10,15]$; however, it holds
for $s_1$ because its value is below the threshold in both intervals.

\subsubsection{Alternative formalizations}
Data assertion properties like \textit{pDA}  can be also
expressed in \stl and \stlstar: 
\[
\begin{aligned}  
         \logiclbl{STL}{\textit{pDA}} \equiv \logiclbl{STL*}{\textit{pDA}}
\quad &\mathsf{G}_{[2,6]} (s < 3)\land\mathsf{G}_{[10,15]} (s < 3)
\end{aligned}        
\]

 \subsection{Spike}
\label{sec:bumps}
A spike\footnote{A spike is also called bump, peak, or pulse in the
  literature.}
can be informally defined as a short-lived, (relatively) large
increase or decrease of the value of a signal. Such a signal
behavior is typically
undesirable~\cite{bartocci2018specification}. However, there are
situations in which a spike characterized by a set of specific
features is desirable, as it is the case for the discovery
pulse~\cite{NickovicMonitHybridTut} in the discovery mode of the DSI3
protocol~\cite{consortium11:_dsi3_bus_stand}.

\begin{figure*}[t]
	\centering
	\subfloat[]
	{\scalebox{.6}{

\begin{tikzpicture}

\begin{axis}[xmin=0, xmax=50, ymin = 0, ymax=3, extra y ticks={0.5,1,2}, extra y tick 
labels={$\mathit{s(VP_{2})}$, $\mathit{s(VP_{1})}$, $\mathit{s(PP)}$ }, extra x ticks={3,13,23,38,48}, 
extra x tick labels={$f$,$\mathit{VP_{1}}$, $\mathit{PP}$, $\mathit{VP_{2}}$,$g$}, xlabel={\emph{time 
(tu)}},xticklabels=\empty, 
every axis x label/.style={at={(current axis.right of origin)},anchor=west},
every axis y label/.style={at={(current axis.north west)},above=2mm},
ylabel={\emph{value}},yticklabels=\empty]

        \node[inner sep=0pt] (n1) at (axis cs:12,1.3) {};
        \node[inner sep=0pt] (n2) at (axis cs:13.5,1) {};
        \draw[] (13.1,0.99) circle (0.08 cm);
        \node[inner sep=0pt] (n3) at (axis cs:23,2) {};
        \draw[] (23,2) circle (0.08 cm);
        \node[inner sep=0pt] (n4) at (axis cs:36.6,0.58) {};
        \node[inner sep=0pt] (n5) at (axis cs:39,0.65) {};
        \draw[] (38,0.52) circle (0.08 cm);
        
\path[name path = GraphCurve, draw, very thick] plot [smooth, very thick] coordinates { (n1) (n2) (n3) (n4) (n5)};

\node[inner sep=0pt] (n1) at (axis cs:6.5,1.5) { $a_1$};
         \node[inner sep=0pt] (n2) at (axis cs:43,1.4) { $a_2$};
          \node[inner sep=0pt] (n3) at (axis cs:25,0.38) { $w$};
          \node[inner sep=0pt] (n4) at (axis cs:16,1.7) { $\mathit{sp}_{1}$};
          \node[inner sep=0pt] (n5) at (axis cs:30,1.7) { $\mathit{sp}_{2}$};

\addplot[name path global=Threshold1, dashed, very thin, domain=0:50] {0.5} ;
  \addplot[name path global=Threshold1, dashed, very thin, domain=0:50] {1} ;
  \addplot[name path global=Threshold3, dashed, very thin, domain=0:50] {2} ;

   \draw [name path = vertical, dashed, very thin] (axis cs:13,0) -- (axis cs:13,3) ; \draw [name path = vertical, dashed, very thin] (axis cs:23,0) -- (axis cs:23,3) ; \draw [name path = vertical, dashed, very thin] (axis cs:38,0) -- (axis cs:38,3) ;

    \draw[<->] [name path = vertical, dashed, blue, thick] (axis cs:8,1) -- (axis cs:8,2) ; \draw[<->] [name path = vertical, dashed, blue, thick] (axis cs:45,0.5) -- (axis cs:45,2) ;

\draw[<->] [name path = vertical, dashed, blue, thick] (axis cs:13,0.3) -- (axis cs:38,0.3) ; 

\draw[<->] [name path = vertical, dashed, blue, thin] (axis cs:14,1) -- (axis cs:22.6,1.98) ;  

\draw[<->] [name path = vertical, dashed, blue, thin] (axis cs:23,1.96) -- (axis cs:37,0.49) ; \draw [name path = vertical, dashed, very thin] (axis cs:3,0) -- (axis cs:3,3) ;
       \draw [name path = vertical, dashed, very thin] (axis cs:48,0) -- (axis cs:48,3) ;

\end{axis}

\end{tikzpicture}

 			\label{fig:bumpBioMedical}
		}
	}
	\hfill
	\subfloat[]
	{\scalebox{.6}{\begin{tikzpicture}

\begin{axis}[xmin=0, xmax=50, ymin = 0, ymax=3.5,xlabel={\emph{time (tu)}}, 
every axis x label/.style={at={(current axis.right of origin)},anchor=west},
every axis y label/.style={at={(current axis.north west)},above=2mm},
ylabel={\emph{value}},yticklabels=\empty, extra y ticks={0.5, 1, 1.5, 2, 2.5, 3}, extra y tick labels={$0.5$, $1$, $1.5$, $2$, $2.5$, $3$},yticklabel shift=2pt
]
\node[inner sep=0pt] (n1) at (axis cs:4,1.17) {};
    \node[inner sep=0pt] (n2) at (axis cs:6,1.17) {};
    \node[inner sep=0pt] (n3) at (axis cs:8,1.17) {};
    \node[inner sep=0pt] (n4) at (axis cs:9,1.09) {};
    \node[inner sep=0pt] (n5) at (axis cs:10,1) {};
    \draw[] (10,1) circle (0.08 cm);
    \node[inner sep=0pt] (n6) at (axis cs:12,1.1) {};
    \node[inner sep=0pt] (n7) at (axis cs:14,1.3) {};
    \node[inner sep=0pt] (n8) at (axis cs:18,1.8) {};
    \node[inner sep=0pt] (n9) at (axis cs:20,2) {};
    \draw[] (20,2) circle (0.08 cm);
    \node[inner sep=0pt] (n10) at (axis cs:21.65,1.88) {};
    \node[inner sep=0pt] (n11) at (axis cs:25,1.5) {};
    \node[inner sep=0pt] (n12) at (axis cs:28,1.2) {};
    \node[inner sep=0pt] (n13) at (axis cs:30,1) {};
    \draw[] (30,1) circle (0.08 cm);
    \node[inner sep=0pt] (n14) at (axis cs:32,1.33) {};
    \node[inner sep=0pt] (n15) at (axis cs:34,1.3) {};
    \node[inner sep=0pt] (n16) at (axis cs:36,1.3) {};
    \node[inner sep=0pt] (n17) at (axis cs:38,1.3) {};
    \node[inner sep=0pt] (n18) at (axis cs:40,1.3) {};
    \node[inner sep=0pt] (n19) at (axis cs:42,1.3) {};
    \node[inner sep=0pt] (n20) at (axis cs:44,1.3) {};
    \node[inner sep=0pt] (n21) at (axis cs:46,1.3) {};

\path[name path = GraphCurve, draw, very thick] plot [smooth, very thick] coordinates { (n1) (n2) (n3) (n4) (n5) (n6) (n7) (n8) (n9) (n10) (n11) 
(n12) (n13) (n14) (n15) (n16) (n17) (n18) (n19) (n20) 
(n21) };

\node[inner sep=0pt] (n1) at (axis cs:3,1.6) {};
       \node[inner sep=0pt] (n2) at (axis cs:5,1.6) {};
       \node[inner sep=0pt] (n3) at (axis cs:7,1.6) {};
       \node[inner sep=0pt] (n4) at (axis cs:9,1.6) {};
       \node[inner sep=0pt] (n5) at (axis cs:10,1.5) {};
       \draw[] (10,1.5) circle (0.08 cm);
       \node[inner sep=0pt] (n6) at (axis cs:12,1.6) {};
       \node[inner sep=0pt] (n7) at (axis cs:16,2) {};
       \node[inner sep=0pt] (n8) at (axis cs:25,3) {};
       \draw[] (25,3) circle (0.08 cm);
       \node[inner sep=0pt] (n9) at (axis cs:35,2) {};
       \draw[] (35,2) circle (0.08 cm);
       \node[inner sep=0pt] (n10) at (axis cs:37,2.6) {};
       \node[inner sep=0pt] (n11) at (axis cs:43,2.5) {};
       \node[inner sep=0pt] (n12) at (axis cs:45,2.5) {};
\path[name path = GraphCurve, draw] plot [smooth] coordinates { (n1) (n2) (n3) (n4) (n5) (n6) (n7) (n8) (n9) (n10) (n11) (n12) 
       };
\end{axis}
\end{tikzpicture} } 
		\label{fig:bumpsBioMedicalTwo}
	}
	\caption{\protect\subref{fig:bumpBioMedical} Main features
          used to define a spike based 
	on~\cite{dumpala1982algorithm}.
        \protect\subref{fig:bumpsBioMedicalTwo} two signals used to evaluate  property 
	\emph{pSPK1}: signal $s_1$ (\protect \veryThickLine) satisfies the
        property, whereas $s_2$ (\protect \thinLine) violates it.}
      \label{fig:bumpsBioMedical}
\end{figure*}
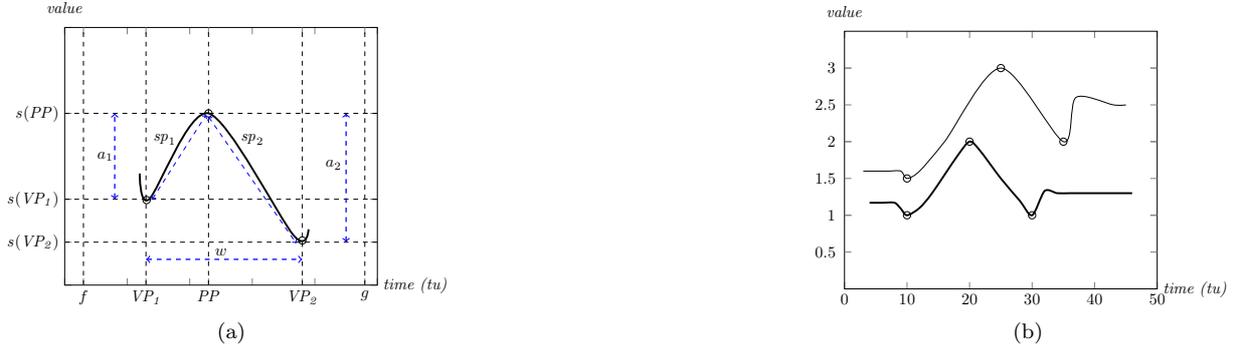

Inspired by the definitions in the bio-medical domain~\cite{dumpala1982algorithm}, we consider
four main features to characterize a spike, based on three extrema of
the function corresponding to the signal shape, which are local
extrema with respect to an observation interval $[f,g] \subset I_s$.
These three points (with their respective coordinates) are: the peak
point $(\mathit{PP}, s(\mathit{PP}))$ representing the local maximum
of the signal and characterizing the actual spike\footnote{In the
  following we only characterize and formalize spikes corresponding to an increase
  of the signal value; the case of a decrease of
  the signal value is the dual.}, and the two
surrounding valley points $(\mathit{VP}_{1},s(\mathit{VP}_{1}))$ and
$(\mathit{VP}_{2},s(\mathit{VP}_{2}))$ representing the local minima (closest to the peak point) of the first and second half of the spike, respectively. These three
local extrema are shown in figure~\ref{fig:bumpBioMedical}; we refer
the reader to reference~\cite{dumpala1982algorithm} for a detailed
description of how to detect these points.
The four features (also shown in figure~\ref{fig:bumpBioMedical}) characterizing a spike are:
\begin{itemize}
\item Amplitude  $a$ of the spike, defined as
  $a=\psi(a_1,a_2)$, where $a_1$ is the amplitude of the first-half of the spike shape 
	$a_1= \abs{(s(\mathit{PP})-s(\mathit{VP}_{1}))}$, 
        $a_2$ is the amplitude  of the second-half of the spike shape 
$a_2= \abs{(s(\mathit{PP})-s(\mathit{VP}_{2}))}$, and $\psi$ is  a  generic amplitude
  function\footnote{This function depends on the application domain;
    for example, in the context of bio-medical
    systems~\cite{dumpala1982algorithm}, $\psi$ is the minimum function.};
	
\item slope  $\mathit{sp}_{1}$ between the peak 
point and the valley point of the first half of the spike shape, 
  $\mathit{sp}_{1} = 
\abs{\left(\frac{s(\mathit{PP})-s(\mathit{VP}_{1})}{\mathit{PP}-\mathit{VP}_{1}}\right)}$;
\item  slope  $\mathit{sp}_{2}$ between the 
peak point and the valley point of the second half of the spike shape, 
  $\mathit{sp}_{2} = 
\abs{\left(\frac{s(\mathit{PP})-s(\mathit{VP}_{2})}{\mathit{PP}-\mathit{VP}_{2}}\right)}$;
\item \label{spike:width} spike width $w$ between the two consecutive valley points, 
$w=\mathit{VP}_{2}-\mathit{VP}_{1}$. Note that the width $w$
can be also defined as 
$w=w_1+w_2$, where $w_1=\mathit{PP}-\mathit{VP}_{1}$ and $w_2=\mathit{VP}_{2}-\mathit{PP}$.
\end{itemize}
The four features $a$, $\mathit{sp}_{1}$, $\mathit{sp}_{2}$, and $w$ can be opportunely 
combined to define a spike of a
particular shape\footnote{Although other spike features have been
  proposed in the spike detection literature---such as different types of width,
  amplitude, and
  slope~\cite{acir2004automatic,acir2005automated,acir2005automatic,liu2002multistage,
    dingle1993multistage}, as well as the area under the
  curve~\cite{hagglund1995control}---we decided not to adopt them
  since the features we have selected are sufficient to describe (and specify) the spike
  behaviors we consider in this paper.}.

A spike property specifies a constraint on the existence of a spike
with certain features; it evaluates to true when the signal
exhibits a spike whose features satisfy certain criteria.
More specifically, when defining a spike property, an engineer has to
specify---for each feature---a
predicate with a \textit{threshold criterion} whose value
depends on the application context. The signal predicates of each
feature are then logically conjoined for characterizing the spike.

Formally, given the threshold criteria for the four features
(specified as \sfo terms over the value domain  of  signal $s$)
$\mathit{\Gamma_{a}, \Gamma_{\mathit{sp}_{1}}, \Gamma_{\mathit{sp}_{2}}, \Gamma_{w}}$, a 
spike property holds on a signal $s$ iff the following \sfo formula evaluates to true:
\begin{equation}\label{bump:bio}
\begin{split}
\exists \mathit{VP}_{1}, \mathit{PP},\mathit{VP}_{2} \in [f,g] \colon \mathit{local\_min}( 
\mathit{VP}_{1},f,\mathit{PP}) \land \\ \mathit{local\_max}( \mathit{PP},\mathit{VP}_{1},g) \land  \\\mathit{local\_min}(\mathit{VP}_{2},\mathit{PP},g)  \land \\
\mathit{a} \bowtie \Gamma_{a} \land \mathit{sp}_1 \bowtie \Gamma_{\mathit{sp}_{1}}
\land \mathit{sp}_2\bowtie \Gamma_{\mathit{sp}_{2}} \land \mathit{w} \bowtie \Gamma_{w}
\end{split}
\end{equation}
where $\bowtie{} \in \mathit{Rel}$, $\mathit{local\_min}$ and $\mathit{local\_max}\in \mathit{Aux}$ are 
predicates identifying local extrema, and 
$a,\mathit{sp}_1, \mathit{sp}_2,w$ are \sfo terms defined as shown
above using the 
three variables $\mathit{VP}_{1}$,$\mathit{VP}_{2}$, and 
$\mathit{PP}$.

In essence, formula~\eqref{bump:bio} requires
\begin{inparaenum}[a)]
\item the existence of the
three local extrema in a proper order characterizing the spike shape
(i.e.,  a local minimum followed by a 
local maximum, followed by another local minimum), and
\item the
  satisfaction of the constraints for all the features.
\end{inparaenum}
More relaxed formulations can be obtained by omitting some  of the
spike features from the above definition.

The predicate  $\mathit{local\_min}(x,y,z)$ (respectively, $\mathit{local\_max}(x,y,z)$) 
returns true if the time point $x$ is a local minimum (respectively,
local maximum)  with respect to the interval $[y,z]$. These predicates
can be defined in several ways; below we provide three possible definitions.

\begin{definition}[local extrema through punctual derivatives]
\label{def:extrema:punctual}
  
Some specification languages allow for
defining expressions corresponding to punctual derivatives. For
example, in \sfo the punctual derivatives can be defined
as language terms as follows:
 \[
 s_{p}^{\prime}(t) \equiv \frac{s(t+\epsilon)-s(t)}{\epsilon}\text{ and } 
 s_{p}^{\prime\prime}(t) \equiv s_{p}^{\prime}(s_{p}^{\prime}(t))
\]        
with $\epsilon$ being an arbitrary, small constant\footnote{In the
  context of a discrete signal, the $\epsilon$ constant can be
  replaced with the sampling interval $\Delta$.}.
The local extrema predicates can then be defined in \sfo as
  follow:

\begin{equation*}
\begin{aligned}
    \mathit{local\_min(x,y,z)} &\equiv \exists x\in[y,z]\colon s_{p}^{\prime}(x)=0 \land	s_{p}^{\prime\prime}(x)>0\\
 \mathit{local\_max(x,y,z)} &\equiv \exists x\in[y,z]\colon s_{p}^{\prime}(x)=0 \land	s_{p}^{\prime\prime}(x)<0
\end{aligned}          
\end{equation*}
\end{definition}

\begin{definition}[local extrema - analytical formulation]
\label{def:extrema:analytical}
  
Another way to characterize local extrema is to write a logical
expression corresponding to their analytical definition; in \sfo we have
\begin{equation*}
\begin{aligned}
\mathit{local\_min(x,y,z)} &\equiv \exists x \in [y,z]\colon\forall t\in[y,z], x \neq t\colon s(x)\le s(t)\\
\mathit{local\_max(x,y,z)} &\equiv \exists x \in [y,z]\colon\forall t\in[y,z], x \neq t\colon s(x)\ge s(t)
\end{aligned}          
\end{equation*}
\end{definition}

\begin{definition}[local extrema through pre-computed derivatives]
\label{def:extrema:pre-computed}
  
  When the first and second order derivatives of a signal are available
  as \emph{(pre-computed), separate signals}, the local extrema can be
  characterized using such signals. Let $s_{c}^{\prime}$ and
  $s_{c}^{\prime\prime}$ be the first and second order derivatives of
  signal $s$; the local extrema predicates can defined in \sfo as
  follow:

\begin{equation*}
\begin{aligned}
    \mathit{local\_min(x,y,z)} &\equiv \exists x\in[y,z]\colon s_{c}^{\prime}(x)=0 \land	s_{c}^{\prime\prime}(x)>0\\
 \mathit{local\_max(x,y,z)} &\equiv \exists x\in[y,z]\colon s_{c}^{\prime}(x)=0 \land	s_{c}^{\prime\prime}(x)<0
\end{aligned}          
\end{equation*}
\end{definition}

The choice of which definition to use for defining local extrema predicates 
 depends on the specification language and 
 the application context; as shown above, all three definitions can be
 used with  \sfo.

For example, let us characterize spikes through features
width $w$ and amplitude $a$, with the latter defined by using the
maximum function as the amplitude function $\psi$; let us consider the evaluation of property
\emph{pSPK1}: ``In a signal, there is a spike with a maximum width of
\SI{20}{\tu} and a maximum amplitude of 1''. For this property, the
parameters of an instance of specification~\eqref{bump:bio} are
$\Gamma_{a}=1$ and $\Gamma_{w}=20$; the resulting \sfo formula is:
\begin{multline*}
  \logiclbl{SFO}{\textit{pSPK1}} \quad
\exists t, t^\prime,t^{\prime\prime} \in [f,g] \colon \mathit{local\_min}( 
t,f,t^\prime) \land \\ \mathit{local\_max}( t^\prime,t,g) \land  \\\mathit{local\_min}( 
t^{\prime\prime},t^\prime,g)
 \land 
\\  \max(\abs{(s(t^{\prime}) - s(t))}, \abs{(s(t^{\prime\prime}) -
s(t^{\prime}))}) \le 1 \land \abs{(t^{\prime\prime}-t)} \le 20 
\end{multline*}
                              
In figure~\ref{fig:bumpsBioMedicalTwo}, we show two signals, $s_1$
plotted with a thick line (\veryThickLine) and $s_2$
plotted with a thin line (\thinLine). To
evaluate property \emph{pSPK1} on these signals, we first need to
evaluate the local extrema predicates in
specification~\eqref{bump:bio} (according to one of the three
definitions above):
 signal
$s_1$ exhibits a spike where $\mathit{VP}_{1}=10$, $\mathit{PP}=20$,
and $\mathit{VP}_{2}=30$, while $s_2$ exhibits a spike where
$\mathit{VP}_{1}=10$, $\mathit{PP}=25$, and $\mathit{VP}_{2}=35$.
In both cases, the three points satisfy the local extrema predicates. The second step is to evaluate the threshold criteria of the spike features.
We calculate the amplitude $a_{s_1}$ and the width $w_{s_1}$ of
the spike in $s_1$ as:
$
a_{s_1}=\max(\abs{(s_1(\mathit{PP})-s_1(\mathit{VP}_1))},\abs{(s_1(\mathit{PP})-s_1(\mathit{VP}_2))})=
\max(\abs{(s_1(20)-s_1(10))},\abs{(s_1(20)-s_1(30))})=\max(\abs{(2-1)},\abs{(2-1)})=1
$
and $w_{s_1}=\mathit{VP}_2-\mathit{VP}_1=30-10=20$.
Signal $s_1$ satisfies property \emph{pSPK1} because the expression  $a_{s_1} \le 
1 \land w_{s_1} \le 20 \equiv 1 \leq 1 \land 20 \leq 20$ evaluates to true.  Following a
similar computation,
the amplitude $a_{s_2}$ and the width $w_{s_2}$ of the spike in $s_2$
are $a_{s_2}=\max(1.5,1)=1.5$ and $w_{s_2}=25$; signal $s_2$ violates
 property \emph{pSPK1} because the expression $a_{s_2} \le 1 \land w_{s_2} \le 20 \equiv 1.5 \leq 1 \land 25 \leq 20$ evaluates to false.

Another definition, proposed in  the context of automotive control
applications~\cite{KapinskyIEE16}, characterizes a spike using 
two parameters,  $w$ and  $m=\frac{a}{w}$, where $w$ is the spike
width and $a$ the spike amplitude. Formally, a signal $s$ exhibits a spike with parameters $m$ and 
$w$ (defined as numerical constants) iff the following \sfo formula evaluates to true: 
\begin{equation}\label{stlib-bump-def}
\exists t \in I_s \colon s^\prime(t) > m  \land  \exists t^\prime
\in [t,t+w]\colon s^\prime(t^\prime) < {-m}
\end{equation}
where $s^{\prime}$, denoting the first order derivative of $s$, can be
either a pre-computed, separated signal  $s_{c}^{\prime}$ or the punctual derivative $s_{p}^{\prime}$ 
introduced above.
This characterization identifies two time instants: the first in which
the signal derivative  is greater than parameter $m$ and another one
in which the signal derivative is less than ${-m}$; the distance
between these two points is the spike width $w$.

\begin{figure}[t]
	\centering
	{\scalebox{.7}{\begin{tikzpicture}

\begin{axis}[xmin=0, xmax=60, ymin = 0, ymax=4,xlabel={\emph{time (tu)}}, 
every axis x label/.style={at={(current axis.right of origin)},anchor=west},
every axis y label/.style={at={(current axis.north west)},above=2mm},
ylabel={\emph{value}},yticklabels=\empty, yticklabels=\empty, extra y ticks={1,2,3}, extra y tick labels={$1$,$2$,$3$}]
  
\node[inner sep=0pt] (n1) at (axis cs:5,0.2) {};
        \node[inner sep=0pt] (n2) at (axis cs:10,0.2) {};
        \node[inner sep=0pt] (n3) at (axis cs:15,0.2) {};
        \node[inner sep=0pt] (n4) at (axis cs:20,1) {};
        \node[inner sep=0pt] (n5) at (axis cs:21,1.2) {};
        \node[inner sep=0pt] (n6) at (axis cs:23,1.65) {};
        \node[inner sep=0pt] (n7) at (axis cs:30,3) {};
        \node[inner sep=0pt] (n8) at (axis cs:36,1.6) {};
        \node[inner sep=0pt] (n9) at (axis cs:39,0.7) {};
        \node[inner sep=0pt] (n10) at (axis cs:41,0.2) {};
        \node[inner sep=0pt] (n11) at (axis cs:45,0.2) {};
        \node[inner sep=0pt] (n12) at (axis cs:52,0.2) {};
            
\path[name path = GraphCurve, draw, color=black] plot [smooth] coordinates { (n1) (n2) (n3) (n4) (n5) (n6) (n7) (n8) (n9) (n10) (n11)
        (n12) };

\node[inner sep=0pt] (n1) at (axis cs:5,0.2) {};
        \node[inner sep=0pt] (n2) at (axis cs:10,0.2) {};
        \node[inner sep=0pt] (n3) at (axis cs:15,0.2) {};
        \node[inner sep=0pt] (n4) at (axis cs:20,1) {};
        \node[inner sep=0pt] (n6) at (axis cs:25,1.65) {};
        \node[inner sep=0pt] (n7) at (axis cs:30,1.9) {};
        \node[inner sep=0pt] (n8) at (axis cs:35,1.4) {};
        \node[inner sep=0pt] (n9) at (axis cs:39,0.7) {};
        \node[inner sep=0pt] (n10) at (axis cs:41,0.2) {};
        \node[inner sep=0pt] (n11) at (axis cs:45,0.2) {};
        \node[inner sep=0pt] (n12) at (axis cs:52,0.2) {};

\path[name path = GraphCurve, draw, color=black, very thick] plot [smooth] coordinates { (n1) (n2) (n3) (n4) (n5)  (n6) (n7) (n8) (n9)
         (n10) (n11) (n12) 
         };
   
\end{axis}
\end{tikzpicture} }
	}
	\caption{Characterization of the spike in two signals $s_1$ (\protect \veryThickLine) and $s_2$ 
		(\protect \thinLine) based on the definition
                in~\cite{KapinskyIEE16}, with parameters 
		$m=0.1$, $w=20$.
	}
	\label{fig:bumpstlib}
      \end{figure}
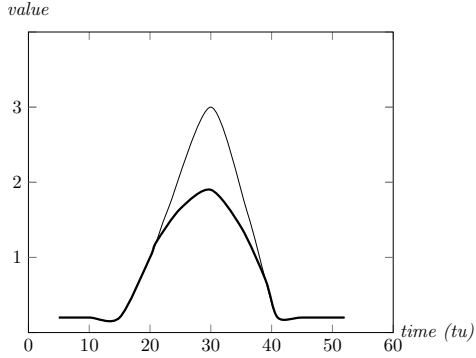

The main limitation of this formulation is that it does not
allow to express precise constraints on the absolute value of the amplitude of a
spike; instead, it uses parameter $m$ that is a quotient between amplitude and width. We illustrate this 
with the example in
figure~\ref{fig:bumpstlib}, with the signals $s_1$ plotted with a thick line (\veryThickLine) and $s_2$ plotted with a thin line (\thinLine). Let us consider the
evaluation of property~\emph{pSPK2}: ``In a signal, there exists a spike with a maximum 
width of \SI{20}{\tu} and an amplitude greater than 2''. This property cannot be captured by 
an instance of specification~\eqref{stlib-bump-def}, since the latter does not
take into account the concept
of amplitude; the property needs to be adapted. 
Based on the desired values of width and amplitude in 
property~\emph{pSPK2}, the parameters of an instance of specification~\eqref{stlib-bump-def} would be 
$m=0.1$, $w=20$. Therefore, instead of property~\emph{pSPK2}, one can consider the following 
alternative~\emph{pSPK3}: ``In a signal, there exists a spike with a maximum
width of \SI{20}{\tu} and parameter $m$ equal to 0.1'', which can be captured by an instance of 
specification~\eqref{stlib-bump-def}; the corresponding \sfo formula is:
\[
\logiclbl{SFO}{\textit{pSPK3}} \quad
 \exists t \in I_s \colon s^\prime (t) > 0.1 
 \land \exists t^\prime \in [t,t+20] \colon s^\prime(t^\prime)<{-0.1}
\]
This formula will evaluate to true for both 
$s_1$ and $s_2$.
However, signal $s_1$ should not satisfy the property, since its peak point does not reach a
magnitude (amplitude) of 2 as was required in the original formulation
of the property (\emph{pSPK2}).
This spurious spike characterization
happens with specification~\eqref{stlib-bump-def} because signal $s_1$
follows the same shape as signal $s_2$ in the points in which
the signal derivative $s^{\prime}$ is compared to $m$. We
remark that the application of specification~\eqref{bump:bio} to the
evaluation of property \emph{pSPK2} 
would correctly characterize the spike only in signal $s_2$.
Given a lack of precision in specification~\eqref{stlib-bump-def}, in the
following we will consider spikes defined according to
specification~\eqref{bump:bio}.

\subsubsection{Alternative formalizations}
\paragraph{\emph{\stl}}
Our characterization of a spike through the \sfo
formulation~\eqref{bump:bio} relies on the existence of three extrema
in the function corresponding to the signal shape.  In \stl, the
existence of these extrema could be formalized through proper nesting
of the ``eventually'' and ``once'' operators, in conjunction with a
constraint on the width of the spike. However, it would not be
possible to include in such a formulation a constraint on the
amplitude or on the slope, since in \stl one cannot refer to the value
of the signal at an arbitrary time point. For all these reasons, we
cannot express a property like \emph{pSPK1} in \stl.

On the other hand, spike properties characterized through the \sfo
formulation~\eqref{stlib-bump-def} can be expressed in \stl when the
pre-computed signal derivatives are available. For example, property
\emph{pSPK3} can be expressed as
 \[
\logiclbl{STL}{\textit{pSPK3}} \quad \mathsf{F}_{[0,|s|)} (s^{\prime} > 0.1 \land 
\mathsf{F}_{[0,20]} s^{\prime} < -0.1) 
\]

\paragraph{\emph{\stlstar}}\label{sec:bumps:stlstar}
Differently from \stl, \stlstar can refer to the value of the signal
at a certain time point in which a local formula holds thanks to the
freeze operator; below we
discuss how it can be used to express properties \emph{pSPK1} and \emph{pSPK3}.

\subparagraph{(Using local extrema expressed through punctual derivatives)}\label{sec:bump:punctual-deriv-test}
Definition~\ref{def:extrema:punctual} for local extrema uses the values of the signal at two
consecutive time points, within a small distance $\epsilon$. However,
in \stlstar it is not possible to explicitly reference the signal value
at time points that are not associated with the evaluation of a local
(sub-)formula; hence, properties defined using punctual
derivatives cannot be specified using \stlstar\footnote{Such a
  restriction could be lifted when using discrete signals, since the
  distance between two consecutive time points is known and is equal
  to the sampling interval $\Delta$.}.

\subparagraph{(Using local extrema expressed through the analytical formulation)}\label{sec:an}

We can characterize local extrema using the analytical formulation
(definition~\ref{def:extrema:analytical})  by
assuming a variant of \stlstar with past
operators\footnote{Although the version of \stlstar presented
  in~\cite{brim2014stl} does not use past operators, the addition of
  such operators would be done along the lines of the definition of \stl
  with past operators in~\cite{maler2013monitoring}.} and using a 3D frozen time vector.

\begin{equation*}
\begin{aligned}
  \logiclbl{STL*}{\textit{pSPK1}} \quad
  &\mathsf{F}_{[f,g]}*_1  ( \mathsf{G}_{[0,w_1]} (s>s^{*_1}) \\
  &\land \mathsf{F}_{[0,w_1]} *_2 (
  \mathsf{H}_{[0,w_1]} (s<s^{*_2}) \\
  & \land \mathsf{F}_{[0,w_2]} *_3 (\mathsf{H}_{[0,w_2]} (s>s^{*_3}) \\
 & \land  \max(\abs(s^{*_1}-s^{*_2}),\abs(s^{*_2}-s^{*_3}))\le 1
 \land w_1+w_2 \leq 20
 ))) 
\end{aligned}
\end{equation*}

In the formula above, the expression in the first row states the
existence of the first local minimum by checking for the existence,
within the observation interval $[f,g]$, of a point (whose time
instant is frozen in the first component of the frozen time vector)
for which the corresponding signal value is smaller than all other
signal values in the interval $[0,w_1]$; this condition is captured by
the sub-formula with the ``globally'' operator.  The expression on the
second row, nesting the ``historically'' operator within the
``eventually'', states the existence of the local maximum (whose time
instant is frozen in the second component of the frozen time vector),
such that all the signal values between the first local minimum and
such a point are indeed smaller than the local maximum. Notice that
the distance between the first local minimum and the local maximum is
equal to $w_1$\footnote{If the spike shape is symmetrical, the
  distance between all local extrema is equal to $\frac{w}{2}$.}.  The
expression on the third row checks in a similar way for the existence
of the second local minimum within an interval $[0,w_2]$ from the
local maximum. The expression on the fourth row checks the constraints
on the spike amplitude and on the spike width. For the former, it uses
the values of the signal in correspondence of the first local minimum
($s^{*_1}$), of the local maximum ($s^{*_2}$), and of the second local
minimum ($s^{*_3}$).

Note that this property relies on a particular sequence of local
extrema (i.e., valley-peak-valley); other variants of this property
can be specified by changing the order of the sub-formulae stating the
existence of a certain extremum. Furthermore, we remark that the
specification of this property assumes the knowledge of the signal
shape, since it uses the two components of the width $w_1$ and $w_2$
as defined on page~\pageref{sec:bumps}. However, making such an
assumption in practice is not reasonable because typically the shape
of a spike is unknown.

\subparagraph{(Using local extrema defined through  pre-computed derivatives)}
Property \emph{pSPK1} can be expressed using
definition~\ref{def:extrema:pre-computed} for local extrema, assuming the existence of signals $s^{\prime}$ and
$s^{\prime\prime}$ and a 3D frozen time vector.

\begin{equation*} 
\begin{split}
  \logiclbl{STL*}{\textit{pSPK1}} \quad \mathsf{F}_{[f,g]} *_1
  \bigl( s^{\prime}=0 \land s^{\prime\prime}>0 \land 
  \mathsf{F}_{[0,w_1]} *_2 (s^{\prime}=0 \land s^{\prime\prime}<0 \land 
  \mathsf{F}_{[0,w_2]} *_3 (s^{\prime}=0 \land s^{\prime\prime}>0
   \\\land  \max(\abs(s^{*_1}-s^{*_2}),\abs(s^{*_2}-s^{*_3})) \le 1  \land w_1+w_2 \leq 20
  ))\bigr)
\end{split}
\end{equation*}

The structure of the formula above is similar to the one for the
case of using definition~\ref{def:extrema:analytical} for local
extrema, except for the direct use of the first and
second order derivatives, available as pre-computed signals. The same remarks
made above in terms of assuming the knowledge of the signal shape also
apply in this case.

Furthermore, pre-computed derivative signals can be used to specify
property \textit{pSPK3} in \stlstar in the same way as it was done
above using \stl.
 
 \subsection{Oscillation}
\label{sec:osc}

An oscillation can be informally described as a repeated variation over
time of the value of a signal, possibly with respect to a reference
value; often, in the context of CPS, oscillations represent an
undesirable signal behavior.

\begin{figure*}[tb]
	\centering
	\begin{tikzpicture}
\begin{axis}[xmin=0, xmax=60, ymin = 0, ymax=3, xlabel={\emph{time (tu)}},xticklabels=\empty, 
ylabel={\emph{value}},yticklabels=\empty,
extra y ticks={1}, 
extra y tick 
labels={$\mathit{ref}$},
every axis x label/.style={at={(current axis.right of origin)},anchor=west},
every axis y label/.style={at={(current axis.north west)},above=2mm},extra x 
ticks={2,58}, extra x tick 
labels={$a$,$b$},
]
\addplot[thick, name path global=GraphCurve, domain=0:60,samples=200] 
{sin(deg(x/3.6))+1};
    \addplot[
          scatter,only marks,scatter src=explicit symbolic,
          scatter/classes={
              a={mark=*,draw=black,fill=black, mark size=1.5pt},
              highlitedCirclesGreen={mark=square, draw=blue, mark size=4pt}
          }
      ]
      table[x=x,y=y,meta=label] { 
      x y label
      5.5 2 highlitedCirclesGreen
		28 2  highlitedCirclesGreen 
      51 2 highlitedCirclesGreen 
      17 0.08 highlitedCirclesGreen
		39.5 0.08 highlitedCirclesGreen 
 };\label{mark:blue}
\node[inner sep=0pt] () at (axis cs:5.5,2.2) { $p_1$};
\node[inner sep=0pt] () at (axis cs:17,0.35) { $p_2$};
\node[inner sep=0pt] () at (axis cs:28,2.2) { $p_3$};
\node[inner sep=0pt] () at (axis cs:51,2.2) { $p_5$};
\node[inner sep=0pt] () at (axis cs:39.5,0.35) { $p_4$};

\draw [name path = vertical, dashed, thin] (axis cs:2,0) -- (axis cs:2,3) ; 
\draw [name path = vertical, dashed, thin] (axis cs:58,0) -- (axis cs:58,3) ;

\addplot[red,name path global=Threshold1,dashed,domain=0:60] {1} ;
\addplot[name path global=Threshold2,dashed,domain=13:29] {2} ;

\draw[<->] [name path = vertical, dashed, blue, thick] (axis cs:13.5,1) -- (axis cs:13.5,2) ; \draw[<->] [name path = vertical, dashed, blue, thick] (axis cs:30,2) -- (axis cs:49,2) ;

\node[inner sep=0pt] () at (axis cs:18,1.8) { $\mathit{oscA}$};
\node[inner sep=0pt] () at (axis cs:40,2.1) { $\mathit{oscP}$};

\end{axis}
\end{tikzpicture}

 	\caption{A signal exhibiting an
oscillatory behavior; the reference value \emph{ref} is shown in red. }
	\label{fig:sine}
\end{figure*}
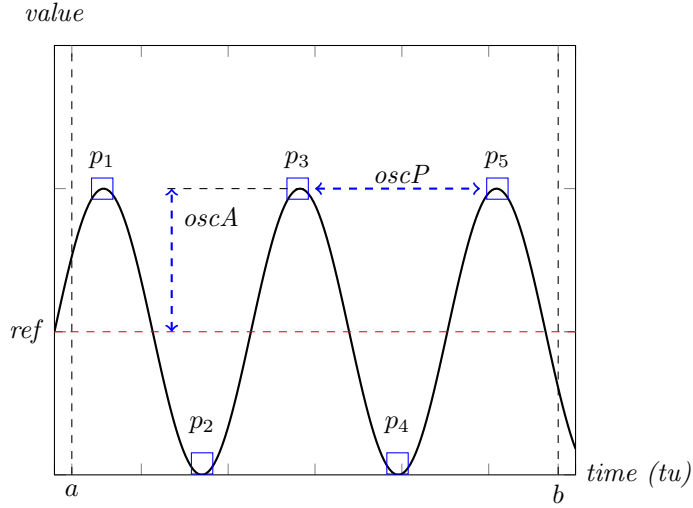

Figure~\ref{fig:sine} depicts an analog signal $s$ exhibiting an
oscillatory behavior with respect to a reference value $\mathit{ref}$,
within an observation interval
$\mathit{oscI} = [a,b] \subset I_{s}$. Such a behavior is
characterized by the existence, within the observation interval, of
$M$ extrema of the function corresponding to the signal
shape; these points are marked with blue squares (\ref{mark:blue}) in
the figure. A \emph{cycle} (i.e., a \emph{complete oscillation}) occurs when
the signal value swings from one extremum to the adjacent extremum of
the same type, by traversing an extremum of the other type; for
example, in the figure there is one complete oscillation when the
signal goes from $p_3$ to $p_5$ (two peak points) through $p_4$ (a
valley point).  The figure also shows two additional features typically
used to characterize oscillations:
\begin{itemize}
\item the \emph{(peak) amplitude}, denoted by $\mathit{oscA}$, is the
  distance between the maximum
  magnitude of the  signal and its reference value;
  
\item the \emph{period}, denoted by $\mathit{oscP}$, is the time
  required to complete one cycle. Its reciprocal, called
  \emph{frequency}, represents the number of complete oscillations
  occurring in a unit of time.

\end{itemize}

An oscillation property specifies a constraint on the existence, in a
signal, of an oscillatory behavior with certain features; it
evaluates to true when the signal exhibits an oscillatory behavior
whose features satisfy certain criteria. More specifically, these
criteria are expressed as relational expressions, on the oscillation
amplitude and/or period,  with an
application-specific threshold. 
More formally, given the \sfo terms representing the threshold
criteria  $\Gamma_{\mathit{oscP}}$ (for the period) and 
$\Gamma_{\mathit{oscA}}$ (for the amplitude), an
oscillation property holds on a signal $s$ in the observation interval $[a,b]$ iff the following 
\sfo formula evaluates to true:
\begin{equation}\label{eq:osc}
\begin{split}
\forall t \in  [a,b]\colon&(\exists t^\prime,
t^{\prime\prime}\in[t,b]\colon \\
&\mathit{local\_min}(t,a, 
t^\prime)\rightarrow\\
&(\mathit{local\_max}(t^\prime,t,b) \land \mathit{local\_min}(t^{\prime\prime},t^\prime,b)\\
&\land \mathit{checkOsc} (t,t^\prime,t^{\prime\prime}, \bowtie_{P},
\Gamma_{oscP}, \bowtie_{A}, \Gamma_{\mathit{oscA}})) \\
&\land \mathit{local\_max}(t,a,t^\prime) \rightarrow\\
&(\mathit{local\_min}(t^\prime,t, b) \land \mathit{local\_max}(t^{\prime\prime},t^\prime,b) \\ &\land 
\mathit{checkOsc}(t,t^\prime,t^{\prime\prime}, \bowtie_{P},
   \Gamma_{oscP},  \bowtie_{A}, \Gamma_{\mathit{oscA}})))
\end{split}
\end{equation}
where $\mathit{local\_min}(x,y,z)$ (respectively,
$\mathit{local\_max}(x,y,z)$) is a predicate that returns true if the
time point $x$ is a local minimum (respectively, local maximum) with
respect to the interval $[y,z]$ (see section~\ref{sec:bumps});
$\mathit{checkOsc}(t,t^\prime,t^{\prime\prime}, \bowtie_{P},
\Gamma_{oscP}, \bowtie_{A}, \Gamma_{\mathit{oscA}})$ is a predicate
that returns whether the expression
$\mathit{oscA} \bowtie_A \Gamma_{\mathit{oscA}} \land \mathit{oscP}
\bowtie_P \Gamma_{\mathit{oscP}}$ evaluates to true for the
oscillation (with amplitude $\mathit{oscA}$ and period
$\mathit{oscP}$) determined by its first three arguments
$t,t^\prime,t^{\prime\prime}$; $\bowtie_{P}$ and $\bowtie_{A}$ are
relational operators in $\mathit{Rel}$ of $\Sigma$.

In essence, formula~\eqref{eq:osc} requires a) the existence of the
three local extrema in a proper order characterizing the complete
oscillation (i.e., either a local minimum followed by a local maximum
followed by another local minimum, or a local maximum followed by a
local minimum followed by another local maximum), and b) the
satisfaction of the constraints on the oscillation features evaluated
in the $\mathit{checkOsc}$ predicate.

\begin{figure*}[tb]
	\centering
        \begin{tikzpicture}
\begin{axis}[xmin=0, xmax=60, ymin = 0, ymax=3, xlabel={\emph{time (tu)}},ylabel={\emph{value}},every axis x label/.style={at={(current axis.right of origin)},anchor=west},
every axis y label/.style={at={(current axis.north west)},above=2mm},]
\addplot[very thick, name path global=GraphCurve, domain=0:60,samples=200] 
{sin(deg(x/2))+1};
\addplot[name path global=GraphCurve, domain=0:60,samples=200] 
{sin(deg(x/6))+1};

\draw [name path = vertical, dashed, thin] (axis cs:3.14,0) -- (axis cs:3.14,3) ; 
\draw [name path = vertical, dashed, thin] (axis cs:15.7,0) -- (axis cs:15.7,3) ;

\draw [name path = vertical, dashed, thin] (axis cs:9.42,0) -- (axis cs:9.42,3) ; 
\draw [name path = vertical, dashed, thin] (axis cs:47.1,0) -- (axis cs:47.1,3) ;

\addplot[red,name path global=Threshold1,dashed,domain=0:60] {1} ;

\draw[<->] [name path = vertical, dashed, blue, thick] (axis cs:3.14,2.3) -- (axis cs:15.7,2.3) ; \draw[<->] [name path = vertical, dashed, blue, thick] (axis cs:9.42,2.5) -- (axis cs:47.1,2.5) ;

\node[inner sep=0pt] () at (axis cs:6,2.4) { $\mathit{4\pi}$};
\node[inner sep=0pt] () at (axis cs:30,2.6) { $\mathit{12\pi}$};

\end{axis}
\end{tikzpicture}

	\caption{Two signals used to evaluate  property 
		\emph{pOSC}: signal $s_1$ (\protect \veryThickLine) satisfies the
		property, whereas $s_2$ (\protect \thinLine) violates it.}
	\label{fig:posc1}
\end{figure*}
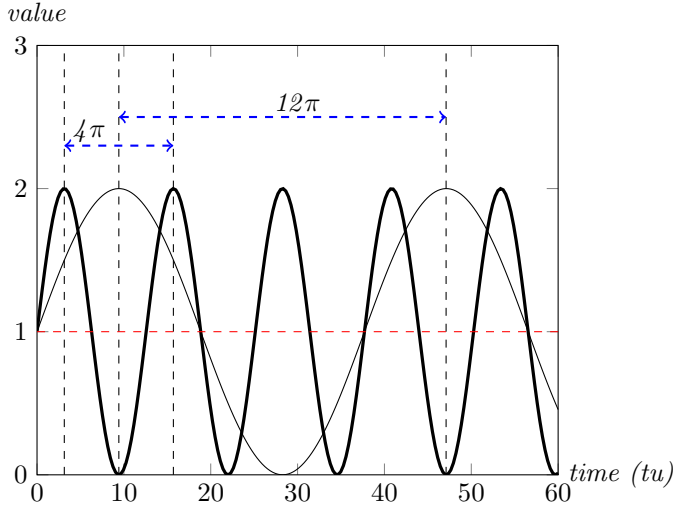

As an example, let us consider property \emph{pOSC}: ``Within an
observation interval of 60 time units (starting from the beginning of
the signal), in the signal there exist oscillations with a period less
than 20 and an amplitude less than 3''. For this property the
parameters of an instance of specification \eqref{eq:osc} are
$a=0, b=60, \Gamma_{\mathit{oscP}}=20$, $\Gamma_{\mathit{oscA}}=3$,
$\bowtie_A=\bowtie_P=<$.  For evaluating the property, we show two
signals in figure~\ref{fig:posc1}: $s_1$ (drawn with a thick line)
corresponds to a sine wave defined as $y=\sin(\frac{x}{2})+1$; $s_2$
(drawn with a thin line) is defined by $y=\sin(\frac{x}{6})+1$.  In
both signals, oscillations have a peak amplitude equal to 1, which
satisfies the constraint on the amplitude.  The period of signal
$s_1$, calculated from its sine definition, is equal to $4\pi$;
similarly, the period of $s_2$ is equal to $12\pi$ (see
figure~\ref{fig:posc1}). Signal $s_1$ satisfies property \emph{pOSC}
because it oscillates by exhibiting alternating local minima and
maxima, with a period and an amplitude satisfying the thresholds
($4\pi < \Gamma_{\mathit{oscP}}$ and $1 <
\Gamma_{\mathit{oscA}}$). However, signal $s_2$ violates the property
 because its period is greater than the threshold value of
20 ($12\pi > {\Gamma_{\mathit{oscP}}}$).

The pure sine wave shown in Figure~\ref{fig:sine} is characterized by
a constant period and by a constant amplitude. However, in the context
of CPSs, signals may be noisy; this means that the amplitude and the period of
their oscillatory behaviors may vary over time. Furthermore, a reference value may be
unknown, making the computation of the oscillation amplitude
challenging. In such cases one may use an aggregation function (e.g.,
average, maximum, minimum) over different amplitude values (e.g.,
peak-to-peak). In the following, we introduce the concepts of
\emph{average amplitude} and \emph{average period}; these definitions
can easily be adapted to take into account other aggregation functions.

To deal with situations in which the reference value is not known, we
will consider the peak-to-peak amplitude, i.e., the difference between
two adjacent extrema, denoted by $\mathit{oscA}_{\mathit{PP}}$. The \emph{average peak-to-peak 
amplitude} $\overline{\mathit{oscA}_{\mathit{PP}}}$ can
then be computed as the arithmetic mean of the peak-to-peak amplitude
between adjacent extrema. More formally, given the sequence
$p_1,\dots,p_{M-1},p_M$ of local extrema,  $\overline{\mathit{oscA}_{\mathit{PP}}}=\dfrac{\sum_{i=1}^{M-1} \abs{(s(p_{i}) - s(p_{i+1} ))}}{M-1}$.
Other definitions of amplitude (such as the root mean square) can be
used too, depending on the application domain.

The \emph{average period} can be defined as the arithmetic mean of the period
of each complete oscillation of the signal, computed over pairs of
extrema of the same type. More formally, given the sequence
$p_1,\dots,p_{M-1},p_M$ of local extrema, we define the number
$\mathit{oscN}$ of complete oscillations within the observation
interval of the signal as
$\mathit{oscN} = \left \lfloor \frac{M-1}{2} \right \rfloor $; the
\emph{average period} $\overline{\mathit{oscP}}$ is then defined as
$\overline{\mathit{oscP}}=\dfrac{\sum_{i=1}^{\mathit{oscN}} \abs{(p_{2i-1} - p_{2i+1})}}{\mathit{oscN}}$.

When the concepts of average amplitude and average period are used to
characterize an oscillatory behavior, specification~\eqref{eq:osc}
has to be adapted accordingly; more precisely, predicate
$\mathit{checkOsc}$ has to be redefined to consider the average
amplitude  $\overline{\mathit{oscA}_{\mathit{PP}}}$ and the average
period  $\overline{\mathit{oscP}}$.

\paragraph{Damped/Driven oscillations}
In the real world, oscillatory behaviors may be subject to various
forces that reduce or increase their amplitude. More precisely, we
distinguish between \emph{damped} and \emph{driven} oscillations: for
the former the amplitude decays monotonically, whereas for the latter
the amplitude increases monotonically.

The characterization of these specific behaviors can be done by
constraining the change of the amplitude of the oscillatory
signal. For example, given the sequence $p_1,\dots,p_{M-1},p_M$ of local extrema, we say that an
oscillatory signal $s$ (formalized according to
specification~\eqref{eq:osc}) exhibits damped oscillations iff the
following \sfo formula evaluates to \emph{true}:
\begin{equation}\label{eq:d1}
  \forall j \in [1,M-2]\colon\abs{(s(p_j)-s(p_{j+1}))} \geq \abs{(s(p_{j+1})-s(p_{j+2}))}
\end{equation}
The case for driven oscillations is similar and can be obtained from
the expression above  by
replacing the relational operator with its dual.

The amplitude of signals may not change monotonically; in such cases, statistical trends (e.g., a linear
trend) in amplitude changes may be observed. We could account for
statistical trends by specifying that, on average, the difference in
amplitude tends to decrease/increase; such a constraint would then be
included in the formula above.

\subsubsection{Alternative formalizations}
\paragraph{\emph{\stl}}
Similar to the case of spike properties (see section~\ref{sec:bumps}),
our formalization in \sfo of oscillation properties relies on the
existence of local extrema in the signal. Converting such
formalization to \stl would rely on the use of properly nested
``eventually'' and ``once'' operators, in conjunction with a
constraint on the oscillation period. However, a constraint on the
amplitude could not be expressed because in \stl one cannot refer to
the value of the signal at an arbitrary time point.

\paragraph{\emph{\stlstar}}
The specification of oscillatory behaviors is one of the main
motivations behind the definition of \stlstar. 
Below, we discuss how to specify property \emph{pOSC1} in \stlstar using 
the three local extrema characterization  
approaches introduced in section~\ref{sec:bumps}.

\subparagraph{(Using local extrema expressed through punctual derivatives)}
As discussed for the case of spike properties (see
page~\pageref{sec:bump:punctual-deriv-test}), properties referring to
local extrema expressed according to definition~\ref{def:extrema:punctual} cannot be specified using
\stlstar because they would require to explicitly reference the signal
value at time points that are not associated with the evaluation of a
local (sub-)formula.

\subparagraph{(Using local extrema expressed through the analytical formulation)}
We can express local extrema using their analytical formulation
(definition~\ref{def:extrema:analytical}) by
assuming a variant of \stlstar with past
operators.  Property \emph{pOSC} can be specified in 
the following way using a 3D frozen time vector:

\begin{equation*} \label{posc11}
\begin{aligned}
\logiclbl{STL*}{\textit{pOSC}} \quad \mathsf{G}_{[a,b]} (\mathsf{F}_{[0,b]*_1} 
(\mathsf{G}_{[0,\frac{\Gamma_\mathit{oscP}}{2}]} 
(s>s^{*_1}) \rightarrow\\
 \mathsf{F}_{[0,\frac{\Gamma_\mathit{oscP}}{2}]*_2} 
(\mathsf{H}_{[0,\frac{\Gamma_\mathit{oscP}}{2}]} 
(s<s^{*_2}) \\
\land \mathsf{F}_{[0,\frac{\Gamma_\mathit{oscP}}{2}]*_3} 
( \mathsf{H}_{[0,\frac{\Gamma_\mathit{oscP}}{2}]} 
(s>s^{*_3}) \\ \land \abs{(s^{*1}-s^{*2})}<3 )))\\ \land 
\mathsf{F}_{[0,b]*_1} (\mathsf{G}_{[0,\frac{\Gamma_\mathit{oscP}}{2}]} (s<s^{*_1}) \rightarrow\\
\mathsf{F}_{[0,\frac{\Gamma_\mathit{oscP}}{2}]*_2} 
(\mathsf{H}_{[0,\frac{\Gamma_\mathit{oscP}}{2}]} 
(s>s^{*_2}) \\
\land \mathsf{F}_{[0,\frac{\Gamma_\mathit{oscP}}{2}]*_3} 
(\mathsf{H}_{[0,\frac{\Gamma_\mathit{oscP}}{2}]} 
(s<s^{*_3}) \\ \land \abs{(s^{*1}-s^{*2})}<3 ))))
\end{aligned} 
\end{equation*}

In the formula above, the expression on the first row prescribes the
existence of the first local minimum, by checking all points within
the observation interval $[a,b]$ for the existence of a point (whose
time instant is frozen in the first component of the frozen time
vector) for which the corresponding signal value is smaller than all
other signal values in the interval
$[0,\frac{\Gamma_\mathit{oscP}}{2}]$; this condition is captured by
the sub-formula with the second ``globally'' operator.  The expression
on the second row, nesting the ``historically'' operator within the
``eventually'', states the presence of a local maximum (whose time
instant is frozen in the second component of the frozen time vector),
such that all the signal values between the first local minimum and
such a point are indeed smaller than the local maximum. Notice that
the distance between two neighboring extrema for an oscillation with
period $\Gamma_\mathit{oscP}$ is equal to
$\frac{\Gamma_\mathit{oscP}}{2}$.  The expression on the third row
checks for the existence of the second local minimum in a similar way;
the expression on the fourth row checks the constraint on the
peak-to-peak amplitude using the values of the signal in
correspondence of the first local minimum and of the local maximum.
The remaining part of the formula has the same structure and considers
the dual case, in which the first extremum in the oscillatory behavior
is a local maximum.

We remark that this specification assumes that the oscillation is
regular, i.e., its period is constant and the constraint on the period
is specified as ``\emph{oscP}=$\Gamma_\mathit{oscP}$". However, making
such an assumption in practice is not reasonable because typically the
shape of oscillations is unknown.

\subparagraph{(Using local extrema defined through  pre-computed derivatives)}
Property \emph{pOSC} can be expressed using definition~\ref{def:extrema:pre-computed} for local extrema, assuming the
existence  of pre-computed derivatives as 
separate signals 
$s_{c}^{\prime}$ and $s_{c}^{\prime\prime}$ and a 3D frozen time vector.
\begin{equation*} \label{posc12}
\begin{aligned}
\logiclbl{STL*}{\textit{pOSC}}  \quad \mathsf{G}_{[a,b]} ( \mathsf{F}_{[0,b]*_1} ((s^{\prime}=0 
\land 
s^{\prime\prime}>0) \rightarrow\\
\mathsf{F}_{[0,\frac{\Gamma_\mathit{oscP}}{2}]*_2} ((s^{\prime}=0 \land 
s^{\prime\prime}<0)\\
\land \mathsf{F}_{[0,\frac{\Gamma_\mathit{oscP}}{2}]*_3} ((s^{\prime}=0 \land 
s^{\prime\prime}>0) 
\\ \land 
\abs{(s^{*1}-s^{*2})}<3 )))\\ \land 
\mathsf{F}_{[0,b]*_1}((s^{\prime}=0 \land 
s^{\prime\prime}<0) \rightarrow\\
\mathsf{F}_{[0,\frac{\Gamma_\mathit{oscP}}{2}]*_2} ((s^{\prime}=0 \land 
s^{\prime\prime}>0)\\
\land \mathsf{F}_{[0,\frac{\Gamma_\mathit{oscP}}{2}]*_3} ((s^{\prime}=0 \land 
s^{\prime\prime}<0 )
\\\land 
\abs{(s^{*1}-s^{*2})}<3 ))))
\end{aligned} 
\end{equation*}

The structure of the formula above is similar to the one for the
case of using definition~\ref{def:extrema:analytical} for local
extrema, except for the direct use of the first and
second order derivatives, available as pre-computed signals.
The signal  values frozen
at the local extrema points are used to compute the peak-to-peak
amplitude of the oscillations. The same remarks made above in terms of
assuming the knowledge of the signal shape also apply in this case.

 \subsection{Relationship between signals}
The property types illustrated in the previous sections deal with only
one signal; in this section we present property types characterizing
\emph{relationships} between two (or more) signals. We
consider two types of signal relationships:
\begin{itemize}
\item \emph{functional}, based on the application of a signal
  transforming function;
\item \emph{order}, describing sequences of events/states related to signal behaviors.
\end{itemize}

\subsubsection{Functional Relationship}

The concept of a functional relationship between two (or more) signals
is captured by the application of a signal transforming function to
the signals, which yields a new signal based on the semantics of the
function. Formally, let
$\xi\colon \mathbb{D}_1 \times \mathbb{D}_2\to \mathbb{D}_3$ (with $\xi \in \mathit{Aux}$) be an
application-dependent signal transforming function\footnote{To keep
  the notation light and without loss of generality, we only consider
  a signal transforming function with arity 2.}  and let $s_1$ and
$s_2$ be two signals (called \emph{source} signals), with value
domains $\mathbb{D}_1$ and $ \mathbb{D}_2$ respectively, and domains
of definition $I_{s_1}=I_{s_2}=I_s$; the application of $\xi$ to
$s_1$ and $s_2$ yields a \emph{target} signal $s_T$ over the value
domain $\mathbb{D}_3$ defined as
$ s_T(t)=\xi\left(s_1(t), s_2(t)\right), \forall t \in I_s$. The
target signal can then be referred to in the specification of other
properties.  More precisely, let $P$ be an
instance of one of the property types seen in the previous subsections
(e.g., a data assertion), with $\xi$ the signal transforming
function defined above for the source signals $s_1$ and $s_2$. We say
that property $P$ holds on the signal representing the functional
relationship between $s_1$ and $s_2$ captured by $\xi$ iff $P$
holds on the target signal $s_T$ returned by the application of
$\xi$.

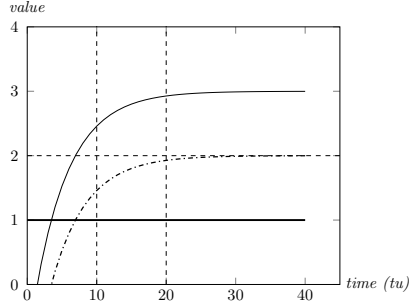
\begin{figure*}[t]
	\centering
{\scalebox{.6}{\begin{tikzpicture}

\begin{axis}[xmin=0, xmax=45, ymin = 0, ymax=4,xlabel={\emph{time (tu)}}, 
every axis x label/.style={at={(current axis.right of origin)},anchor=west},
every axis y label/.style={at={(current axis.north west)},above=2mm},
ylabel={\emph{value}},
 yticklabel shift=2pt]

\addplot[domain=0:40,samples at={0,1,...,40},line width = 0.5pt
] { -4*exp(-0.2*x)+3};
\addplot[domain=0:40,samples at={0,1,...,40}, line width = 0.5pt, thick, dash dot] { -4*exp(-0.2*x)+2};
\addplot[domain=0:40,samples at={0,1,...,40},
, very thick] {(-4*exp(-0.2*x)+3)-(-4*exp(-0.2*x)+2)};

\addplot[name path global=Threshold1, dashed,domain=0:50] {2} ;
\draw [name path = vertical, dashed, thin] (axis cs:20,0) -- (axis cs:20,50) ;
\draw [name path = vertical, dashed, thin] (axis cs:10,0) -- (axis cs:10,50) ;
\end{axis}
\end{tikzpicture} 			\label{satSignalsf}
		}
	}
 \caption{Signals used to evaluate property \emph{pRSH-F}: the
   source signals are $s_{1}$ (\protect \frshF) and $s_{2}$
     (\protect \frshS), the target signal is $s_T$
     (\protect \veryThickLine); Signal $s_T$ satisfies the
   property.}
 \label{RSHfigures2}
\end{figure*}

For example, let us consider property \emph{pRSH-F}:``The difference
between the values of signal $s_1$ and signal $s_2$ shall be equal to
1'', which contains two parts: a functional relationship part  ``The
difference between the values of signal $s_1$ and signal $s_2$\dots''
and a data assertion part  ``The [difference \dots] shall be equal to 1''.
This property is expressed in \sfo as follows:
\begin{equation} \label{p12}       
\logiclbl{SFO}{\textit{pRSH-F}} \quad \forall t \in [0,|s|): \abs(s_1(t) - s_2(t)) = 1
    \end{equation}

Figure~\ref{RSHfigures2} shows the two source signals, \signalss, as well as the target signal
$s_T$, plotted with a thick line
(\veryThickLine). Signal $s_T$ is obtained by the
application of the signal transforming function $\xi$ defined as
$\xi(s_1(t),s_2(t))\equiv\abs{(s_1(t)-s_2(t))}, \forall t \in I_s$. This signal is then used for
the actual evaluation of the data assertion contained in property
\emph{pRSH-F}, as if the latter was rewritten as ``The value of signal
$s_T$ shall be equal to 1''; since signal $s_T$ is equal to 1 across
its domain of definition, property \emph{pRSH-F} evaluates to
\emph{true}.

\subsubsection{Order Relationship}
\label{sec:order-relationships}
This type of signal relationships prescribes a sequence of events/states
corresponding to signal behaviors; in practice, it captures the
\emph{precedence} and \emph{response} temporal specification patterns
proposed in the literature~\cite{dwyer1999:patterns-in-pro}, including
their real-time extension~\cite{konrad2005:real-time-speci}. More
specifically, a precedence property specifies that an event/state (cause)
\emph{precedes} another event/state (effect); dually, a response
property requires that an event/state (effect) \emph{responds to} the occurrence
of another event/state (cause). Notice that a response property allows
effects to occur without causes, whereas a precedence property allows
causes to occur without subsequent effects. Furthermore, in the
context of real-time systems, both a precedence and a
response property  can include an additional constraint on the
temporal distance between a cause and an effect.

When dealing with signals, the events/states used to express order
relationships correspond to specific signal behaviors, which can be
further expressed (and identified) using one of the property types
seen above. More specifically, we define a \emph{signal event} as a
change in the signal value~\cite{chechik1999:events-in-prope}
occurring at a specific time instant, whereas a \emph{signal state} is
a signal behavior that holds over an interval delimited by two time
boundaries or by the occurrence of two events. In the following,
we discuss the concepts of signal events/states in the context of the
property types described in the previous sections.

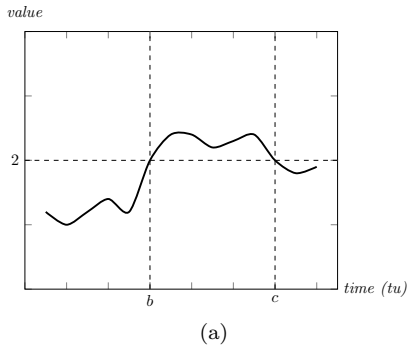
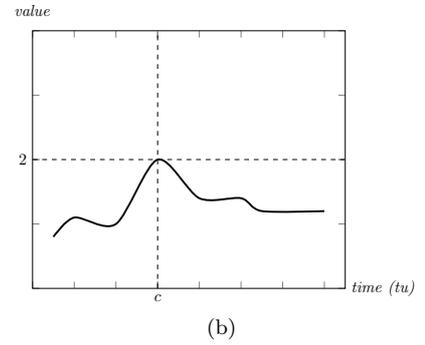
\begin{figure*}[t]
	\centering
	\subfloat[]
        {\scalebox{.6}{\begin{tikzpicture}

\begin{axis}[xmin=0, xmax=15, ymin = 0, ymax=2, xlabel={\emph{time (tu)}}, ylabel={\emph{value}}, 
yticklabels=\empty, every axis x label/.style={at={(current axis.right of origin)},anchor=west}, 
xticklabels=\empty, every axis y label/.style={at={(current axis.north west)},above=2mm},
extra y ticks={1}, extra y tick labels={$2$}, extra x 
ticks={6,12}, extra x tick 
labels={ $b$,$c$}
 ]

\node[inner sep=0pt] (n1) at (axis cs:1,0.6) {};
    \node[inner sep=0pt] (n2) at (axis cs:2,0.5) {};
    \node[inner sep=0pt] (n3) at (axis cs:3,0.6) {};
    \node[inner sep=0pt] (n4) at (axis cs:4,0.7) {};
    \node[inner sep=0pt] (n5) at (axis cs:5,0.6) {};
    \node[inner sep=0pt] (n6) at (axis cs:6,1.0) {};
    \node[inner sep=0pt] (n7) at (axis cs:7,1.2) {};
    \node[inner sep=0pt] (n8) at (axis cs:8,1.2) {};
    \node[inner sep=0pt] (n9) at (axis cs:9,1.1) {};
    \node[inner sep=0pt] (n10) at (axis cs:10,1.15) {};
    \node[inner sep=0pt] (n11) at (axis cs:11,1.2) {};
    \node[inner sep=0pt] (n12) at (axis cs:12,1.0) {};
    \node[inner sep=0pt] (n13) at (axis cs:13,0.9) {};
    \node[inner sep=0pt] (n14) at (axis cs:14,0.95) {};
\path[name path = GraphCurve, draw, very thick] plot [smooth, very thick] coordinates { (n1) (n2) (n3) (n4) (n5)
           (n6) (n7) (n8) (n9) (n10) (n11) (n12) (n13) (n14)};

       \addplot [dashed] coordinates {(6, 0) (6, 2)};
       \addplot [dashed] coordinates {(12, 0) (12, 2)};
\addplot [dashed,domain=0:15] {1};

\end{axis}
\end{tikzpicture} 
 }
                  \label{fig:daBool}}
	\hfill
	\subfloat[]
        	{\scalebox{.6}{\begin{tikzpicture}

\begin{axis}[xmin=0, xmax=15, ymin = 0, ymax=2, xlabel={\emph{time (tu)}}, ylabel={\emph{value}}, 
yticklabels=\empty, every axis x label/.style={at={(current axis.right of origin)},anchor=west}, 
xticklabels=\empty, every axis y label/.style={at={(current axis.north west)},above=2mm},
extra y ticks={1}, extra y tick labels={$2$}, extra x 
ticks={6}, extra x tick 
labels={ $c$}
 ]
  
\node[inner sep=0pt] (n1) at (axis cs:1,0.4) {};
    \node[inner sep=0pt] (n2) at (axis cs:2,0.55) {};
    \node[inner sep=0pt] (n3) at (axis cs:4,0.5) {};
    \node[inner sep=0pt] (n4) at (axis cs:6,1.0) {};
    \node[inner sep=0pt] (n5) at (axis cs:8,0.7) {};
    \node[inner sep=0pt] (n6) at (axis cs:10,0.7) {};
    \node[inner sep=0pt] (n7) at (axis cs:11,0.6) {};
    \node[inner sep=0pt] (n8) at (axis cs:14,0.6) {};

\path[name path = GraphCurve, draw, very thick] plot [smooth, very thick] coordinates { (n1) (n2) (n3) (n4) (n5) (n6) (n7) (n8) };
    
      \addplot [dashed] coordinates {(6, 0) (6, 2)};
      \addplot [dashed,domain=0:15] {1};
       
\end{axis}
\end{tikzpicture} 

 }
	\label{fig:instantBool}	}

	\caption{\protect\subref{fig:daBool} 
        A signal being in the state characterized by property \emph{pDAs} in the
	interval 
	[$b$,$c$].~\protect\subref{fig:instantBool} A signal changing
          its value to 2 at time instant $c$, satisfying property \emph{pDAe}. 
	}
	\label{fig:edges}
      \end{figure*}

\paragraph{Data assertions}

The typical use of data assertions\footnote{For simplicity, in the
  following we consider data assertion properties defined on one time
  interval.} is to represent signal states, as in property
\emph{pDAs}: ``The signal value shall be greater than or equal to
2''. For example, figure~\ref{fig:daBool} shows a signal that
satisfies this property in the interval $[b$,$c]$.

Another formulation of this type of properties corresponds to signal
events. As an example, let us consider property \emph{pDAe}: ``The
signal value shall become equal to 2''. Informally, this property
corresponds to a predicate that captures the event of the signal
\emph{becoming} equal to 2, i.e., changing from a value different from
2 to the actual value of 2. This behavior can be seen in the signal
plotted in figure~\ref{fig:instantBool}: property \emph{pDAe} holds at
time instant $c$.

Notice that signal events can be used to characterize the boundaries
of a signal state: for example, the time instants delimiting the
interval in which the state represented by property \emph{pDAs} holds
correspond to the time instants in which the event represented by
property \emph{pDAe} and by its negation (i.e., ``signal $s$ becoming
different from 2'') occur.

\paragraph{Spike}
When a signal satisfies a spike property following the
specification template~\eqref{bump:bio} on page~\pageref{bump:bio},
the spike behavior of the signal can be associated with three
different events, corresponding to the time instants in which the peak
point and the two valley points of the spike shape (see
section~\ref{sec:bumps}) occur. The actual choice of the most relevant
event among these three is application-specific.  Furthermore, the
state induced by such a property type is defined over the interval
$[\mathit{VP}_{1},\mathit{VP}_{2}]$; such a state lasts for a duration corresponding to
the spike width $w$.

\paragraph{Oscillation}
When a signal satisfies an oscillation property following the
specification template~\eqref{eq:osc} in section~\ref{sec:osc}, the
oscillatory behavior of the signal can be associated with distinct
events, corresponding to the time instants in which the extrema points
of the oscillations occur. The choice among these events is
application-specific.  Moreover, the state induced by such a property
type is defined over the interval bounded by the first and last
observed extrema of the oscillation.

\paragraph{Functional relationship between signals}
Similar to data assertions, functional relationship between signals
can represent either signal events (captured by a predicate
``\emph{becomes}'') or signal states.

\paragraph{Formalization}  After defining the concepts of events and states associated with
signal property types, we are now ready to formalize the concept of
order relationship between signal behaviors.

Given a signal $s$ and an instance $P$ of one of the signal property
types described above, we define the \emph{signal event boolean
  projection} of $P$ on $s$ as the predicate
$\sigma^{\mathbb{B}e}_{s,P}(t)$, which evaluates to true iff the event
associated with the signal behavior specified in $P$ occurs in signal
$s$ at time instant $t$; similarly, we define the \emph{signal state
  boolean projection} of $P$ on $s$ as the predicate
$\sigma^{\mathbb{B}s}_{s,P}(t)$, which evaluates to true iff the state
associated with the signal behavior specified in $P$ holds on signal
$s$ at time instant $t$.

Given two signals $s_1$ and $s_2$ with domains of definition
$I_{s_1}=I_{s_2}=[0,r)$ and lengths  $|s_1|=|s_2|$ denoted with $|s|$, and two signal-based properties 
$P_1$ and
$P_2$, we say that the event captured by $P_2$ in $s_2$ \emph{responds
  to} (following the ``response''
pattern in~\cite{dwyer1999:patterns-in-pro}) the event captured by $P_1$ in $s_1$  iff the following
\sfo formula evaluates to \emph{true}:
\begin{equation} \label{eq:respe}
\forall t \in [0,|s|) \colon  \uparrow \sigma^{\mathbb{B}e}_{s_1,P_1}(t)   \rightarrow
 \left(\exists k \in (t,|s|)\colon   \uparrow \sigma^{\mathbb{B}e}_{s_2,P_2}(k)  \right)
\end{equation}
where $\uparrow$ denotes the rising edge operator, defined as
$\uparrow s(t) \equiv s(t)=1 \land \exists c \in (0,t): \forall
c^{\prime} \in (0,c): s(t-c^{\prime})=0$. 

If the relevant behavior captured by a property results in a state
instead of an event, the formula above becomes:
\begin{equation} \label{eq:resps}
  \forall t \in [0,|s|) \colon  \sigma^{\mathbb{B}s}_{s_1,P_1}(t)   \rightarrow
 \left(\exists k \in(t,|s|)\colon   \sigma^{\mathbb{B}s}_{s_2,P_2}(k)  \right)
\end{equation}

Similarly,  we say that the event captured by $P_1$ in $s_1$
\emph{precedes} (following the ``precedence''
pattern in~\cite{dwyer1999:patterns-in-pro}) the event captured 
by $P_2$ in $s_2$ iff the following formula evaluates to \emph{true}:
\begin{equation} \label{eq:prece}
\forall t \in [0,|s|) \colon \uparrow \sigma^{\mathbb{B}e}_{s_2,P_2}(t) 
\rightarrow
\left(\exists k \in[0,t)\colon \uparrow \sigma^{\mathbb{B}e}_{s_1,P_1}(k) \right)
\end{equation}
When the relevant behavior captured by a property results in a state
instead of an event, the formula above becomes:
\begin{equation} \label{eq:precs}
  \forall t \in [0,|s|)\colon  \sigma^{\mathbb{B}s}_{s_2,P_2}(t)  \rightarrow
\left(\exists k \in[0,t) \colon  \sigma^{\mathbb{B}s}_{s_1,P_1}(k) \right)
\end{equation}

In some cases, an order relationship may prescribe a temporal distance
between the cause and the effect. We assume this distance to be
specified as a bound of the form $\bowtie n$, where $ \bowtie{} \in \mathit{Rel}$
 and $n \in \mathbb{R}$. In this case the formulae above have
to be extended to take the distance into account, by conjoining the clause
$\abs{(k-t)}\bowtie n$ to the consequent. For example,
formula~\eqref{eq:respe} will become:
\begin{equation} \label{eq:respe-t}
  \forall t \in [0,|s|) \colon  \uparrow \sigma^{\mathbb{B}e}_{s_1,P_1}(t)   \rightarrow
 \left(\exists k \in (t,|s|) \colon  \uparrow \sigma^{\mathbb{B}e}_{s_2,P_2}(k)  \land \abs{(k-t)} \bowtie n \right)
\end{equation}
Notice that when one property induces a state and the other induces an event, the
resulting formula for the corresponding  order relationship is obtained by
opportunely combining  the occurrences of  the signal boolean
projection functions for states and events, following one of the above
templates.

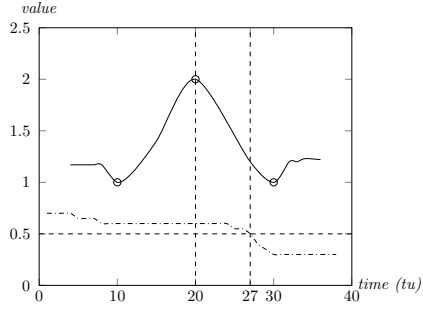
\begin{figure*}[tb]
 	\centering
 	\scalebox{.6}{\begin{tikzpicture}

\begin{axis}[xmin=0, xmax=40, ymin = 0, ymax=2.5,xlabel={\emph{time (tu)}}, 
every axis x label/.style={at={(current axis.right of origin)},anchor=west},
every axis y label/.style={at={(current axis.north west)},above=2mm},
ylabel={\emph{value}},extra x ticks={27}, extra x tick labels={$27$},
yticklabel shift=2pt
]

\node[inner sep=0pt] (n1) at (axis cs:4,1.17) {};
    \node[inner sep=0pt] (n2) at (axis cs:5,1.17) {};
    \node[inner sep=0pt] (n3) at (axis cs:6,1.17) {};
    \node[inner sep=0pt] (n4) at (axis cs:7,1.17) {};
    \node[inner sep=0pt] (n5) at (axis cs:8,1.17) {};
    \node[inner sep=0pt] (n6) at (axis cs:10,1) {};
    \draw[] (10,1) circle (0.08 cm);
    \node[inner sep=0pt] (n7) at (axis cs: 12,1.1) {};
    \node[inner sep=0pt] (n8) at (axis cs:15,1.4) {};
    \node[inner sep=0pt] (n9) at (axis cs:20,2) {};
    \draw[] (20,2) circle (0.08 cm);
    \node[inner sep=0pt] (n10) at (axis cs:27,1.2) {};
    \node[inner sep=0pt] (n11) at (axis cs:30,1) {};
    \draw[] (30,1) circle (0.08 cm);
    \node[inner sep=0pt] (n12) at (axis cs:32,1.2) {};
    \node[inner sep=0pt] (n13) at (axis cs:33,1.2) {};
    \node[inner sep=0pt] (n14) at (axis cs:34,1.23) {};
    \node[inner sep=0pt] (n15) at (axis cs:36,1.22) {};

\path[name path = GraphCurve, draw] plot [smooth, thick, line width = 0.5pt] coordinates { (n1) (n2) (n3) (n4) (n5) (n6) 
      (n7) (n8) (n9) (n10) (n11) (n12) (n13) (n14) (n15) };

    \node[inner sep=0pt] (n1) at (axis cs:1,0.7) {};
    \node[inner sep=0pt] (n2) at (axis cs:2,0.7) {};
    \node[inner sep=0pt] (n3) at (axis cs:3,0.7) {};
    \node[inner sep=0pt] (n4) at (axis cs:4,0.7) {};
    \node[inner sep=0pt] (n5) at (axis cs:5,0.65) {};
    \node[inner sep=0pt] (n6) at (axis cs:6,0.65) {};
    \node[inner sep=0pt] (n7) at (axis cs:7,0.65) {};
    \node[inner sep=0pt] (n8) at (axis cs:8,0.6) {};
    \node[inner sep=0pt] (n9) at (axis cs:9,0.6) {};
    \node[inner sep=0pt] (n10) at (axis cs:10,0.6) {};
    \node[inner sep=0pt] (n11) at (axis cs:11,0.6) {};
    \node[inner sep=0pt] (n12) at (axis cs:12,0.6) {};
    \node[inner sep=0pt] (n13) at (axis cs:13,0.6) {};
    \node[inner sep=0pt] (n14) at (axis cs:14,0.6) {};
    \node[inner sep=0pt] (n15) at (axis cs:15,0.6) {};
    \node[inner sep=0pt] (n16) at (axis cs:16,0.6) {};
    \node[inner sep=0pt] (n17) at (axis cs:17,0.6) {};
    \node[inner sep=0pt] (n18) at (axis cs:18,0.6) {};
    \node[inner sep=0pt] (n19) at (axis cs:19,0.6) {};
    \node[inner sep=0pt] (n20) at (axis cs:20,0.6) {};
    \node[inner sep=0pt] (n21) at (axis cs:21,0.6) {};
    \node[inner sep=0pt] (n22) at (axis cs:22,0.6) {};
    \node[inner sep=0pt] (n23) at (axis cs:23,0.6) {};
    \node[inner sep=0pt] (n24) at (axis cs:24,0.6) {};
    \node[inner sep=0pt] (n25) at (axis cs:25,0.55) {};
    \node[inner sep=0pt] (n26) at (axis cs:26,0.55) {};
    \node[inner sep=0pt] (n27) at (axis cs:27,0.5) {};
    \node[inner sep=0pt] (n28) at (axis cs:28,0.4) {};
    \node[inner sep=0pt] (n29) at (axis cs:29,0.35) {};
    \node[inner sep=0pt] (n30) at (axis cs:30,0.3) {};
    \node[inner sep=0pt] (n31) at (axis cs:31,0.3) {};
    \node[inner sep=0pt] (n32) at (axis cs:32,0.3) {};
    \node[inner sep=0pt] (n33) at (axis cs:33,0.3) {};
    \node[inner sep=0pt] (n34) at (axis cs:34,0.3) {};
    \node[inner sep=0pt] (n35) at (axis cs:35,0.3) {};
    \node[inner sep=0pt] (n36) at (axis cs:36,0.3) {};
    \node[inner sep=0pt] (n37) at (axis cs:37,0.3) {};
    \node[inner sep=0pt] (n38) at (axis cs:38,0.3) {};

\path[name path = GraphCurve, draw, line width = 0.5pt, dash dot] plot [smooth, thick] coordinates { (n1) (n2) (n3) (n4) (n5) (n6) (n7) (n8) (n9) (n10) (n11) (n12) (n13) (n14) (n15) (n16) (n17) (n18) (n19) (n20) (n21) (n22) (n23) (n24) (n25) (n26) (n27) (n28) (n29) (n30) (n31) (n32) (n33) (n34) (n35) (n36) (n37) (n38) 
}; 

    \draw [name path = vertical, dashed, very thin] (axis cs:20,0) -- (axis cs:20,2.5) ;
     \draw [name path = vertical, dashed, very thin] (axis cs:27,0) -- (axis cs:27,2.5) ;
    \addplot[name path global=Threshold1, dashed, very thin, domain=0:50] {0.5} ;
\end{axis}
\end{tikzpicture}

 } 
 		\caption{Signals $s_1$ (\protect \frshF) and $s_2$ (\protect \frshS) used to evaluate  
 		property \emph{pRSH-O}; the property holds. 
               }
               	\label{fig:bumpsOrder}
\end{figure*}

Order relationship properties can be defined recursively, i.e., when
the cause and/or effect sub-property is also an order relationship. In
these cases, we consider an event-based interpretation of the
cause/effect sub-property.

As an example of order relationship property, let us consider the
following response property \emph{pRSH-O}: ``If in signal $s_1$ there
is a spike with a maximum width of \SI{30}{\tu} and a
maximum amplitude of 1, then---within \SI{10}{tu}---the value of
signal $s_2$ shall become less than 0.5''.  Assuming we use an event-based
interpretation of both cause and effect sub-properties, we can rewrite the
property as \emph{pRSH-O$^\prime$} : ``If there is an event 
corresponding to [signal $s_1$ having a spike with a maximum
width of \SI{30}{\tu} and a maximum amplitude of 1] then---within
\SI{10}{tu}---there shall be an event corresponding to
[signal $s_2$ becoming less than 0.5]''.  In this
instance of the response pattern, the cause is represented by the
spike property ``In signal $s_1$ there is a spike with a
maximum width of \SI{30}{\tu} and a maximum amplitude of 1'', whereas
the effect is represented by the data assertion property ``Signal
$s_2$ shall become less than 0.5''; furthermore, the temporal distance
between the cause and the effect can be at most \SI{10}{tu}.  We
refer to the cause and effect sub-properties as $P_1$ and $P_2$, respectively.

The specification of property \emph{pRSH-O} in \sfo is the following:
\begin{equation} \label{p13}
\begin{split}
 \logiclbl{SFO}{\textit{pRSH-O}} \quad
\forall t \in [0,|s_1|) \colon
\uparrow \sigma^{\mathbb{B}e}_{s_1,P_1}(t) \\ \rightarrow
\left(\exists k \in (t,|s_2|) \colon \uparrow \sigma^{\mathbb{B}e}_{s_2,P_2}(k) \land \abs{(k-t)} \le 10 \right)
\end{split}
\end{equation}
where $\sigma^{\mathbb{B}e}_{s_1,P_1}$ and
$\sigma^{\mathbb{B}e}_{s_2,P_2}$ are the signal event boolean
projection predicates.

We evaluate the property with respect to the
two signals shown in figure~\ref{fig:bumpsOrder}, \signalss.  In this
example, we assume that the signal boolean projection predicate for
spike properties (used for the evaluation of the cause sub-property) is defined
such that it is true at the actual time instant at which the spike
peak point occurs (i.e., \SI{20}{tu}).
By looking at figure~\ref{fig:bumpsOrder}, we see that
property \emph{pRSH-O} holds on $s_1$ and $s_2$ because the event
captured by the effect sub-property (the change of value of $s_2$
happening at time instant \SI{27}{tu}) responds to the occurrence of
the event associated with the cause sub-property within the prescribed time
bound (since $\abs{(\SI{27}{tu}-\SI{20}{tu})} = \SI{7}{tu} < \SI{10}{tu}$).

\subsubsection{Transient Behaviors}
We consider transient signal behaviors (i.e., behaviors of a signal
when changing from the current value to its target value) as a special
case of order relationship. This category includes \emph{rise time}
(and \emph{fall time}) and \emph{overshoot} (and \emph{undershoot}) properties.

\paragraph{\textit{Rise time} (\textit{Fall time})} 
We say that a signal exhibits a \emph{rising} (dually, \emph{falling}) behavior
when its value increases (decreases) towards a target value.  Informally speaking, a property on the
\emph{rise (fall) time} defines a constraint on the time by which the
signal reaches the target value. More specifically, it defines a constraint on the temporal distance between 
two events:
\begin{inparaenum}[1)]
\item a (generic) cause event, also called \emph{trigger event}, that coincides
  with the signal starting to manifest a transient behavior;
\item an effect event that represents the signal reaching the target value.
\end{inparaenum}

Figure~\ref{fig:risetimeDef} depicts a signal exhibiting a rising
behavior starting from time instant $\mathit{st}$. The signal rises
monotonically from the value $s(\mathit{st})$ and reaches the target
value $s_{\mathit{target}}$ at time instant $c$; the time interval
$[\mathit{st},c]$ is called \emph{rise interval}.  The left bound of
the rise interval, also called \emph{trigger time}, corresponds to the
time instant at which the trigger event occurs. The right bound of the
rise interval corresponds to the occurrence of the effect event, in
which signal $s$ reaches the target value.  The trigger time can also
be expressed in terms of an absolute time reference value; in such a
case, the trigger event is the event in which a special \emph{clock}
signal reaches a certain value.

A \rt property defines a constraint on the right bound of the rise
interval. More formally, given two signals $s_{\mathit{tr}}$ and $s$
with domains of definition $I_{s_{\mathit{tr}}}=I_{s}=[0,r)$, let $P_{\mathit{tr}}$
and $P$ be two signal-based properties.  Property $P_{\mathit{tr}}$ captures the
trigger event defined in terms of the behavior of $s_{\mathit{tr}}$; property $P$
captures the event of $s$ reaching the target value. A \rt property
bounds the \rt of $s$ by a threshold $\mathit{RT} \in \mathbb{N}$
(indicated by the end-user);
such a property holds iff the following \sfo formula evaluates to
\emph{true}:
\begin{equation} \label{eq:rise}
\begin{split}
\forall \mathit{st} \in [0,|s_{\mathit{tr}}|)\colon 
\uparrow \sigma^{\mathbb{B}e}_{s_{\mathit{tr}},P_{\mathit{tr}}}(\mathit{st}) \rightarrow
\left(\exists k \in[\mathit{st},\mathit{st}+\mathit{RT}]\colon  \uparrow \sigma^{\mathbb{B}e}_{s,P}(k) \right )
\end{split}
\end{equation}
A stricter definition requiring signal $s$ to 
rise (strictly) monotonically can be expressed by adding the
conjunct $ \forall j \in [\mathit{st},\mathit{st}+k)\colon 
\forall j^{\prime} \in (j,\mathit{st}+k] : s(j) < s(j^{\prime}) $ to the
consequent in the formula
above.

A \ft constraint can be expressed
in a similar way, replacing the relational operators with their duals.

As an example, let us consider the \rt property \emph{pRT}: ``If
signal $s_{\mathit{tr}}$ becomes greater than 1, then signal $s$ shall
reach the target value of 2 within at most \SI{8}{\tu}''. The trigger
event in this property is represented by the data assertion property
$P_{\mathit{tr}}$: ``The value of signal $s_{\mathit{tr}}$ becomes greater than 1''.  The
effect sub-property of this order relationship property can be
specified with the data assertion property $P$: ``The value of
signal $s$ shall become greater than 2''. The constraint on the rise
time is \SI{8}{\tu}.  Property \textit{pRT} can be expressed in \sfo
as:
\begin{equation} \label{eq:rise-ex}
\begin{split}
 \logiclbl{SFO}{\textit{pRT}}\quad
\forall \mathit{st} \in [0,|s_\mathit{tr}|)\colon \uparrow \sigma^{\mathbb{B}e}_{s_{\mathit{tr}},P_{\mathit{tr}}}(\mathit{st}) \rightarrow
\left(\exists k \in[\mathit{st},\mathit{st}+8]\colon  \uparrow \sigma^{\mathbb{B}e}_{s,P}(k)
\right )
\end{split}
\end{equation}

We evaluate  property  \emph{pRT} with respect to signal $s$ on the two signals shown in Figure~\ref{fig:risetime}:
\signals.  In the figure, an arrow at timestamp \SI{4}{\tu} denotes
the trigger time $\mathit{st}$ corresponding to the trigger event
captured by property $P_{\mathit{tr}}$ for signal $s_\mathit{tr}$ 
drawn with a dash-dotted line (\protect \frshS). The maximum allowed value for the right 
bound of the
rise interval ($\mathit{st}+\mathit{RT}=4+8=\SI{12}{\tu}$) is indicated with 
a red, vertical dashed line.
Signal $s_1$ satisfies the property because it reaches the target
value (2) at time instant $\SI{9}{\tu} < \mathit{st}+\mathit{RT}$.
Signal $s_2$ violates the property
because it does not reach the target value by time
instant $\mathit{st}+\mathit{RT}=\SI{12}{\tu}$.

The variant \emph{pRT-monot} of property \emph{pRT} with a
monotonicity constraint can be expressed in \sfo as:
\begin{equation} \label{eq:risem}
\begin{split}
\logiclbl{SFO}{\textit{pRT-monot}}\quad
\forall \mathit{st} \in [0,|s_\mathit{tr}|)\colon \uparrow
\sigma^{\mathbb{B}e}_{s_{\mathit{tr}},P_{\mathit{tr}}}(\mathit{st}) \rightarrow\\
\left(\exists k \in[\mathit{st},\mathit{st}+8]\colon  \uparrow \sigma^{\mathbb{B}e}_{s,P}(k) \land \forall j \in [\mathit{st},\mathit{st}+k)\colon 
\forall j^{\prime} \in (j,\mathit{st}+k] : s(j) < s(j^{\prime}) \right )
\end{split}
\end{equation}

\begin{figure*}[tb]
	\centering
	\subfloat[]
	{\scalebox{.6}{
\begin{tikzpicture}

\begin{axis}[xmin=0, xmax=15, ymin = 0, ymax=3, xlabel={\emph{time (tu)}}, ylabel={\emph{value}}, yticklabels=\empty, every axis x label/.style={at={(current axis.right of origin)},anchor=west}, xticklabels=\empty, every axis y label/.style={at={(current axis.north west)},above=2mm},
extra y ticks={2}, extra y tick labels={$\mathit{s_{target}}$}, extra x ticks={2,9, 12}, extra x tick 
labels={$\mathit{st}$, $c$,$\mathit{st+RT}$}
 ]
   
\node[inner sep=0pt] (n1) at (axis cs:1,0.5) {};
        \node[inner sep=0pt] (n2) at (axis cs:2,0.42) {};
        \node[inner sep=0pt] (n3) at (axis cs:4,0.58) {};
        \node[inner sep=0pt] (n4) at (axis cs:6,1) {};
        \node[inner sep=0pt] (n5) at (axis cs:9,2) {};
        \node[inner sep=0pt] (n6) at (axis cs:10.5,2.35) {};
        \node[inner sep=0pt] (n7) at (axis cs:12,2.5) {};
        \node[inner sep=0pt] (n8) at (axis cs:14,2.5) {};
        \node[inner sep=0pt] (n9) at (axis cs:14.5,2.5) {};

\path[name path = GraphCurve, draw, color=black, very thick] plot [smooth] coordinates { (n1) (n2) (n3) (n4) (n5) (n6) (n7) (n8) (n9)};

        \addplot [dashed] coordinates {(2, 0) (2, 3)};
        \addplot [dashed] coordinates {(9, 0) (9, 3)};
        \addplot [dashed] coordinates {(12, 0) (12, 3)};
        \addplot [dashed,domain=0:15] {2};

\end{axis}
\end{tikzpicture}

 			\label{fig:risetimeDef}
		}
	}
	\hfill
	\subfloat[]
	{\scalebox{.6}{\begin{tikzpicture}

\begin{axis}[xmin=0, xmax=15, ymin = 0, ymax=3, xlabel={\emph{time (tu)}}, ylabel={\emph{value}}, 
every axis x label/.style={at={(current axis.right of origin)},anchor=west}, every axis y 
label/.style={at={(current axis.north west)},above=2mm},extra x ticks={9},
 ]
  
\node[inner sep=0pt] (n1) at (axis cs:0.5,0.6) {};
        \node[inner sep=0pt] (n2) at (axis cs:1,0.2) {};
        \node[inner sep=0pt] (n3) at (axis cs:2,0.7) {};
\node[inner sep=0pt] (n5) at (axis cs:4,0.4) {};
        \node[inner sep=0pt] (n6) at (axis cs:6,0.8) {};
        \node[inner sep=0pt] (n7) at (axis cs:9,1.5) {};
        \node[inner sep=0pt] (n8) at (axis cs:10,1.6) {};
        \node[inner sep=0pt] (n9) at (axis cs:11,1.85) {};
        \node[inner sep=0pt] (n10) at (axis cs:12,1.9) {};
        \node[inner sep=0pt] (n11) at (axis cs:12.7,2.1) {};
        \node[inner sep=0pt] (n12) at (axis cs:13,1.9) {};
        \node[inner sep=0pt] (n13) at (axis cs:14,2.1) {};
        \node[inner sep=0pt] (n14) at (axis cs:14.5,2.1) {};

\path[name path = GraphCurve, draw, color=black] plot [smooth] coordinates { (n1) (n2) (n3) (n5)  (n6) (n7) (n8)
        (n9) (n10) (n11) (n12) (n13) (n14)};

\node[inner sep=0pt] (n1) at (axis cs:1,1) {};
        \node[inner sep=0pt] (n2) at (axis cs:2,1) {};
        \node[inner sep=0pt] (n3) at (axis cs:4,0.58) {};
        \node[inner sep=0pt] (n4) at (axis cs:6,1) {};
        \node[inner sep=0pt] (n5) at (axis cs:9,2) {};
        \node[inner sep=0pt] (n6) at (axis cs:10.5,2.35) {};
        \node[inner sep=0pt] (n7) at (axis cs:12,2.5) {};
        \node[inner sep=0pt] (n8) at (axis cs:14,2.5) {};
        \node[inner sep=0pt] (n9) at (axis cs:14.5,2.5) {};

\path[name path = GraphCurve, draw, color=black, very thick] plot [smooth] coordinates { (n1) (n2) (n3) 
        (n4) (n5) (n6) (n7) (n8) (n9)};

         \node[inner sep=0pt] (n1) at (axis cs:1,0.7) {};
        \node[inner sep=0pt] (n2) at (axis cs:3,0.7) {};
        \node[inner sep=0pt] (n3) at (axis cs:4,1) {};
        \node[inner sep=0pt] (n4) at (axis cs:7,1.7) {};
        \node[inner sep=0pt] (n5) at (axis cs:9,2.5) {};
        \node[inner sep=0pt] (n6) at (axis cs:11,3) {};
        \node[inner sep=0pt] (n7) at (axis cs:12,4) {};
        \node[inner sep=0pt] (n8) at (axis cs:14,5) {};

\path[name path = GraphCurve, draw, line width = 0.5pt, dash dot] plot [smooth, thick] coordinates { 
        (n1) (n2) (n3) (n4) (n5) (n6) (n7) (n8) };

        \addplot [dashed] coordinates {(4, 0) (4, 3)};
        
\addplot [dashed] coordinates {(9, 0) (9, 3)};
        \addplot [dashed, red,very thick] coordinates {(12, 0) (12, 3)};
        \addplot [dashed,domain=0:15] {2};
        \draw[line width=0.3mm,->] (axis cs:3,1) -- (axis cs:3.9,1) node[above left, font=\tiny] {} ;

\end{axis}
\end{tikzpicture}

 			\label{fig:risetime}
		}
	}
	\caption{\protect\subref{fig:risetimeDef} Main concepts related to
		the specification of \rt. \protect\subref{fig:risetime} two signals 
		used to evaluate property 
		\emph{pRT}: signal \protect\so satisfies the property, whereas \protect\st violates 
		it.} 
	\label{fig:risetime-full}
\end{figure*}

\paragraph{\textit{Overshoot} (\textit{Undershoot})}

We say that a signal exhibits an \emph{overshoot} (dually, \emph{undershoot}) behavior when it exceeds (goes below) its target value\footnote{Other definitions of overshoot also constrain the behavior of the signal after 
it exceeds (goes below) the target value, e.g., by requiring it to converge back to the target value.}. Informally 
speaking, an overshoot  property specifies the maximum signal value, above  the target value, 
that a signal can reach when overshooting within a certain time
interval; an undershoot property is defined dually. 

Figure~\ref{fig:overshootDef} depicts a signal exhibiting an overshoot
behavior starting from time instant $\mathit{st}$. This time instant
is the \emph{trigger time} and can be specified in different ways, as
discussed above in the context of rise time properties. The signal
rises from the value $s(\mathit{st})$ and overshoots the target value
$s_{\mathit{target}}$ after time instant $c$, reaching the maximum
magnitude $s_{\mathit{max}}$ at time instant $b$. The time interval
$[c, c+\mathit{OI}]$ is called \emph{overshoot interval}; its width
$\mathit{OI}$ is specified by the end-user. This signal overshoots the
target value $s_{\mathit{target}}$ by an \emph{overshoot value}
$O_s =s_{\mathit{max}}-s_{\mathit{target}}$.  An overshoot property
defines a boundary on the overshoot value within the overshoot
interval; such a boundary is expressed either with an absolute value
or with a relative value with respect to the target value.

\begin{figure*}[tb]
  \centering
  \subfloat[]
  {\scalebox{.6}{
\begin{tikzpicture}

\begin{axis}[xmin=0, xmax=15, ymin = 0, ymax=3, xlabel={\emph{time (tu)}}, ylabel={\emph{value}}, yticklabels=\empty, every axis x label/.style={at={(current axis.right of origin)},anchor=west}, xticklabels=\empty, every axis y label/.style={at={(current axis.north west)},above=2mm},
extra y ticks={2,2.5}, extra y tick labels={$\mathit{s_{target}}$,$\mathit{s_{max}}$}, extra x 
ticks={2,9,11, 13}, extra x tick 
labels={$\mathit{st}$, $c$,$b$, $c+\emph{OI}$}
 ]

\node[inner sep=0pt] (n1) at (axis cs:1,0.6) {};
    \node[inner sep=0pt] (n2) at (axis cs:2,0.5) {};
    \node[inner sep=0pt] (n3) at (axis cs:4,0.7) {};
    \node[inner sep=0pt] (n4) at (axis cs:6,1.1) {};
    \node[inner sep=0pt] (n5) at (axis cs:8,1.7) {};
    \node[inner sep=0pt] (n6) at (axis cs:9,2) {};
    \node[inner sep=0pt] (n7) at (axis cs:10,2.3) {};
    \node[inner sep=0pt] (n8) at (axis cs:11,2.5) {};
    \node[inner sep=0pt] (n9) at (axis cs:12,2.3) {};
    \node[inner sep=0pt] (n10) at (axis cs:13,2.25) {};
    \node[inner sep=0pt] (n11) at (axis cs:14,2.25) {};

\path[name path = GraphCurve, draw, very thick] plot [smooth, very thick] coordinates { (n1) (n2) (n3) (n4) (n5) (n6) (n7)(n8) (n9) (n10) (n11) 
      };    

        \addplot [dashed] coordinates {(2, 0) (2, 3)};
        \addplot [dashed] coordinates {(9, 0) (9, 3)};
        \addplot [dashed] coordinates {(11, 0) (11, 3)};
        \addplot [dashed] coordinates {(13, 0) (13, 4)};
\addplot [dashed,domain=0:15] {2};
        \addplot [dashed,domain=0:15] {2.5};

       \draw[<->] [name path = vertical, dashed] (axis cs:9,0.5) -- (axis cs:13,0.5) ;
       \node[inner sep=0pt] (n1) at (axis cs:10.4,0.62) { $\emph{OI}$};

\end{axis}
\end{tikzpicture}

 }
    \label{fig:overshootDef}
  }
  \hfill
  \subfloat[]
  {\scalebox{.6}{
\begin{tikzpicture}

\begin{axis}[xmin=0, xmax=15, ymin = 0, ymax=4, xlabel={\emph{time (tu)}}, ylabel={\emph{value}}, 
every axis x label/.style={at={(current axis.right of origin)},anchor=west}, every axis y 
label/.style={at={(current axis.north west)},above=2mm},extra x ticks={2,5,7,11,13}
 ]
   
\node[inner sep=0pt] (n1) at (axis cs:1,0.5) {}; 
    \node[inner sep=0pt] (n2) at (axis cs:2,0.2) {}; 
    \node[inner sep=0pt] (n3) at (axis cs:4,0.4) {}; 
    \node[inner sep=0pt] (n4) at (axis cs:6,0.8) {}; 
    \node[inner sep=0pt] (n5) at (axis cs:7,1) {}; 
    \node[inner sep=0pt] (n6) at (axis cs:10,1.8) {}; 
    \node[inner sep=0pt] (n7) at (axis cs:11,1.9) {}; 
    \node[inner sep=0pt] (n8) at (axis cs:12,1.9) {}; 
    \node[inner sep=0pt] (n9) at (axis cs:13,1.9) {};
    \node[inner sep=0pt] (n10) at (axis cs:14,1.9) {};  

\path[name path = GraphCurve, draw, very thick] plot [smooth, very thick] coordinates { (n1) (n2) (n3) (n4) (n5) (n6) (n7) (n8) (n9) (n10) };

\node[inner sep=0pt] (n1) at (axis cs:1,0.8) {}; 
    \node[inner sep=0pt] (n2) at (axis cs:2,0.5) {};
    \node[inner sep=0pt] (n3) at (axis cs:3,0.6) {};
\node[inner sep=0pt] (n5) at (axis cs:5,1) {};  
    \node[inner sep=0pt] (n6) at (axis cs:6,1.35) {}; 
    \node[inner sep=0pt] (n7) at (axis cs:7,1.7) {};
    \node[inner sep=0pt] (n8) at (axis cs:10,2.75) {};
    \node[inner sep=0pt] (n9) at (axis cs:11,3.2) {}; 
    \node[inner sep=0pt] (n10) at (axis cs:12,3.3) {};
    \node[inner sep=0pt] (n11) at (axis cs:13,3.3) {};  
    \node[inner sep=0pt] (n12) at (axis cs:14,3.3) {}; 

\path[name path = GraphCurve, draw] plot [smooth] coordinates { (n1) (n2) (n3) 
(n5) (n6) (n7) (n8) (n9) (n10) (n11) (n12) };   
       
        \addplot [dashed] coordinates {(2, 0) (2, 4)};
        \addplot [dashed] coordinates {(5, 0) (5, 4)};
        \addplot [dashed] coordinates {(7, 0) (7, 4)};
        \addplot [dotted, very thick, blue] coordinates {(11, 0) (11, 4)};
        \addplot [dashed, very thick, red] coordinates {(13, 0) (13, 4)};
        \addplot [dashed,domain=0:15] {3};
        \addplot [dashed,domain=0:15] {1};
        \draw[line width=0.3mm,->] (axis cs:1,1.2) -- (axis cs:1.9,1.2) node[above left, font=\tiny] {} ;

\node[inner sep=0pt] (n1) at (axis cs:1,0.85) {};
        \node[inner sep=0pt] (n2) at (axis cs:1.4,0.75) {};
        \node[inner sep=0pt] (n3) at (axis cs:2,1) {};
        \node[inner sep=0pt] (n4) at (axis cs:3,1.2) {};
        \node[inner sep=0pt] (n5) at (axis cs:4,1.18) {};
        \node[inner sep=0pt] (n6) at (axis cs:6,1.7) {};
        \node[inner sep=0pt] (n7) at (axis cs:11,3.4) {};
        \node[inner sep=0pt] (n8) at (axis cs:14,3.7) {};

\path[name path = GraphCurve, draw, line width = 0.5pt, dash dot] plot [smooth, thick] coordinates { 
        (n1) (n2) (n3) (n4) (n5) (n6) (n7) (n8) };

\end{axis}
\end{tikzpicture}

 }
    \label{fig:overshoot}
  }
  \caption{\protect\subref{fig:overshootDef} Main concepts
          related to the specification of \osh. \protect\subref{fig:overshoot} two signals used to evaluate  property 
    \emph{pOSH}: signal \protect \so satisfies the property, whereas \protect \st violates 
    it.
  }
  \label{fig:overshoot-full}
\end{figure*}

Similarly to the case of rise time specification, given two
signals $s_{\mathit{tr}}$ and $s$, let $P_{\mathit{tr}}$ and $P$ be
two signal-based properties. Property $P_{\mathit{tr}}$ captures the
trigger event defined in terms of the behavior of $s_{\mathit{tr}}$;
property $P$ captures the event of signal $s$ reaching the target
value. An \osh property bounds the \osh of $s$ by a threshold
$\mathit{OI} \in \mathbb{N}$; such a property holds iff the following
\sfo formula evaluates to \emph{true}:
\begin{equation} \label{eq:osh}
\begin{split}
\forall \mathit{st} \in [0,|s_{\mathit{tr}}|)\colon 
\uparrow \sigma^{\mathbb{B}e}_{s_{\mathit{tr}},P_{\mathit{tr}}}(\mathit{st}) \rightarrow
(\exists k \in[\mathit{st},|s|)\colon  \uparrow \sigma^{\mathbb{B}e}_{s,P}(k) \\ \land  \forall i \in [k,k+ \mathit{OI}]: s(i) \le s_{\mathit{max}}  )
\end{split}
\end{equation}

A monotonicity constraint can be added to the formula above in the
same way as done for the case of rise time properties.
An \ush constraint can be expressed
in a similar way, replacing the relational operators with their duals.

As an example, let us consider property \emph{pOSH}: ``If signal
$s_{\mathit{tr}}$ becomes greater than 1, then signal $s$ may
overshoot the target value of 1 by at most 2 within an overshoot
interval of at most \SI{6}{\tu}''.  As we did above for the \emph{pRT}
property, the trigger event in \emph{pOSH} is represented by the data
assertion property $P_{tr}$. The remaining part of the property
represents the effect sub-property.  The corresponding \sfo formula
is the following:
\begin{equation} \label{posh1}
    \begin{split}
      \logiclbl{SFO}{\textit{pOSH}}\quad \forall \mathit{st} \in [0,|s_{\mathit{tr}}|)\colon 
\uparrow \sigma^{\mathbb{B}e}_{s_{\mathit{tr}},P_{\mathit{tr}}}(\mathit{st}) \rightarrow
(\exists k \in[\mathit{st},\mathit{st}+|s|)\colon  \uparrow \sigma^{\mathbb{B}e}_{s,P}(k)  \\ \land  \forall i \in [k,k+6]: s(i) \le 3  )
    \end{split}
    \end{equation}

The variant of property \emph{pOSH-monot} with a monotonicity constraint can be expressed in \sfo as:
  \begin{equation} \label{posh2}
    \begin{split}
     \logiclbl{SFO}{\textit{pOSH-monot}}\quad \forall \mathit{st} \in [0,|s_{\mathit{tr}}|)\colon 
\uparrow \sigma^{\mathbb{B}e}_{s_{\mathit{tr}},P_{\mathit{tr}}}(\mathit{st}) \rightarrow
(\exists k \in[\mathit{st},\mathit{st}+|s|)\colon  \uparrow \sigma^{\mathbb{B}e}_{s,P}(k) \\ \land \forall i \in [k,k+6]: s(i) \le 3  \land  \forall j \in [\mathit{st},\mathit{st}+k)\colon 
\forall j^{\prime} \in (j,\mathit{st}+k] : s(j) < s(j^{\prime}) )
    \end{split}
  \end{equation}

We evaluate property \emph{pOSH} with respect to signal $s$ on the two
signals shown in figure~\ref{fig:overshoot}: \signals.  In the
figure, an arrow at timestamp \SI{2}{\tu} denotes the trigger time
$\mathit{st}$ corresponding to the trigger event captured by property 
$P_{\mathit{tr}}$ for signal $s_{\mathit{tr}}$, drawn with a dash-dotted line
(\protect \frshS). After
this time instant, both $s_1$ and $s_2$ rise reaching the target
value of 1 at time instants \SI{7}{\tu} and \SI{5}{\tu},
respectively. We consider a threshold expressed as a relative value
with respect to the target value; i.e.,
$s_{\mathit{max}}= s_{\mathit{target}} +2=1+2=3$. The maximum allowed
value for the right bound of the overshoot interval for $s_1$
($\SI{7}{\tu}+\mathit{OI}=\SI{7}{\tu}+\SI{6}{\tu}=\SI{13}{\tu}$) is
indicated with a red, vertical dashed line. Similarly, in the case of $s_2$, the
right bound for the overshoot interval
($\SI{5}{\tu}+\mathit{OI}=\SI{5}{\tu}+\SI{6}{\tu}=\SI{11}{\tu}$) is
drawn with a blue, dotted vertical line. Signal $s_1$ satisfies the
property because its overshoot value is below the threshold within the
overshoot interval $[\SI{7}{\tu}, \SI{13}{\tu}]$; signal $s_2$
violates the property as its overshoot value exceeds the threshold
within the overshoot interval $[\SI{5}{\tu}, \SI{11}{\tu}]$.

\subsubsection{Alternative formalizations}
The capability of expressing functional relationship properties in
\stl and \stlstar depends on the possibility, in the chosen language,
of expressing a certain property type on the target signal resulting
from the transforming function.

Similarly, expressing  order relationship
properties in \stl and \stlstar requires that the cause and effect
sub-properties can be expressed in the chosen
formalism. For example, the cause sub-property of property
\emph{pRSH-O} cannot be expressed in \stl; however, it can be expressed in
\stlstar as explained in section~\ref{sec:bumps} (page~\pageref{sec:an}).

The same remarks made above for the general case of order
relationships apply also to the case of rise time and overshoot
properties. In addition, we remark that the specification of such
properties containing a monotonicity constraint requires keeping track
of the signal values seen throughout the rise/overshoot interval; this is not
supported in \stl but can be expressed in \stlstar using the freeze
operator.

 \section{Expressiveness}
\label{sec:expr}

Another challenge in using signal-based temporal properties for
expressing requirements of CPSs is the expressiveness of the
specification languages used for defining such properties.  Starting
from the seminal work on \stl, there have been several proposals of
languages that extend more traditional temporal logics like LTL to
support the specification of signal-based behaviors. For example, in
the previous section, we formally specified all property types
included in our taxonomy using \sfo and, when applicable, also using
\stl and \stlstar.  All these languages have different levels of
expressiveness when it comes to describing certain signal behaviors.

In this section, we summarize and discuss the expressiveness of these
state-of-the-art temporal logics \emph{with respect to the property types
  included in our taxonomy}. We remark that we do not aim to provide a
complete and formal treatment of the expressiveness of these temporal
logics; our main goal is to guide engineers to choose a
specification formalism based on their needs in terms of the property
types to express.

Table~\ref{expMatrix} provides an overview of the expressiveness of
\stl, \stlstar, and \sfo with respect to the property types included
in the taxonomy.  The ``$+$'' and ``$-$'' symbols denote,
respectively, support (or lack of support) for a certain property
type; the ``$\pm$'' symbol indicates that the property type can be
expressed under certain assumptions. Note that in the table, we also
list property subtypes based on a particular feature. For example,
``SPK with amplitude'' indicates a spike property type (see
figure~\ref{fig:taxonomy} for the acronyms) with a constraint on the
amplitude.  In addition, we list as property subtypes (e.g., ``SPK
pre-computed derivatives'') the three definitions to express the predicates for
local extrema for spikes and oscillations (introduced in
section~\ref{sec:bumps}, page~\pageref{def:extrema:punctual}). In the second column, we provide examples of
properties corresponding to the property (sub)type indicated in the
first column.

At a 
glance, the table shows that
\sfo can be used to express all the property types considered in this
paper. \stlstar can be used to express most of the property types
included in our taxonomy, provided that some assumptions are
made (see below). \stl cannot be used to express all the property types; this is
due to the lack of support for referring to signal values at an instant
in which a certain property was satisfied. This limitation impacts on
the specification of properties that constrain signal values at
different time instants, such as spike and oscillation properties.
In the following, we discuss the expressiveness for the various
property types in details, mainly focusing on \stl and \stlstar.

\begin{table}[t]
  \centering
   \caption{Expressiveness of  \stl, \stlstar, and \sfo with respect to the
  property types included in the taxonomy in Fig.~\ref{fig:taxonomy}}
\begin{tabular}{lclccc}
 \toprule
     \multicolumn{2}{l}{\multirow{2}{*}{Property Type}} &\multirow{2}{*}{Example} &
      \multicolumn{2}{c}{\hspace{1cm}Formalism} \\
    & & & \stl & \stlstar & \sfo \\
      \midrule
 \multicolumn{2}{l}{\textbf{Data assertions (DA)}} & \emph{pDA} &$+$ &$+$&$+$   \\ 
\midrule
 \multicolumn{2}{l}{\textbf{Spikes}} &&&&   \\ 
&\multicolumn{1}{r}{SPK with amplitude}& \emph{pSPK1} &$-$ &$+$&$+$   \\ 
&\multicolumn{1}{r}{SPK with slope}& \emph{n/a} &$-$ &$+$&$+$   \\ 
&\multicolumn{1}{r}{SPK with width}& \emph{pSPK1} &$-$ &$\pm$&$+$   \\ 
&\multicolumn{1}{r}{SPK - punctual derivatives}& &$-$ &$-$&$+$   \\ 
&\multicolumn{1}{r}{SPK analytical formulation}&  &$-$ &$+$&$+$   \\ 
&\multicolumn{1}{r}{SPK pre-computed derivatives}&\emph{pSPK3} &$+$ &$+$&$+$   \\ 
\multicolumn{2}{l}{\textbf{Oscillations}} &&&&   \\ 
&\multicolumn{1}{r}{OSC with amplitude}&  \emph{pOSC} &$-$ &  $\pm$&$+$\\   
&\multicolumn{1}{r}{OSC with period}&  \emph{pOSC} &$\pm$ &  $\pm$&$+$\\   
&\multicolumn{1}{r}{OSC punctual derivatives}&   &$-$ &  $-$&$+$\\   
&\multicolumn{1}{r}{OSC analytical formulation}& &$-$ &  $+$&$+$\\   
&\multicolumn{1}{r}{OSC pre-computed derivatives}&   &$+$ &  $+$&$+$\\   
  \midrule
 \multicolumn{2}{l}{\textbf{Relationship between signals}} &&&&   \\ 
&\multicolumn{1}{r}{RSH-F}&  \emph{pRSH-F} &$\pm$&$\pm$&$+$  \\ 
&\multicolumn{1}{r}{RSH-O}&  \emph{pRSH-O} &$\pm$ & $\pm$ &$+$\\  
\multicolumn{2}{l}{\textbf{Transient Behaviors}} &&&&   \\ 
&\multicolumn{1}{r}{RT (FT) with monotonicity}&  \emph{pRT-monot}  &$-$ &$+$ &$+$\\
&\multicolumn{1}{r}{RT (FT)}&  \emph{pRT}   &$+$ &$+$ &$+$\\
&\multicolumn{1}{r}{OSH (USH) with monotonicity}&  \emph{pOSH-monot}  &$-$&$+$&$+$\\  
&\multicolumn{1}{r}{OSH (USH)}&  \emph{pOSH}  &$+$&$+$&$+$\\  
\bottomrule
\end{tabular}
\label{expMatrix}
\end{table}

\paragraph{Data assertion}
All three formalisms can express data assertion properties. This is
expected since the three logics  we have considered were proposed with the goal of 
expressing predicates on a signal value.

\paragraph{Spike}
A formalism supports our definition of spike properties if it allows
for the definition of
\begin{inparaenum}[1)]
  \item two predicates for detecting local extrema, and
  \item constraints on features of the signal shape (e.g., amplitude).
\end{inparaenum}

\stl can be used to define the predicates for detecting local extrema
only through definition~\ref{def:extrema:pre-computed} (as indicated with the ``$+$" mark in the 
table), which
assumes the availability of the first and second order derivatives of
a signal.  Furthermore, it cannot be used to express spike properties that constrain
the spike amplitude or slope, since they refer to signal values at different points in the
signal timeline. 
For example, the  only spike property among those presented in the
previous section that can be expressed in \stl
is \textit{pSPK3}, because it uses pre-computed derivative signals and
does not constrain the spike amplitude. 

\stlstar can be used to define the predicates for detecting local
extrema using two of the definitions we propose
(definition~\ref{def:extrema:analytical} - analytical formulation, and
definition~\ref{def:extrema:pre-computed} - pre-computed derivatives). Furthermore, it can be used to express
constraints on the different features of the signal shape. However, to
do so, one has to assume the knowledge of the signal shape, since it
uses the two components of the width $w_1$ and $w_2$ as defined on
page~\pageref{spike:width}. However, making such an assumption in
practice is not reasonable because typically the shape of a spike is
unknown.  Finally, since \stlstar (and \stl) cannot refer to the value
of the signal at arbitrary time points, properties defined using local
extrema expressed according to definition~\ref{def:extrema:punctual} (punctual derivatives) cannot be specified.

\paragraph{Oscillation}
The expressiveness results in terms of oscillation properties mirror
those for spike properties, since the former  property type can be
seen as an extension of the latter.

\stl can be used to express oscillation properties when the
oscillatory behavior is defined through the sequence of alternating
local extrema, in which the latter are expressed using definition~\ref{def:extrema:pre-computed}. However, as in the case of spike properties, \stl
cannot be used to express constraints on the oscillation amplitude. 

Again, similarly to the case of spike properties, \stlstar supports
definition~\ref{def:extrema:analytical} and
definition~\ref{def:extrema:pre-computed} for defining local extrema  and can be 
used
to express constraints on the different features of an oscillatory
behavior. However, such formulations (including the one based on definition~\ref{def:extrema:pre-computed} for \stl) require to assume
that
\begin{inparaenum}[1)]
\item the oscillation is regular;
\item its period is known a priori.
\end{inparaenum}
These assumptions are required to express distance constraints between
local extrema.  Once again, in practice these assumptions are not
realistic because typically the shape of an oscillatory behavior is
unknown.

\paragraph{Relationship between signals}
Expressing functional relationship properties boils down to expressing
a certain property type on the target signal resulting from the
transforming function. The type of the property
in which the target signal is used ultimately affects (e.g., in case
of a spike property) the expressiveness for this type
of properties. Furthermore, one has to consider whether the
transformed (target) signal is available as a pre-computed signal or as
function of other signals; in the latter case, only SFO supports
function symbols.

A necessary requirement to express order relationship properties is the support
for temporal operators that can capture the \emph{precedence} and
\emph{response} temporal specification
patterns~\cite{dwyer1999:patterns-in-pro}. This is possible in \stl 
and \stlstar through the ``\emph{Until}'' operator and in \sfo by
means of explicit quantification on the time variable.  Another
requirement is that the properties corresponding to the ``cause'' and
``effect'' of an order relationship can be expressed in the chosen
formalism; as shown in Table~\ref{expMatrix}, only SFO fulfills such a
requirement.

\paragraph{Transient behaviors}

Transient behavior properties \emph{without monotonicity constraints}
can be expressed with all three formalisms, assuming the trigger
property can be expressed in the chosen formalism. When a monotonicity
constraint is used (as it is the case in properties \emph{pRT-monot}
and \emph{pOSH-monot}), properties cannot be expressed in \stl
because one cannot compare the value of the signals at two different
time instants.

\paragraph{Monitoring algorithms and tools}
When discussing the expressiveness of specification languages, it is
also important to review the complexity of the corresponding
verification algorithms and the availability of tools implementing
them. Below we discuss the computational complexity of  tools for
\emph{(offline) monitoring} of \stl, \stlstar, and \sfo properties; we focus on
monitoring because it is one of the most used V\&V techniques for CPSs~\cite{bartocci2018specification}.

The complexity of monitoring \stl is $O(k \cdot n)$ where $k$ is the
number of sub-formulae and $n$ is the number of intervals on which the
signal is defined~\cite{maler2004monitoring}.  For \stlstar, the
monitoring complexity is (similarly to \stl) polynomial in the number
of intervals on which the signal is defined and the size of the
syntactic parse tree of the formula; however, it is exponential in the
number of nested freeze operators in the formula~\cite{brim2014stl}.
The monitoring complexity of \sfo is $2^{(m+n)^{2^{O(k+l)}}}$, where
$n$ is the length of the trace, $m$ is the length of the formula, $k$
is the number of quantifiers in the formula, and $l$ is the number of
occurrences of function symbols in the formula; for a fragment of \sfo
in which intervals have bounded duration, the complexity is
$n \cdot 2^{(m+j)^{2^{O(k+l)}}}$, where $n, m, k, l$ are defined as above, and
$j$ is the maximum number of linear segments in the trace during any
time period as long as the sum of the absolute values of all time
constants in the formula~\cite{bakhirkin2018first}. In general, one
can see that the complexity of the monitoring problem becomes harder
for more expressive languages like \stlstar and \sfo.

\rev{In terms of monitoring tools, \stl is supported both by \emph{offline}
tools---such as
\textit{AMT}~\cite{nickovic2018amt,nickovic2007:amt:-a-property} (a
stand-alone GUI tool with qualitative semantics),
\textit{Breach}~\cite{donze2010breach} and
\textit{S-Taliro}~\cite{fainekos2012verification} (two \textit{Matlab}\,\textsuperscript{\tiny\textregistered}
plugins with quantitative semantics)---and by \emph{online} tools,
such as the \texttt{rtamt} library~\cite{nickovic20:_rtamt}, which
automatically generates online monitors with robustness semantics from
\stl specifications.
}
For \stlstar, a prototype implementation in \textit{Matlab} is
mentioned in the original paper~\cite{brim2014stl} but it has not been
made available; furthermore, robustness analysis is supported by an
extension of the \textit{Parasim} tool~\cite{EPTCS125.2}.  No tool
implementation is available for \sfo at the time of writing this
paper.

\rev{Recently, some of the authors have developed
  \emph{SB-TemPsy}~\cite{sbtpASE2020}, a model-driven trace checking
  approach for the property types included in the taxonomy proposed in
  this paper. \emph{SB-TemPsy} includes \emph{SB-TemPsy-DSL}, a
  domain-specific specification language for signal-based properties,
  as well as the corresponding monitoring algorithm and tool, called
  \emph{SBTemPsy-Check}. The complexity of the pattern-specific trace
  checking algorithm implemented in \emph{SBTemPsy-Check} is
  polynomial in the size of the trace for all property types included in
  this taxonomy except for data assertions, for which the complexity
  is linear (in the size of the trace).
}

In conclusion, \emph{with respect to the property types identified in
  our taxonomy}, \stl has limited expressiveness, restricting its
application in practice to simple property types (e.g., data
assertion); nevertheless, it has a good support from a number of tools.
\stlstar is more expressive than \stl provided that some assumptions
(e.g., on the signal shape) are made; however, such assumptions are
impractical. In addition, \stlstar suffers from the limited tool
support. \sfo is the most expressive language for the property types
defined in our taxonomy; however, its application in V\&V activities
is still challenging given the computational complexity of associated
monitoring algorithms and the lack of tools.

 \section{Application to an Industrial Case Study}
\label{sec:casestudy}

We applied our taxonomy of signal-based properties to classify the
requirements specifications of a case study provided by our industrial
partner \emph{LuxSpace Sàrl}\footnote{https://luxspace.lu/}, a system
integrator of micro-satellites.  Our goal is to show
\begin{inparaenum}[(1)]
\item the feasibility of expressing 
requirements specifications of a real-world CPS using the property types included in
our taxonomy;
\item the completeness of our taxonomy, so that all requirements
  specifications of the case study can be defined using the property
  types included in our taxonomy.
\end{inparaenum}

The case study deals with a satellite sub-system called \emph{Attitude
  Determination and Control System} (ADCS), which is responsible for
autonomously controlling the attitude of the satellite, i.e., its orientation with
respect to some reference point.  The ADCS is
mainly composed of sensors (e.g., gyroscope, sun sensors), actuators
(e.g., reaction wheels, magnetic torquer),
and on-board software (e.g., control algorithms).  During flight, the
ADCS can be in four different modes (represented with an enumeration as integer values), which determine the capabilities
of the satellite: \emph{idle} (IDLE), \emph{Safe Mode} (SM), \emph{Normal
  Mode Coarse} (NMC), and \emph{Normal Mode Fine} (NMF); the logic controlling the
switch among modes is encoded in a state machine. Overall, this
sub-system has the typical characteristics of a CPS, with a deep
intertwining of hardware and software.

The documentation of the ADCS includes \theTotal~specifications
written in English.  Two of the authors carefully analyzed these
specifications, discussed and (in some cases) refined them with a
domain expert, and finally classified them using one of the property
types in our taxonomy; the resulting classification was then validated by the domain expert. 
 Table~\ref{dist} shows the number of
specifications classified for each property type (column ``Total
(Main)''); since properties of type functional and order relationship
include additional properties as sub-properties (e.g., the type of the
``cause'' or ``effect'' sub-property in an order relationship), we
indicate their number separately under column ``Total (Sub)''.  From
the table we can conclude that \emph{all requirements
  specifications of the case study could be classified using the
  property types included in our taxonomy}; this is an indication of
the completeness of our taxonomy.  In the following we provide some
insights for each property type, derived from our classification
exercise.
\rev{We remark that the signal names used in the specifications
  correspond to the signals of a FES (Functional Engineering
  Simulator) in Matlab; when possible, we preserved the original
  signal name}.

\begin{table}[tb]
	\centering
	\caption{Distribution of property types in the case study} \label{dist}
	\begin{tabular}{lrr}
          \toprule
          Property Type & Total (Main) & Total (Sub)\\ \midrule
Data assertion  &\theTotalDA &\theSDA\\
		Spike &\theSPK & \theSSPK\\
                Oscillation  &\theOSC &0\\
		Functional relationship & \theFRSH &0\\
                Order relationship  & \theRESP &0\\
                \hspace{1ex} $\vartriangleright$ Fall Time & 0 & 1\\
		  \bottomrule                 
	\end{tabular}
\end{table}

\begin{table}[tb]
	
	\caption{Data assertion properties in the case study} \label{das}
	\begin{tabular}{p{.03\textwidth}p{.94\textwidth}}
	\toprule	ID&Property                    \\ \midrule
			\multicolumn{2}{c}{Untimed Data Assertions}               \\\midrule
\DTLforeach
{cs}
{\pt=Property,\ty=Type}{\ifthenelse{\DTLiseq{\ty}{DA}}
	{P\rownumber  &\pt  \\}
	{}}&\\[-5pt] \midrule
\multicolumn{2}{c}{Time-Constrained Data Assertions}               \\\midrule
\DTLforeach
{cs}
{\pt=Property,\ty=Type}{\ifthenelse{\DTLiseq{\ty}{DAT}}
	{P\rownumber  &\pt  \\}
	{}}\\[-5pt]\bottomrule                 
	\end{tabular}
      \end{table}

\begin{table}[]
	
	\caption{Spike and oscillation properties in the case study} \label{spkosc}
	\begin{tabular}{p{.03\textwidth}p{.9\textwidth}}
			\toprule	ID&Property                    \\ \midrule
		\multicolumn{2}{c}{Spike}               \\\midrule
		\DTLforeach
{cs}
		{\pt=Property,\ty=Type}{\ifthenelse{\DTLiseq{\ty}{SPK}}
			{P\rownumber  &\pt  \\}
			{}}&\\[-5pt] \midrule
		\multicolumn{2}{c}{Oscillation}               \\\midrule
		\DTLforeach
{cs}
		{\pt=Property,\ty=Type}{\ifthenelse{\DTLiseq{\ty}{OSC}}
			{P\rownumber  &\pt  \\}
			{}}\\[-5pt]\bottomrule                 
	\end{tabular}
\end{table}      

\begin{table}[]
	
	\caption{Properties of type ``functional relationship''
           in the case study} \label{frsh}
	\begin{tabular}{p{.03\textwidth}p{.8\textwidth}p{.1\textwidth}}
	\toprule	ID&Property&Subtype                 \\ \midrule
		\DTLforeach
{cs}
		{\pt=Property,\ty=Type,\sty=SubType}{\ifthenelse{\DTLiseq{\ty}{FRSH}}
			{P\rownumber  &\pt  &\sty \\}
			{}}\\[-5pt]\bottomrule                 
	\end{tabular}
     
\end{table}

\begin{table}[]
	
	\caption{Properties of type ``order relationship'' in the case study} \label{resp}
	\begin{tabular}{p{.03\textwidth}p{.84\textwidth}p{.15\textwidth}}
		\toprule	ID&Property&SubType                 \\ \midrule
		\DTLforeach
{cs}
		{\pt=Property,\ty=Type,\sty=SubType}{\ifthenelse{\DTLiseq{\ty}{RESP}}
			{P\rownumber  &\pt&\sty  \\}
			{}}\\[-5pt]\bottomrule                 
	\end{tabular}
\end{table}

\paragraph{Data assertion properties (Table~\ref{das})}
This is the most represented category, if one considers the
sub-properties included in the properties of type functional and order
relationship.  The \numberstring{DAT} time-constrained data assertions
show different interval types used in such properties. For example, in
property P6 both boundaries of the interval are explicitly
mentioned. In property P5, only the left boundary is explicitly
indicated (with the expression ``Starting from $\SI{2000}{s}$''),
whereas the right boundary is implicit and is assumed to be the end of
the (finite) signal. Finally, in property P7 the interval is singular
(i.e., the two boundaries coincide) and corresponds to a single time
point (as in the expression ``At \SI{2000}{s}''). To express the
latter using one of the logic-based formalizations illustrated above,
which does not allow singular intervals (e.g., \stl), one has to
rewrite a singular interval $[a,a]$ as $[a-\epsilon, a+\epsilon]$, for
a small $\epsilon > 0$.

We remark that time-constrained data assertions can be used to specify
system-level properties such as system stabilization.  For example,
property P5 was originally expressed as ``The stabilization time of
signal \emph{pointing\_error}, when stabilizing below \SI{2}{ degrees},
shall be under \SI{2000}{s}''; through the interaction with the domain
expert, we further refined it into the version shown in
Table~\ref{das}.  The refinement step was straightforward and
consisted of rewriting the system-level property (i.e., stabilization)
into a low-level one (of type ``data assertion''), by expanding the
definitions of domain concepts.

\paragraph{Spike and oscillation properties (Table~\ref{spkosc})}
We identified \numberstring{SPK}~spike property (P8); furthermore an
additional spike property is included in an order relationship
property (P41).  Both spike properties refer to one feature
(``width'').

We also identified \numberstring{OSC} oscillation property (P9), which
refers to the ``period'' feature.  Initially, the property was defined
in the frequency domain (which we did not discuss in this
paper). After discussing it with the domain expert, we converted it
into a property defined on the time domain by changing the
corresponding constraint. This type of transformation is straightforward as it only
requires to convert the units in the property (e.g., a 
\SI{100}{\hertz} frequency is converted into a \SI{0.01}{\s} period).

All three properties include an observation interval. In properties P8
and P9, it is defined explicitly using absolute time boundaries (with
the expression ``between \SI{2000}{s} and \SI{7400}{s}''). In property
P41, the observation interval is defined through the event
representing the left boundary (denoted with ``the value of signal
\emph{pointing\_error} after \SI{16200}{\s} goes below the pointing
accuracy threshold of \ang{2}'') and the duration (\SI{5400}{s})
representing the right boundary.

\paragraph{Functional relationship properties (Table~\ref{frsh})}
These properties were expressed using several signal transforming
functions, such as modulus (P10--P20), vector elements sum (P21),
angular difference (P22--P23), scalar difference (P24--P26), and
differentiation (P26). Notice that property P26 contains nested
applications of signal transforming functions (i.e., the second
operand of the scalar difference is the result of the application of
the derivative).

In all properties, the signal resulting from the application of the
transforming function is used in a data assertion property (see column
``Subtype'' in Table~\ref{frsh}\footnote{\rev{See 
Figure~\ref{fig:taxonomy} for the acronyms used in column ``Subtype''
of Tables~\ref{frsh} and~\ref{resp}}.}).

\paragraph{Order relationship properties (Table~\ref{resp})}

All the order relationship properties we classified were instances of
the ``response'' pattern (see section~\ref{sec:order-relationships});
we did not encounter any instance of the ``precedence'' pattern.

Some properties (P35--P38) contain nested properties of type
``order relationship'', meaning that the effect of the response
pattern is represented by another property of type ``order
relationship''. For example, in property P36, the top-level response
property has ``the value of signal \emph{currentADCSMode} is equal to
\emph{NF}'' as cause and ``if the value of signal \emph{RWs\_command}
becomes greater than 0, then the value of signal
\emph{pointing\_error} shall be less than \ang{2}'' as effect. The
latter is another response property that can be further decomposed
into the cause ``the value of signal \emph{RWs\_command} becomes
greater than 0'' and the effect ``the value of signal
\emph{pointing\_error} shall be less than \ang{2}''.
The same group of 
properties also includes a temporal distance constraint (expressed with
``within'') as part of the nested response property.

As shown in column ``Subtype'' of Table~\ref{resp},
all the sub-properties used as ``cause'' and the vast majority
of the sub-properties used as ``effect'' were data
assertions.  For example, in property P27 both the cause ``the value
of signal \emph{not\_Eclipse} is equal to 0'' and the effect ``the
value of signal \emph{sun\_currents} shall be equal to 0'' are data
assertions. This is reflected in the third column of Table~\ref{resp},
with the notation ``DA-DA''.

Regarding transient behaviors, we only encountered \numberstring{SFT}
property of type ``fall time'', used as effect of the response
property P30. Other types of properties (e.g., rise time, overshoot)
were not present in this case study.  \vspace{1ex}

Summing up, through this case study we have shown the
\emph{feasibility} of expressing requirements specifications of a
real-world CPS using the property types included in our taxonomy. In
the vast majority of the cases, the mapping from a specification written in
English to its corresponding property type defined in the taxonomy was
straightforward. In two cases, the specifications had to be refined,
either by expressing a system-level property into a low-level one (e.g.,
stabilization being expressed as a (time-constrained) data assertion)
or by converting a property defined in the frequency domain into the
corresponding one defined in the time domain (e.g., in the case of an oscillation
property); both types of refinement were simple and intuitive (with
the help of a domain expert). Furthermore, the case study has shown
the \emph{completeness} of our taxonomy, since all requirements
specifications of the case study could be classified using the
property types included in our taxonomy.

Guided by the mapping to one of the property types included in our
taxonomy, and by means of the formalization presented in
section~\ref{sec:taxonomy}, an engineer can obtain a formal
specification of a property (e.g, in \sfo), which can then be
used in the context of V\&V activities (e.g., as test oracle).

\rev{\paragraph{Threats to validity} The results regarding the 
  \emph{feasibility} of expressing requirements specifications of a
  real-world CPS and the \emph{completeness} of our taxonomy, have been
  obtained through one large industrial case study, involving a domain expert; 
  this is a threat to the generalization of the results. 
  We tried to mitigate this threat
  by selecting a case study with a rich set of requirements extracted
  from the documentation of a complex, production-grade system. Such requirements are
  representative, in many ways, of those defined in the
  satellite and other cyber-physical domains. Nevertheless, some CPS
  domains (e.g., healthcare) may have specific types of requirements
  (e.g., supporting frequency-domain in the temporal specifications),
  which could lead to different results.
}

 \section{\rev{Applications}}
\label{sec:discussion}

In this section, we discuss how the main contributions of the
papers can support the research community and practitioners working
in the CPS domain.

\paragraph{Application of the taxonomy} The taxonomy of signal-based
temporal properties can be used by researchers to \emph{design new
  specification languages}, whose constructs can be directly mapped to
the main property types identified in the taxonomy.  This type of
impact has been already observed for similar contributions in the
literature, such as the seminal work of
\citet{dwyer1999:patterns-in-pro} on temporal specification patterns,
which has influenced the design of many domain-specific languages for
temporal specifications (e.g., Temporal OCL~\cite{kanso2013temporal},
OCLR~\cite{dbb-ecmfa2014}, VISPEC - graphical
formalism~\cite{7353863}, TemPsy~\cite{dou2017model}, ProMoboBox -
property language~\cite{8419296},
FRETISH~\cite{giannakopoulou20:_gener_formal_requir_struc_natur_languag}),
and the work on service provisioning patterns~\cite{bianculli2012specification}, which has led to the
design of new specification languages and
tools~\cite{bianculli13:tale-solois,bbgks-fase2014,bgs-sefm2014,bgks-soca2014,
  boufaied19:_model_approac_trace_check_tempor_proper_aggreg}.
For instance, as mentioned in section~\ref{sec:expr}, some of the
authors have already developed
  \emph{SB-TemPsy-DSL}~\cite{sbtpASE2020}, a domain-specific
  specification language for signal-based properties based on the
  taxonomy proposed in this paper.

The property types included in our taxonomy can also be used to
\emph{assess the expressiveness} of existing languages, in a way
similar to what we have done in section~\ref{sec:expr}. By doing so,
researchers can identify expressiveness gaps in existing languages,
which could then be extended to support specific constructs. For instance,
the motivating example for the development of
\stlstar~\cite{brim2014stl} was the impossibility of expressing
oscillatory behaviors in \stl.

Furthermore, practitioners can use the taxonomy as a reference guide
to systematically \emph{identify and
characterize signal behaviors}, so that the latter can be defined precisely
and used correctly during the development process of CPSs (e.g., when
defining system requirements or test oracles).

\paragraph{Application of the logic-based characterization}
Researchers can leverage the logic-based characterization of
the property types included in our taxonomy to
\emph{define the formal semantics} of the constructs of a
new language, which has been inspired by the taxonomy itself. In this
sense, the logic-based characterization can \emph{guide the
  implementation} of the core, pattern-specific algorithms of a
verification tool, which can be used for checking properties expressed
in a language containing constructs derived from the property types included in our
taxonomy.

For instance, the formal semantics of the aforementioned
\emph{SB-TemPsy-DSL} language and the corresponding trace checking
algorithm implemented in  \emph{SBTemPsy-Check}~\cite{sbtpASE2020}
have been developed based on the logic-based characterization
introduced in this paper.

\paragraph{Expressiveness results}
The expressiveness results of state-of-the-art temporal logics with
respect to the property types   included in our taxonomy, presented
in section~\ref{sec:expr}, can be used by practitioners to carefully \emph{select the language to use} for defining
signal-based properties, based on the type of requirements they are
going to define, the expressiveness of the candidate specification
language(s), and the availability of suitable tools.

 \section{Related Work}
\label{sec:related-work}

To the best of our knowledge, this is the first paper that presents a
comprehensive taxonomy of signal-based temporal properties 
describing signal behaviors in the CPS domain. The closest work is the
taxonomy of automotive controller behaviors presented
in~\cite{kapinski2016st}, in which behaviors are captured in ST-Lib, a
catalogue of formal requirements written in \stl. Although the ST-Lib
catalogue contains several types of signal-based temporal properties
(e.g., spike, overshoot, rise time), the treatment of some property
types is limited (e.g., oscillatory behaviors are only discussed for
the case of short-period behaviors, i.e., ringing). Furthermore, as we
have shown in section~\ref{sec:bumps}, the formalization of spike
properties proposed in~\cite{kapinski2016st} has some limitations.  A
specific type of signal-based temporal properties (i.e., oscillations)
is discussed in~\cite{brim2014stl} and used as a motivation for
introducing \stlstar.

Similarly to what we did in section~\ref{sec:taxonomy}, most of the
papers dealing with the specification or verification of signal-based
temporal properties also include examples of such properties written
using a specific temporal logic.  We systematically reviewed the example
properties used throughout all the papers dealing with specification,
verification, and monitoring of CPS, cited in a recent survey on these
topics~\cite{bartocci2018specification}; we excluded papers using
spatio-temporal and frequency domain properties since they are out of
the scope of this work. 
Table~\ref{tab:relatedWork} shows, for each
of the reviewed papers, the property types (from our taxonomy) to
which the examples included in the paper correspond, as well as the
temporal logic used for their specification; treatment or lack thereof
of a property type is denoted by a ``+'' or ``-'' symbol,
respectively.  One can see that data assertion and relationship
between signals are the most common property types covered in the
literature, whereas transient behaviors (e.g., \rt, \osh) properties are
the least common; spike and oscillation properties have a similar coverage.

To summarize, we propose in this paper the first comprehensive
taxonomy of signal-based properties, formalized in a consistent and
precise manner, which accounts for all reported property types in the
literature.

\setcounter{counterDA}{0}
\newcommand{\DAOcc}{\ifnum\pdfstrcmp{\DA}{+}=0
    \stepcounter{counterDA}\fi
}

\setcounter{counterSpk}{0}
\newcommand{\SPKOcc}{\ifnum\pdfstrcmp{\SPK}{+}=0
    \stepcounter{counterSpk}\fi
}

\setcounter{counterOsc}{0}
\newcommand{\OscOcc}{\ifnum\pdfstrcmp{\OSC}{+}=0
    \stepcounter{counterOsc}\fi
}

\setcounter{counterRFT}{0}
\newcommand{\rftOcc}{\ifnum\pdfstrcmp{\RTFT}{+}=0
    \stepcounter{counterRFT}\fi
}

\setcounter{counterOUSH}{0}
\newcommand{\OUOcc}{\ifnum\pdfstrcmp{\OSHUSH}{+}=0
    \stepcounter{counterOUSH}\fi
}

\setcounter{counterRSH}{0}
\newcommand{\RSHOcc}{\ifnum\pdfstrcmp{\RSH}{+}=0
    \stepcounter{counterRSH}\fi
}

\afterpage{\begin{longtable}{ccccccccc} 
\caption{Coverage of property types (from our taxonomy, see figure~\ref{fig:taxonomy} for 
acronyms) in 
  example specifications from the
  literature.\label{tab:relatedWork}}\\
 \toprule\bfseries Reference & \bfseries Formalism & \bfseries DA & \bfseries SPK & \bfseries RT (FT) & \bfseries OSH (USH) & \bfseries OSC & \bfseries RSH\\ \midrule
\endfirsthead
\caption[]{(continued)}\\
  \toprule\bfseries Reference & \bfseries Formalism & \bfseries DA & \bfseries SPK & \bfseries RT (FT) & \bfseries OSH (USH) & \bfseries OSC & \bfseries RSH\\ \midrule\endhead
  \begin{filecontents*}{properties.csv}
Reference,Formalism,DA,SPK,RTFT,OSHUSH,OSC,RSH
maler2004monitoring,STL,+,-,-,-,-,+
maler2013monitoring,STL/PSL,+,-,+,-,-,+
kapinski2016st,PSTL,+,+,+,+,-,+
brim2014stl,STL*,+,-,-,-,+,-
5675195,MTL,+,-,-,-,-,+
abbas2013probabilistic,MTL,+,-,-,-,-,+
abbas2017quantitative,CTMTL,+,-,-,-,-,-
abbas2017quantitative,XCTL,+,-,-,-,-,+
abbas2017quantitative,CLTL,+,-,-,-,-,+
10.1007/978-3-642-35632-2_18,STL,+,+,-,-,-,+
akazaki2015time,STL,+,-,-,-,-,+
akazaki2015time,AVSTL,+,-,-,-,-,+
bartocci2013temporal,STL,+,-,-,-,-,+
bartocci2015system,STL,+,+,-,-,+,+
bartocci2009model,KSL,+,-,-,-,-,+
10.1007/978-3-319-22264-6_6,MITL,+,-,-,-,-,-
10.1007/978-3-662-45231-8_30,MITL,+,-,-,-,-,-
Deshmukh2017,STL,+,-,-,-,+,+
10.1007/978-3-319-23820-3_4,STL,+,-,-,-,+,+
7318257,MTL,+,-,-,-,-,-
10.1007/978-3-642-15297-9_9,STL,+,-,-,-,-,+
10.1007/978-3-319-17524-9_10,STL,+,-,-,-,-,-
ferre2016,TRE,+,+,-,-,-,+
7353863,STL,+,-,-,-,-,+
QuantSTL,STL,+,-,-,-,-,+
6986132,STL,+,-,-,-,-,+
6986132,PSTL,+,-,-,-,-,-
cameron2015towards,MTL,+,-,-,-,-,+
Nghiem:2010:MTF:1755952.1755983,MTL,+,-,-,-,-,+
dokhanchi2014line,MTL,+,-,-,-,-,+
dokhanchi2015metric,MITL,+,-,-,-,+,+
dokhanchi2015metric,STL,+,-,-,-,-,+
Nguyen:2016:AMP:2895233.2895493,STL,+,+,-,-,-,+
jin2015mining,STL,+,+,-,-,-,+
jin2015mining,PSTL,+,+,-,+,-,+
donze2010breach,MITL,+,-,-,-,-,+
donze2011robustness,STL,+,-,-,-,+,+
eisner2007practical,PSL,+,-,-,-,-,+
fainekos2006robustness,MTL,+,-,-,-,-,+
fainekos2006robustness,MITL,+,-,-,-,-,-
fainekos2006robustness,MTL,+,-,-,-,-,+
fainekos2012verification,MTL,+,-,-,-,-,+
ferrere2015measuring,TRE,+,+,-,-,-,-
hoxha2014towards,PMTL,+,-,-,-,-,-
hoxha2014towards,MTL,+,-,-,-,-,+
hoxha2018mining,MTL,+,-,-,-,-,+
hoxha2018mining,PMTL,+,-,-,-,-,+
jakvsic2015signal,STL,+,-,-,-,+,+
kane2015runtime,BMTL,+,-,-,-,-,+
maler2008checking,STL,+,-,-,-,-,+
nickovic2008checking,MITL,+,-,-,-,-,+
nickovic2008checking,STL,+,-,-,-,-,+
nickovic2008checking,STL/PSL,+,-,+,-,-,+
nickovic2008checking,MTL-B,+,-,-,-,-,+
nickovic2007:amt:-a-property,STL/PSL,+,-,-,-,-,+
pajic2014model,CTL,+,-,-,-,-,+
rizk2008continuous,LTL(R),+,-,-,-,+,-
rizk2008continuous,QFLTL(R),+,-,-,-,+,-
sankaranarayanan2012falsification,MTL,+,-,-,-,-,+
selyunin2017runtime,STL,+,+,+,-,-,-
selyunin2017runtime,TRE,+,+,+,-,-,-
stoma2013stl,STL,+,-,-,-,-,-
ulustimed,TRE,+,-,-,-,+,+
yang2012querying,PMTL,+,-,-,-,-,-
\end{filecontents*}
\csvreader[/csv/head to column names=true]{properties.csv}{
  1=\one, 2=\two, 3=\three, 4=\four,
  5=\five, 6=\six, 7=\seven, 8=\eight}
{\\ \DAOcc \SPKOcc \OscOcc \rftOcc \OUOcc \RSHOcc \cite{\one} & \two & \three & \four & \five & \six & \seven & \eight} \\ 
\midrule
  \multicolumn{2}{c}{\bfseries Total} & \thecounterDA & \thecounterSpk & \thecounterRFT & \thecounterOUSH & \thecounterOsc
    & \thecounterRSH \\
      \bottomrule
\end{longtable}
}

 \section{Conclusion and Future Work}
\label{sec:concl-future-work}

Requirements of cyber-physical systems are usually expressed using
signal-based temporal properties, which characterize the expected
behaviors of input and output signals processed by sensors and actuators. Expressing such requirements is 
challenging
because of the many ways to characterize a signal behavior (e.g.,
using certain features). To avoid ambiguous or inconsistent
specifications, we argue that engineers need precise definitions of
such features and proper guidelines for selecting the features most
appropriate in a certain context.  Furthermore, given the broad variation in expressiveness of the 
specification languages used for defining
signal-based temporal properties, our experience indicates that engineers need guidance for selecting the 
most appropriate specification language,
based on the type of requirements they are going to define and the
expressiveness of each language.

To tackle these challenges, in this paper we have presented a taxonomy
of the most common types of signal-based temporal properties,
accompanied by a comprehensive and detailed description of
signal-based behaviors and their precise characterization in terms of
a temporal logic (\sfo). Engineers can rely on such characterization
to derive---from informal requirements specifications---formal
specifications to be used in various V\&V activities.  

Furthermore, we have reviewed the expressiveness of state-of-the-art
signal-based temporal logics (i.e., \stl, \stlstar, \sfo) in terms of
the property types identified in the taxonomy, while also taking into
account the complexity of monitoring algorithms and the availability
of the corresponding tools. Our analysis indicates that \sfo \emph{is the
most expressive language for the property types of our taxonomy};
however, the application of \sfo in V\&V activities is still
challenging given the computational complexity of the corresponding
monitoring algorithm and the lack of tools.

We have also applied our taxonomy to classify the requirement
specifications of an industrial case study in the aerospace
domain. The case study has shown the feasibility of expressing
requirements specifications of a real-world CPS using the property
types included in our taxonomy, and has provided evidence of the
completeness of our taxonomy.

As part of future work, we plan to assess the expressiveness of other temporal
logics (such as SCL - Signal Convolution
Logic~\cite{silvetti2018signal}, the extension of STL proposed
in~\cite{bakhirkin2019specification}, and the \emph{shape expressions}
formalism~\cite{nivckovic2019shape}) in terms of the property types
identified in our taxonomy.  Moreover, we plan to collect feedback
from practitioners (i.e., software and system engineers) to assess the
usefulness of our taxonomy and of the proposed property formalizations
for the verification of CPS.

 \section*{Acknowledgments}
This work has received funding from the European Research Council
under the European Union's Horizon 2020 research and innovation
programme (grant agreement No~694277), from the University of
Luxembourg (grant ``MOVIDA''), and from the NSERC Discovery and Canada Research Chair programmes. We also wish to thank Claudio Menghi
and Dejan Ni{\v{c}}kovi{\'c} for their feedback on the paper.

\bibliographystyle{elsarticle-num-names}

\end{document}